\documentclass[fleqn,usenatbib]{mnras}

\usepackage{newtxtext,newtxmath}

\usepackage[T1]{fontenc}

\DeclareRobustCommand{\VAN}[3]{#2}
\let\VANthebibliography\thebibliography
\def\thebibliography{\DeclareRobustCommand{\VAN}[3]{##3}\VANthebibliography}

\usepackage{graphicx}	
\usepackage{adjustbox}  
\usepackage{amsmath}	
\usepackage{wasysym}    
\usepackage{bm}         
\usepackage{xspace}
\usepackage{xcolor}
\usepackage{multirow}
\makeatletter 
  \patchcmd{\NAT@citex}
    {\@citea\NAT@hyper@{%
      \NAT@nmfmt{\NAT@nm}%
      \hyper@natlinkbreak{\NAT@aysep\NAT@spacechar}{\@citeb\@extra@b@citeb}%
      \NAT@date}}
    {\@citea\NAT@nmfmt{\NAT@nm}%
    \NAT@aysep\NAT@spacechar\NAT@hyper@{\NAT@date}}{}{}

  \patchcmd{\NAT@citex}
    {\@citea\NAT@hyper@{%
      \NAT@nmfmt{\NAT@nm}%
      \hyper@natlinkbreak{\NAT@spacechar\NAT@@open\if*#1*\else#1\NAT@spacechar\fi}%
        {\@citeb\@extra@b@citeb}%
      \NAT@date}}
    {\@citea\NAT@nmfmt{\NAT@nm}%
    \NAT@spacechar\NAT@@open\if*#1*\else#1\NAT@spacechar\fi\NAT@hyper@{\NAT@date}}
    {}{}
\makeatother

\newtoggle{referee}            %
\toggletrue{referee}         %
\iftoggle{referee}{
    \usepackage{color,soul}    %
}{
}                              %

\newcommand\HI{\ion{H}{I}\xspace} 

\newcommand\orcid[1]{\href{http://orcid.org/#1}{\adjustbox{trim={-.15\width} {0\height} {-.15\width} {0\height},clip}{\includegraphics[height=12pt]{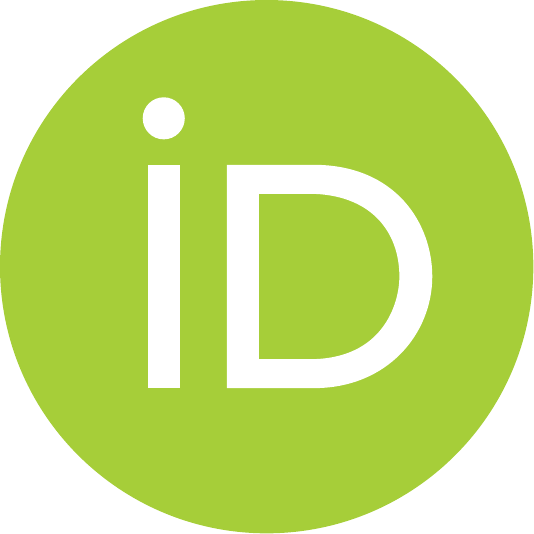}}}}

\title[Ly$\alpha$ radiative transfer in expanding media]{Lyman-$\boldsymbol{\alpha}$ resonant-line radiative transfer in expanding media}

\author[A.~Smith et al.]{%
Aaron~Smith\orcid{0000-0002-2838-9033},$^{1}$\thanks{E-mail: \href{mailto:asmith@utdallas.edu}{asmith@utdallas.edu} (AS); \href{mailto:Kevin.Lorinc@UTDallas.edu}{Kevin.Lorinc@UTDallas.edu} (KL)}
Kevin~Lorinc\orcid{0009-0005-3827-8774},$^{1}$
Olof~Nebrin\orcid{0000-0003-3877-360X}$^{2}$
and
Bing-Xin~Lao\orcid{0000-0002-6320-185X}$^{3}$
\\%
\\%
$^{1}$Department of Physics, The University of Texas at Dallas, Richardson, Texas 75080, USA\\%
$^{2}$Department of Astronomy \& Oskar Klein Centre for Cosmoparticle Physics, AlbaNova, Stockholm University, SE-106 91 Stockholm, Sweden\\%
$^{3}$Department of Physics, Princeton University, Princeton, NJ 08544, USA%
}

\date{Accepted XXX. Received YYY; in original form ZZZ}

\pubyear{2025}

\begin{document}
\label{firstpage}
\pagerange{\pageref{firstpage}--\pageref{lastpage}}
\maketitle

\begin{abstract}
  The Lyman~$\alpha$ (Ly$\alpha$) line of atomic hydrogen encodes crucial information about both the intrinsic sources and surrounding environments of star-forming regions throughout the Universe. Due to the complexity of resonant scattering, analytic solutions remain scarce, with most studies focusing on idealized, static configurations. However, observations of Ly$\alpha$ emitting galaxies consistently reveal signatures of outflows, imprinted through red-peak dominance in spectral line profiles. We derive novel analytic solutions for resonant-line radiative transfer in moving media, specifically homologous-like cloud expansion and unbounded cosmological flows, which capture the main physics of velocity gradients. Our framework accounts for spatial diffusion, partial frequency redistribution, and kinematic effects. To validate these analytic solutions and identify regimes where diffusion-based assumptions hold, we introduce a robust Gridless Monte Carlo Radiative Transfer (GMCRT) method. By integrating optical depths exactly in the comoving frame, GMCRT updates photon frequencies continuously to account for Doppler shifts induced by velocity gradients. We demonstrate excellent consistency between GMCRT and our analytic solutions in regimes where diffusion approximations apply. At higher velocities or lower optical depths, discrepancies highlight the limitations of simplified formalisms. We also provide scaling relations for a point source in a cloud with a maximum-to-thermal velocity ratio $\beta = V_\text{max} / v_\text{th}$, as modifying the standard dependence on line centre optical depth of $\propto (a\tau_0)^{1/3}$ by additional factors, e.g. characteristic escape frequency scale as $x_\text{esc} \propto \beta^{1/3}$, force multipliers as $M_\text{F} \propto \beta^{-1/3}$, and trapping time as $t_\text{trap} \propto \beta^{-2/3}$. Our work complements numerical simulations by improving physical intuition about nonstatic environments when interpreting Ly$\alpha$ observations and guiding subgrid prescriptions in galaxy formation models.
\end{abstract}




\section{Introduction}
\label{sec:intro}
The Lyman~$\alpha$ (Ly$\alpha$) transition of neutral hydrogen is a powerful probe of the physical conditions and dynamical processes in galaxies across cosmic time \citep{Partridge1967}. The main advantage of Ly$\alpha$ arises from the intrinsic brightness of the line and the ubiquitous presence of hydrogen in astrophysical and cosmological environments, encoding signatures from young star-forming regions to the intergalactic medium. However, interpreting observations for understanding galaxy formation and evolution has proven challenging due to the complex nature of the resonant scattering of Ly$\alpha$ photons in optically thick environments \citep{Dijkstra2014}. The goal of this paper is to further develop the rich theory of resonant-line radiative transfer in moving media.

Ly$\alpha$ photons experience repeated scattering events, with the large scattering cross-section in the wings of the line profile leading to a double diffusion process in frequency and space. Valuable insights into the physical mechanisms regulating Ly$\alpha$ escape can be obtained from back of the envelope calculations \citep[e.g][]{Osterbrock1962,Adams1972,Adams1975,HansenOh2006}, while more rigorous understanding comes from analytic solutions derived under simplifying assumptions. These idealized models generally rely on the Fokker-Planck approximation to simplify local frequency diffusion in the wings of the line profile compared to a full treatment of partial redistribution \citep{Unno1952,Hummer1962,Rybicki1994}. Important milestones include diffusion-based solutions for uniform slab, sphere, and cube geometries \citep{Harrington1973,Neufeld1990,Dijkstra2006,Tasitsiomi2006b}, extensions for cosmological flows \citep{LoebRybicki1999,Higgins2012,Smith2017}, and explorations of boundary conditions and time-dependence \citep{GeWise2017,Smith2018,Tomaselli2021,McClellan2022}. We now have closed-form solutions for the internal radiation field and a modern concordance framework for further generalizations \citep{LaoSmith2020,Seon2020,Nebrin2024}. Building on these previous works, we generalize known analytic solutions to include bulk velocity fields, particularly homologous-like expansion, which are known to increase the red-to-blue flux ratio in otherwise symmetric double-peaked profiles emerging from static clouds.

The interplay between Ly$\alpha$ scattering and gas motion has been recognized for decades. Early explorations in the context of planetary nebulae determined that expansion induces Doppler shifts that lower the Ly$\alpha$ opacity to significantly reduce scattering and photon trapping \citep[e.g.][]{Chandrasekhar1945}. Velocity effects have also been studied to understand a wide range of other phenomena, including the Fermi acceleration of Ly$\alpha$ photons by shocks \citep[e.g.][]{Neufeld1988,Chung2016} and the formation of diffuse Ly$\alpha$ haloes prior to cosmic reionization \citep[e.g.][]{LoebRybicki1999,Padmanabhan2024}. Additional processes such as (micro)turbulence, atomic recoil, and inhomogeneous or anisotropic velocity fields further complicate Ly$\alpha$ radiation transport \citep[e.g.][]{Field1959,Gray1973,Basko1981,Rybicki2006,Nebrin2024}. Modern hydrodynamical simulations also reveal that complex configurations of turbulent and columnated outflows can significantly influence emergent Ly$\alpha$ spectra \citep[e.g.][]{Kimm2019,Kimm2022,Kakiichi2021}.

It is often insightful to isolate the role of outflows in settings that still capture the underlying physics, e.g. idealized expanding clouds can be used to validate Ly$\alpha$ radiative transfer codes \citep[e.g.][]{Zheng2002,Dijkstra2006,Verhamme2006,Tasitsiomi2006,Semelin2007,Laursen2009,Yajima2012,GronkeShell2015,Smith2015}. More complex geometric setups, including clumpy or anisotropic media, provide further intuition on how velocity gradients and inhomogeneities affect Ly$\alpha$ escape \citep[e.g.][]{Laursen2013,Behrens2014,Zheng2014,GronkeDijkstra2016,Song2020}. In all these cases, analytic progress provides benchmarks, guides numerical efforts, and allows greater connections with simulations and observational data.

In this paper, we present a follow-up study of \citet{LaoSmith2020} where we derived generalized solutions for the physically motivated case of power-law density and emissivity profiles. We intend to be as thorough and general as possible while highlighting the most salient features of our results. In the spirit of \citet{LaoSmith2020}, we introduce a Gridless Monte Carlo Radiative Transfer (GMCRT) technique for the exact integration of the optical depth along each photon path, fully accounting for continuous frequency changes induced by Doppler shifts \citep{Smith2022}. This accurately captures changes on sub-grid scales and avoids numerical issues with line skipping when generalizing beyond the Sobolev approximation. Comparisons between our analytic results and these GMCRT simulations help determine the domain of accuracy for the diffusion approximation and other assumptions.

This paper is also a companion study with \citet{Nebrin2024}, which is more ambitious in scope and focuses more broadly on the importance of Ly$\alpha$ radiation pressure feedback in the early Universe. \citet{Nebrin2024} found that Ly$\alpha$ photons can inject up to $\sim 100$ times more momentum than UV continuum radiation and stellar winds \citep[see also][]{Smith2017,Kimm2018}. This dominance becomes especially significant in dust-poor environments, common during the formation of the first stars and galaxies. \citet{Nebrin2024} presented generalized analytic Ly$\alpha$ radiative transfer solutions including multiple suppression mechanisms such as continuum absorption, velocity gradients, destruction mechanisms (such as $2p \rightarrow 2s$ transitions), turbulence, and atomic recoil. Our work is complementary, with specific attention and discussion related to expanding media solutions and numerical methods.

This paper is organized as follows. In Section~\ref{sec:Lyman-alpha radiation transport}, we review the fundamentals of Ly$\alpha$ radiative transfer and outline the framework for including bulk expansion in back-of-the-envelope calculations and formal partial differential equations. In Section~\ref{sec:homologous_expansion}, we derive new analytic solutions for a spherically symmetric, homogeneous cloud undergoing homologous-like expansion. In Section~\ref{sec:cosmological_expansion}, we consider the infinite, boundary-free case relevant to cosmological expansion and introduce finite-temperature effects. Previously solutions for zero-temperature, coherent scattering were derived by \citet{LoebRybicki1999}. In Section~\ref{sec:Gridless_MCRT}, we compare our analytical solutions with numerical simulations that employ the GMCRT method. In Section~\ref{sec:summary}, we present a summary of our findings and comment on when analytic approximations succeed and where more complex numerical treatments are necessary.

\section{Resonant-line radiation transport}
\label{sec:Lyman-alpha radiation transport}
The specific intensity $I_\nu(\bmath{r},\bmath{n},t)$ characterizes the radiation field taking into account frequency $\nu$, spatial position $\bmath{r}$, propagation direction $\bmath{n}$, and time $t$. For a slowly moving medium with velocity $|\bmath{v}| \ll c$, the general radiative transfer equation in the comoving frame is
\begin{align} \label{eq:general}
  \frac{1}{c} \frac{\partial I_\nu}{\partial t}\ +\ &\bmath{n} \bmath{\cdot} \bmath{\nabla} I_\nu - \bmath{n} \bmath{\cdot} \bmath{\nabla} (\bmath{n} \bmath{\cdot} \bmath{v}) \frac{\nu}{c} \frac{\partial I_\nu}{\partial \nu} = \notag \\
  &j_\nu - k_\nu I_\nu + \iint k_{\nu'} I_{\nu'} R_{\nu', \bmath{n}' \rightarrow \nu, \bmath{n}} \, \text{d}\Omega' \text{d}\nu' \, ,
\end{align}
where $j_\nu$ and $k_\nu$ are the emission and scattering coefficients, respectively.\footnote{Ly$\alpha$ scattering occurs because absorption is almost immediately followed by decay to the ground state and the re-emission of a Ly$\alpha$ photon.} The last term accounts for frequency redistribution due to partially coherent scattering \citep{Dijkstra2014}. The redistribution function $R_{\nu', \bmath{n}' \rightarrow \nu, \bmath{n}}$ is the differential probability per unit initial photon frequency $\nu'$ and per unit initial directional solid angle $\Omega'$ that the scattering of such a photon traveling in direction $\bmath{n}'$ would place the scattered photon at frequency $\nu$ and directional unit vector $\bmath{n}$. It is convenient to convert to the dimensionless frequency
\begin{equation} \label{eq:x}
  x \equiv \frac{\nu - \nu_0}{\Delta \nu_\text{D}} \, ,
\end{equation}
where $\nu_0$ denotes the frequency at line centre, $\Delta \nu_\text{D} \equiv (v_\text{th}/c)\nu_0$ the Doppler width of the profile, and $v_\text{th} \equiv (2 k_\text{B} T / m_\text{H})^{1/2}$ the thermal velocity. The frequency dependence of the scattering coefficient is given by the Voigt profile $\phi_\text{Voigt}$. For convenience we define the Hjerting-Voigt function $H(a,x) = \sqrt{\pi} \Delta \nu_\text{D} \phi_\text{Voigt}(\nu)$ as the dimensionless convolution of Lorentzian and Maxwellian distributions,
\begin{equation} \label{eq:H}
  H(a,x) = \frac{a}{\pi} \int_{-\infty}^\infty \frac{e^{-y^2}\text{d}y}{a^2+(y-x)^2} \approx
    \begin{cases}
      e^{-x^2} & \quad \text{`core'} \\
      {\displaystyle \frac{a}{\sqrt{\pi} x^2} } & \quad \text{`wing'}
    \end{cases} \, .
\end{equation}
The `damping parameter', $a \equiv \Delta \nu_{\rm L} /2 \Delta \nu_{\rm D}$, describes the relative broadening compared to the natural line width $\Delta \nu_\text{L}$. In isothermal gas, $a$ is simply a parameter that represents the temperature. For Ly$\alpha$ in particular, $a = 4.7 \times 10^{-4} \, (T/10^4 \, \rm K)^{-1/2}$. The cross-section at dimensionless frequency $x$ is simply $\sigma_0  H(a,x)$, where $\sigma_0 = 5.88 \times 10^{-14} \, (T/10^4 \, \rm K)^{-1/2} \, \rm cm^2$ is the cross-section at line centre in the case of Ly$\alpha$. We adopt constant temperature at the cloud or cell level so drop the local dependence on $a$ such that $H(x) \equiv H(a,x)$.

\subsection{Scaling relations}
\label{sec:scaling_relations}
To provide context for the results that follow, we review important scaling relations and consider how non-static effects influence Ly$\alpha$ escape. For a spherical cloud of radius $R$ and comoving line-centre optical depth\footnote{Although the medium is expanding, $\tau_0$ is defined with respect to the line centre in the comoving frame. One could instead adopt the Sobolev optical depth, reducing the effective value due to the velocity gradient. Such a parametrization is helpful to indicate whether line interactions occur but obscures the intuition for resonant scattering in optically-thick clouds.} $\tau_0 \equiv \int_0^R n_{\HI} \sigma_0\,\text{d}r$, the escape of resonance photons in extremely optically thick media ($a\tau_0 \gtrsim 10^3$) can be thought of as a diffusion process in both space and frequency \citep{Adams1972}. The resonant scattering is characterized by random walks in the wings of the line profile, with photons escaping in a single excursion when the rms spatial displacement is comparable to the radius. In addition to the standard drift back to the core \citep{Osterbrock1962}, homologous expansion induces a Doppler shift with each path traversal biasing photons toward the red side of the line ($x < 0$).

Specifically, for a medium characterized by $\beta = V_\text{max} / v_\text{th}$, where $V_\text{max}$ is the maximum expansion velocity, the average frequency shift per traversal becomes $\langle \Delta x | x \rangle \approx -1/x - \beta \lambda_\text{mfp} / R$, where $\lambda_\text{mfp} \approx R / \tau_0 H(x) \approx \sqrt{\pi} x^2 R / a\tau_0$ is the mean-free-path between scatterings. Therefore, red wing photons tend to return to the core after approximately $N_\text{scat} \approx x^2 / (1 + x \beta \lambda_\text{mfp}/R)$ scattering events. Standard random walk arguments also give the number of scatterings as $N_\text{scat} \approx (R / \lambda_\text{mfp})^2$. Equating these expressions gives the following solution for the characteristic escape frequency:
\begin{equation} \label{eq:scaling}
  x_\text{esc} \approx -\left(\frac{a\tau_0}{\sqrt{\pi}}\right)^{1/3} \left(\frac{\beta}{2} + \sqrt{1+\beta^2/4}\right)^{1/3} \approx -\left(\frac{a\tau_0 \beta}{\sqrt{\pi}}\right)^{1/3} \, ,
\end{equation}
where the final approximation is valid for $V_\text{max} \gtrsim v_\text{th}$. We note that this expression reduces to the static version when $V_\text{max} \lesssim v_\text{th}$. From Eq.~(\ref{eq:scaling}) we also estimate the trapping time in units of the light crossing time, $t_\text{light} = R / c$ \citep{Adams1975}, as
\begin{equation} \label{eq:trap_scaling}
  \frac{t_\text{trap}}{t_\text{light}} \approx \frac{a\tau_0}{\sqrt{\pi} x_\text{esc}^2} \approx \left(\frac{a\tau_0}{\sqrt{\pi} \beta^2}\right)^{1/3} \, .
\end{equation}
This is smaller than the static value because velocity gradients aide the escape of photons. Therefore, we must refine the conventional definition of optical thickness for expanding clouds with $V_\text{max} \gtrsim v_\text{th}$ to incorporate dynamical information. Specifically, Ly$\alpha$ analytical solutions are accurate when $a\tau_0/\beta^2 \gtrsim 10^3$.

As an important caveat, in the above derivation we assumed that the scattering coefficient is constant during traversal. However, for rapid expansion this is violated and the optical depth becomes:
\begin{equation}
  \tau(\ell) \approx \int_0^\ell \frac{a\tau_0}{\sqrt{\pi} R (x - \beta \ell' / R)^2}\,\text{d}\ell' = \frac{a\tau_0}{\sqrt{\pi} x^2 (R/\ell - \beta / x)} \, .
\end{equation}
See Section~\ref{sec:Gridless_MCRT} for further details. Setting $\tau(\lambda_\text{mfp}) = 1$ gives a mean-free-path of $\lambda_\text{mfp} \approx \sqrt{\pi} x^2 R / (a\tau_0 + \sqrt{\pi} x \beta)$, which generalizes Eq.~(\ref{eq:scaling}) via a higher-order polynomial equation that can be solved numerically. However, the limiting behaviour is already revealed by taking $\lambda_\text{mfp} \rightarrow \infty$. The critical escape frequency turns over as
\begin{equation} \label{eq:scaling_fast}
  x_\text{esc} \approx -\frac{a\tau_0}{\sqrt{\pi}\beta} \qquad \text{when} \quad \beta \gtrsim \left(\frac{a\tau_0}{\sqrt{\pi}}\right)^{1/2} \, ,
\end{equation}
such that for fast enough velocities ($a\tau_0/\beta^2 \lesssim 10^2$) the peak moves closer to line centre. The light crossing time is also well-behaved with $t_\text{trap} / t_\text{light} \approx \sqrt{\pi} \beta^2 / a\tau_0 > 1$. These scaling arguments agree qualitatively with more accurate analytic solutions derived in Section~\ref{sec:sphere-PS} and MCRT solutions presented in Section~\ref{sec:GMCRT-validation}.

\subsection{Diffusion approximation}
In general, it is only possible to solve Eq.~(\ref{eq:general}) numerically, so we apply further simplifications before attempting to find analytic solutions. We define the first three angular moments of the radiation intensity as $J_\nu \equiv \frac{1}{4\pi} \int \text{d} \Omega \, I_\nu$, $\bmath{H}_\nu \equiv \frac{1}{4\pi} \int \text{d} \Omega \, I_\nu \bmath{n}$, and $\mathbfss{K}_\nu \equiv \frac{1}{4\pi} \int \text{d} \Omega \, I_\nu \bmath{n} \otimes \bmath{n}$. The angular-averaged form of Eq.~(\ref{eq:general}) is the zeroth-order moment equation \citep[see eq.~6.49 in][]{Castor2004}:
\begin{align} \label{eq:RTE-moment}
  \frac{1}{c} \frac{\partial J_\nu}{\partial t}\ +\ &\bmath{\nabla} \bmath{\cdot} \left( \bmath{H}_\nu + \frac{\bmath{v}}{c} J_\nu \right) - \frac{\nu}{c} \frac{\partial \mathbfss{K}_\nu}{\partial \nu} \, \mathbfss{:} \, \bmath{\nabla} \bmath{v} = \notag \\
  &\int \frac{j_\nu}{4\pi} \, \text{d} \Omega - k_\nu J_\nu + \int k_{\nu'} J_{\nu'} R_{\nu' \rightarrow \nu} \, \text{d}\nu' \, ,
\end{align}
where $R_{\nu' \rightarrow \nu} \equiv (4\pi)^{-2} \iint \text{d}\Omega' \text{d}\Omega \, R_{\nu', \bmath{n}' \rightarrow \nu, \bmath{n}}$. The symbol ``$\mathbfss{:}$'' denotes the trace of the matrix product, i.e. $\mathbfss{A} \, \mathbfss{:} \, \mathbfss{B} = \sum_i \sum_j A_{ij} B_{ij}$. The first velocity term is the relativistic correction for the advection of the radiation field with the gas, which we ignore in all applications considered in this paper \citep[see appendix A in][]{Nebrin2024}. The second velocity term is due to the Doppler shift, which is important for resonance lines. To simplify this term we assume the Eddington approximation in which the radiation pressure is isotropic with $\mathbfss{K}_\nu \approx J_\nu \mathbfss{I}/3$, where $\mathbfss{I}$ is the identity matrix, such that $3 \mathbfss{K}_\nu \boldsymbol{:} \bmath{\nabla}\bmath{v} \approx J_\nu \bmath{\nabla} \bmath{\cdot} \bmath{v}$.
We next note that $J_x = \Delta \nu_\text{D} J_\nu$, and from Eq.~(\ref{eq:x}) we get $\partial J_x / \partial \nu = (\Delta \nu_\text{D})^{-1} \partial J_x / \partial x$. Finally, defining $\bmath{u} \equiv \bmath{v}/v_\text{th}$ implies $\bmath{v} = (c/\nu_0) \Delta \nu_\text{D} \bmath{u}$ resulting in:
\begin{equation} \label{eq:sec2_Doppler}
  \frac{\nu}{3 c} \frac{\partial J_x}{\partial \nu} \bmath{\nabla} \bmath{\cdot} \bmath{v} \approx \frac{\nu_0}{3 c} \frac{\partial J_x}{\partial \nu} \bmath{\nabla} \bmath{\cdot} \bmath{v} = \frac{1}{3} \frac{\partial J_x}{\partial x} \bmath{\nabla} \bmath{\cdot} \bmath{u} \, .
\end{equation}
In optically-thick environments, we may also apply Fick's law as a closure relation to the moment equations:
\begin{equation} \label{eq:sec2_Ficks-Law}
  \bmath{H}_x \approx -\frac{\bmath{\nabla} J_x}{3 k_x} \, .
\end{equation}
Likewise, using the Fokker-Planck approximation for the redistribution integral gives \citep[e.g.][]{Rybicki1994, Rybicki2006}:
\begin{equation} \label{eq:Fokker-Planck}
  -k_x J_x + \int k_{x'} J_{x'} R_{x' \rightarrow x} \, \text{d}x' \approx \frac{\partial}{\partial x} \left( \frac{k_x}{2} \frac{\partial J_x}{\partial x} \right) \, ,
\end{equation}
where we have neglected atomic recoil for simplicity.\footnote{In \cite{Nebrin2024} we found an analytical solution with recoil, but at the expense of making the math significantly more complicated and harder to interpret. For simplicity, we ignore recoil in this paper, which is expected to be a good approximation at moderate to high gas temperatures.} Thus, after incorporating Eqs.~(\ref{eq:sec2_Doppler}--\ref{eq:Fokker-Planck}) into Eq.~(\ref{eq:RTE-moment}) we have
\begin{equation} \label{eq:intial_in_any_coordinate}
  \frac{1}{c}\frac{\partial J_x}{\partial t} = \int \frac{j_x}{4\pi} \, \text{d} \Omega + \bmath{\nabla} \bmath{\cdot} \left(\frac{\bmath{\nabla} J_x}{3 k_x} \right) + \frac{1}{3} \frac{\partial J_x}{\partial x} \bmath{\nabla} \bmath{\cdot} \bmath{u} + \frac{\partial}{\partial x} \left( \frac{k_x}{2} \frac{\partial J_x}{\partial x} \right) \, .
\end{equation}
In this paper we focus on steady-state solutions with $\partial J_x / \partial t \approx 0$.
Furthermore, we assume an isothermal environment, which implies spatial-frequency independence for the scattering coefficient $k_x = k(\bmath{r}) H(x)$. We also break the source, of constant luminosity $\mathcal{L}$, into separable components, i.e. $\iiint j_x \, \text{d}V \text{d}x \text{d}\Omega = \mathcal{L}$, with the spatial, frequency, and angular dependence isolated as $\eta(\bmath{r})$, $H(x)/\sqrt{\pi}$, and $1/(4\pi)$, respectively (each normalized to unity):
\begin{equation} \label{eq:sec2_moving_initial}
   \frac{1}{k} \bmath{\nabla} \bmath{\cdot} \left(\frac{\bmath{\nabla} J}{k} \right) + \frac{H}{k} \frac{\partial J}{\partial x} \bmath{\nabla} \bmath{\cdot} \bmath{u} + \frac{3}{2} H \frac{\partial}{\partial x} \left( H \frac{\partial J}{\partial x} \right) = -\frac{3\mathcal{L}}{4\pi} \frac{\eta}{k} \frac{H^2}{\sqrt{\pi}} \, .
\end{equation}
We then apply a standard change of variables with
\begin{equation}
  \text{d}\tilde{x} = \sqrt{\frac{2}{3}}\frac{\text{d}x}{\tau_0 H(x)} \quad \text{such that} \quad \tilde{x} \approx \sqrt{\frac{2\pi}{27}} \frac{x^3}{a \tau_0} \, ,
\end{equation}
where $\tau_0$ denotes the optical depth at line centre. The transformation is based on the wing approximation from Eq.~(\ref{eq:H}), and maps onto the same domain $\tilde{x} \in (-\infty, +\infty)$. In this paper we focus on spherical geometries with uniform density, so we similarly transform from real space to normalized coordinates according to $\tilde{r} = r/R \in (0,1)$, such that $\tilde{\bmath{\nabla}} \equiv R\,\bmath{\nabla}$. In terms of the overall width of the line, $H^2(x)$ is sharply peaked at $x = 0$, so we can replace it with a Dirac delta function \citep{Harrington1973}. To preserve normalization, we note that $\int 3 \tau_0 H^2 \text{d}\tilde{x} = \int \sqrt{6} H\,\text{d}x = \sqrt{6 \pi}$, which allows a replacement of $3 \tau_0 H^2(x) \approx \sqrt{6\pi} \delta(\tilde{x})$.
Finally, if we set $J = \tilde{J} \mathcal{L} \tau_0 \sqrt{6} / (4\pi)$
along with $\bmath{u} = \tilde{\bmath{u}} \sqrt{6}$
then the equation describing our general setup is
\begin{equation} \label{eq:final_x}
  \tilde{\nabla}^2 \tilde{J} + \frac{\partial^2 \tilde{J}}{\partial \tilde{x}^2} + 2 \frac{\partial \tilde{J}}{\partial \tilde{x}} \tilde{\bmath{\nabla}} \bmath{\cdot} \tilde{\bmath{u}} = - \frac{\eta}{k}\delta(\tilde{x}) \, .
\end{equation}

\section{Homologous-like expansion}
\label{sec:homologous_expansion}
Although the static case ($\bmath{u} = \bmath{0}$) already admits a general analytic solution \citep{LaoSmith2020}, here we must first specify a velocity field to proceed further. Of course, there are many possibilities so we choose to focus our attention on the elementary but non-trivial case of homologous-like expansion for which $\tilde{\bmath{\nabla}} \bmath{\cdot} \tilde{\bmath{u}} = w > 0$, a positive constant. The case of homologous-like contraction for which $\tilde{\bmath{\nabla}} \bmath{\cdot} \tilde{\bmath{u}} = w < 0$ results in a nearly identical derivation. In fact, the final results are symmetric under the transformation $\tilde{x} \mapsto -\tilde{x}$, giving rise to the well-known but interesting property that the red versus blue peak dominance swaps depending on the sign of the velocity gradient. Thus, the primary equation we consider is
\begin{equation} \label{eq:final_w}
  \tilde{\nabla}^2 \tilde{J} + \frac{\partial^2 \tilde{J}}{\partial \tilde{x}^2} + 2 w \frac{\partial \tilde{J}}{\partial \tilde{x}} = - \frac{\eta}{k}\delta(\tilde{x}) \, ,
\end{equation}
which only requires suitable boundary conditions for a unique and stable solution. For spherical clouds the velocity gradient parameter is $w \equiv 3 V_\text{max} / \sqrt{6}v_\text{th}$. Following previous studies we require the solution to be finite throughout the sphere, zero as $x \rightarrow \pm\infty$, and the surface intensity to be proportional to the outward flux. If $\bmath{s}$ represents the finite optical depth surface and $\tilde{\partial}_{\bmath{s}}$ is the gradient in the normal direction of the surface, then the boundary conditions are
\begin{equation} \label{eq:sec2_boundary condition}
  \left[\tilde{\partial}_{\bmath{s}} \tilde{J} + f \tau_0 H(\tilde{x}) \tilde{J} \right]_{\bmath{s}} = 0 \qquad \text{and} \qquad \lim_{\tilde{x} \rightarrow \pm \infty} \tilde{J} = 0 \, ,
\end{equation}
where $f$ is a positive constant of order unity. For example, in the two-stream approximation and the Eddington approximation one has $f = \sqrt{3}$ and $f = 3/2$, respectively \citep[e.g.][]{RybickiBook, Dijkstra2006}. Here we leave $f$ unspecified, but note that our results below are insensitive to the specific choice of $f$ \citep[for discussion, see][]{LaoSmith2020, Tomaselli2021}.

The domain of $\tilde{\bmath{r}}$ is compact so we employ an eigenfunction expansion with separable space and frequency components:
\begin{equation} \label{eq:general_decomposition}
  \tilde{J}(\tilde{\bmath{r}},\tilde{x}) = \sum_{n=1}^\infty \vartheta_n(\tilde{\bmath{r}}) \varphi_n(\tilde{x}) \, .
\end{equation}
The solutions of the homogeneous equation
\begin{equation} \label{eq:general_homogeneous}
  \tilde{\nabla}^2 \vartheta_n + \lambda_n^2 \vartheta_n = 0 \, ,
\end{equation}
form an orthonormal basis with eigenvalues $\lambda_n$, requiring that the volume integrals satisfy the relation\footnote{In general, $\vartheta_n$ may be complex. However, as all solutions considered in this work are real, the complex notation is omitted.} $\int \vartheta_n \vartheta_m\,\text{d}\tilde{V} = \delta_{nm}$. Upon substitution of Eq.~(\ref{eq:general_decomposition}) into Eq.~(\ref{eq:final_w}), multiplying by $\vartheta_m$, and integrating over the volume $\tilde{V}$ we obtain
\begin{equation} \label{eq:general_w_ODE}
  \frac{\text{d}^2 \varphi_n}{\text{d}\tilde{x}^2} + 2 w \frac{\text{d} \varphi_n}{\text{d}\tilde{x}} - \lambda_n^2 \varphi_n = -\frac{Q_n}{\tau_0} \delta(\tilde{x}) \, ,
\end{equation}
where the source term coefficients in our convention are
\begin{equation} \label{eq:general_Qn}
  Q_n = \int \eta(\bmath{r}) \vartheta_n(\bmath{r})\,\text{d}V
  = \tau_0 \int \frac{\eta(\tilde{\bmath{r}})}{k(\tilde{\bmath{r}})} \vartheta_n(\tilde{\bmath{r}})\,\text{d}\tilde{V} \, .
\end{equation}
The solution satisfying $\lim_{\tilde{x}\rightarrow\pm\infty} \tilde{J} = 0$, and the jump condition $\Delta (\text{d}\varphi_n/\text{d}\tilde{x})_{\tilde{x} = 0} = -Q_n / \tau_0$ derived from integrating Eq.~(\ref{eq:general_w_ODE}) is
\begin{equation}
  \varphi_n = \frac{Q_n}{2 \tau_0 \chi_n} e^{-\chi_n |\tilde{x}| - w \tilde{x}}  \qquad \text{where} \quad \chi_n = \sqrt{w^2 + \lambda_n^2} \, .
\end{equation}
Putting this all together we have a final solution of
\begin{equation} \label{eq:general_full_solution}
  J(\tilde{\bmath{r}},\tilde{x}) = \frac{\mathcal{L} \sqrt{6}}{8 \pi} e^{-w \tilde{x}} \sum_{n=1}^\infty \frac{Q_n}{\chi_n} e^{-\chi_n |\tilde{x}|} \vartheta_n(\tilde{\bmath{r}}) \, .
\end{equation}
The radiation energy density can be derived as $u(\tilde{\bmath{r}}) = \frac{4\pi}{c} \int J\,\text{d}x$:
\begin{equation} \label{eq:u(r)}
  u(\tilde{\bmath{r}}) = \frac{\mathcal{L}}{c} \Gamma\left(\frac{1}{3}\right) \left(\frac{2 a \tau_0}{\sqrt{\pi}}\right)^{1/3} \sum_{n=1}^\infty Q_n \bar{\chi}_n \vartheta_n(\tilde{\bmath{r}}) \, ,
\end{equation}
where $\bar{\chi}_n \equiv (1/2) \, \lambda_n^{-2/3} \chi_n^{-1} [(\chi_n + w)^{1/3} + (\chi_n - w)^{1/3}] $, and $\Gamma(z) \equiv \int_0^{\infty}\textrm{d}t \, t^{z-1} e^{-t}$ is the gamma function.
Furthermore, we define volume-weighted averages as
\begin{equation} \label{eq:sec2_k-weight}
  \langle f \rangle \equiv \frac{\int f(\bmath{r})\,\text{d}V}{\int \text{d}V}
  = \frac{\int f(\tilde{\bmath{r}})\,\text{d} \tilde{V}/k(\tilde{\bmath{r}})}{\int \text{d}\tilde{V}/k(\tilde{\bmath{r}})} \, ,
\end{equation}
and for convenience we define eigenfunction averages by $T_n \equiv \langle \vartheta_n \rangle$. Thus, a general expression for the average internal spectrum is
\begin{equation} \label{eq:sec2_Jxk}
  \langle J(\tilde{x}) \rangle \equiv \frac{\mathcal{L} \sqrt{6}}{8\pi} e^{-w \tilde{x}} \sum_{n=1}^\infty \frac{Q_n T_n}{\chi_n} e^{-\chi_n |\tilde{x}|} \, ,
\end{equation}
and the average radiation energy density is
\begin{equation} \label{eq:sec2_uk}
  \langle u \rangle \equiv \frac{\mathcal{L}}{c} \Gamma\left(\frac{1}{3}\right) \left(\frac{2 a \tau_0}{\sqrt{\pi}}\right)^{1/3} \sum_{n=1}^\infty Q_n T_n \bar{\chi}_n \, ,
\end{equation}
which is related to the trapping time, defined by the expression $t_\text{trap} = \mathcal{L}^{-1} \int u(\tilde{\bmath{r}})\,\text{d}V$. In spherical geometry we derive the outward force multiplier by using $F \approx -c \nabla u / 3k$ for the flux (which follows from Eq.~\ref{eq:sec2_Ficks-Law}):
\begin{align} \label{eq:sec2_M_F}
  M_\text{F}
  &\equiv \mathcal{L}^{-1}  \iint k(r) F \,\text{d}x\,\text{d}V
  = -\frac{c}{3\mathcal{L}}  \int \nabla u(r)\,\text{d}V  \notag \\
  &= -\Gamma\left(\frac{4}{3}\right) \left(\frac{2 a \tau_0}{\sqrt{\pi}}\right)^{1/3} \sum_{n=1}^\infty Q_n \bar{\chi}_n \int \nabla\vartheta_n(r)\,\text{d}V \, .
\end{align}
The force multiplier quantifies the enhancement of the momentum coupling compared to the single scattering limit, $\mathcal{L}/c$.
This is related to the radial force per unit volume relative to $\mathcal{L}/c$ by:
\begin{equation} \label{eq:a(r)}
  \frac{\rho a(\tilde{r})}{\mathcal{L}/c} = -\Gamma\left(\frac{4}{3}\right) \left(\frac{2 a \tau_0}{\sqrt{\pi}}\right)^{1/3} \sum_{n=1}^\infty  Q_n \bar{\chi}_n \nabla\vartheta_n(\tilde{r}) \, .
\end{equation}
In addition, we define a characteristic radius as the volume-weighted expectation value of position
\begin{equation} \label{eq:sec2_k-weighted-radius}
  r_c \equiv \frac{\langle r u \rangle}{\langle u \rangle}
  = \frac{\sum_{n=1}^\infty Q_n R_n \bar{\chi}_n}{\sum_{n=1}^\infty Q_n T_n \bar{\chi}_n} \, ,
\end{equation}
where for convenience we let $R_n \equiv \langle r \vartheta_n \rangle$.
Finally, the average number of scatterings photons undergo from emission to escape is
\begin{align} \label{eq:sec2_N_scat}
  N_\text{scat} &\approx 4 \pi^{3/2} \mathcal{L}^{-1} \int J(\bmath{r},0) k(\bmath{r})\,\text{d}V \notag \\
  &= \tau_0 \sqrt{\frac{3 \pi}{2}} \sum_{n=1}^\infty \frac{Q_n}{\chi_n} \int \vartheta_n(\tilde{\bmath{r}})\,\text{d}\tilde{V} \, .
\end{align}
Further progress requires specifying the geometry (see Section~\ref{sec:spherical_clouds}).

\begin{figure}
    \centering
    \includegraphics[width=\columnwidth]
    {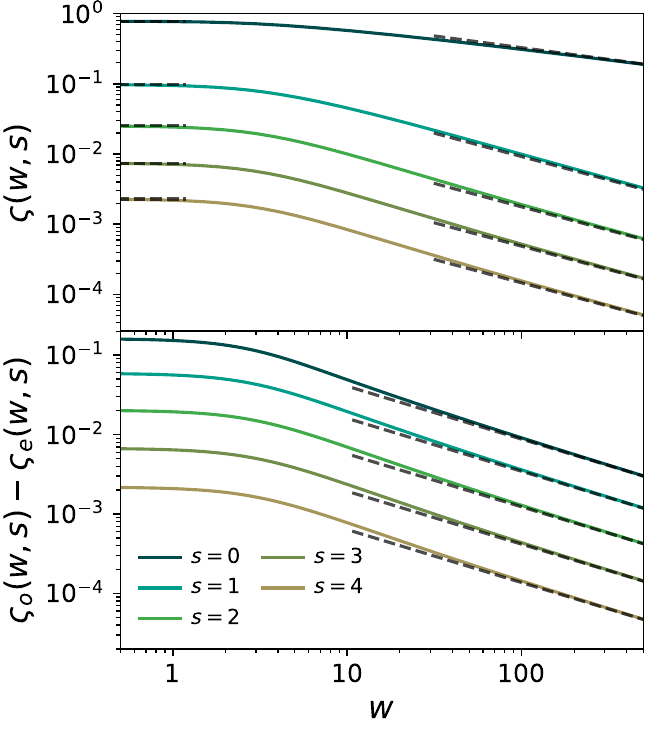}
    \caption{\textit{Top panel:} The Sigma function $\varsigma(w,s)$ for different values of $s$, a series that frequently appears throughout this work. The nearly horizontal dashed lines represent the Riemann Zeta function values derived from Taylor expanding the terms inside the series. The negatively sloped lines show the asymptotic power-law scaling as $w \rightarrow \infty$, showing that for $s = 0$, $\varsigma \propto w^{-1/3}$, while for $s > 1/3$, $\varsigma \propto w^{-2/3}$. \textit{Bottom panel:} The difference between the odd and the even terms of the Sigma function, $\varsigma_\text{o} - \varsigma_\text{e}$. Unlike the full series, the asymptotic scaling for $s = 0$ goes as $w^{-2/3}$ as opposed to $w^{-1/3}$.}
    \label{fig:sigma_function_scaling}
\end{figure}

We now introduce a special function to simplify later results. The following series appears frequently in derivations for spherical geometry (where $\lambda_n \approx \pi n$), which we define as the Sigma function:
\begin{equation} \label{eq:sigma_function}
    \varsigma(w,s) \equiv \sum_{n=1}^{\infty} \bar{\chi}_n \lambda_n^{-s} = \sum_{n=1}^{\infty} \frac{(\chi_n + w)^{1/3} + (\chi_n - w)^{1/3}}{2 \lambda_n^{s + 2/3} \chi_n} \, .
\end{equation}
We provide further details in Appendix~\ref{app:sigma_function}. For convenience, we also define separate series containing only the odd or even terms:
\begin{equation}
    \varsigma_\text{o}(w,s) = \sum_{k=1}^{\infty} \bar{\chi}_{2k-1}\lambda_{2k-1}^{-s}
    \quad \text{and} \quad \varsigma_\text{e}(w,s) = \sum_{p=1}^{\infty} \bar{\chi}_{2k}\lambda_{2k}^{-s} \, .
\end{equation}
To build some intuition for this function, in Fig.~\ref{fig:sigma_function_scaling} we plot $\varsigma$ along with the Taylor expansion at small $w$ and asymptotic scaling at large $w$. Certain quantities depend only on the difference between the odd and even terms of this series, so we also present these scalings as well. In the large-$w$ limit, we show in Appendix~\ref{app:sigma_function} that $\varsigma \propto w^{-1/3}$ for $s = 0$, while for $s > 1/3$, the scaling changes to $\varsigma \propto w^{-2/3}$. Interestingly, the difference between the odd and even terms has $\varsigma_\text{o} - \varsigma_\text{e} \propto w^{-2/3}$ regardless of the value of $s$. The horizontal lines are the values as $w \rightarrow 0$, which has the simple limiting form $\varsigma(0, s) = \pi^{-(s + \, 4/3)} \, \zeta(s + 4/3)$, where the Riemann zeta function is $\zeta(s) = \sum_{n=1}^\infty n^{-s}$. These asymptotic and limiting behaviours provide valuable checks against previous analytic solutions, such as those in \citet{LaoSmith2020}. We discuss additional properties of the Sigma function in Appendix~\ref{app:sigma_function}.

\subsection{Spherical clouds}
\label{sec:spherical_clouds}
We now specialize to homologous-like expansion within homogeneous, spherical clouds. This scenario is astrophysically relevant and captures the salient features of moving media models. In this configuration, the radiation field depends on radius and frequency, $J = J(r,x)$, and at constant density the transformation to normalized optical depth coordinates is $k(r) = k_0$, $\tau_0 = k_0 R$, and $\tilde{r} = r/R \in (0,1)$. Following \citet{LaoSmith2020}, we use the volume element $\text{d}\tilde{V} \rightarrow 2 \tilde{r}^2\,\text{d}\tilde{r}$ and reserve an extra factor of $2\pi R^2$ for volume integrals related to $N_\text{scat}$ and $M_\text{F}$. Eq.~(\ref{eq:general_homogeneous}) reduces to the homogeneous equation, $\vartheta_n'' + 2 \vartheta_n'/\tilde{r} + \lambda_n^2 \vartheta_n = 0$, which has solutions of the form $\vartheta_n = \sin(\lambda_n \tilde{r}) / \tilde{r}$ where $n \in \mathbb{N}$. The boundary conditions require the eigenvalues to satisfy $\lambda_n \cot(\lambda_n) \approx -f \tau_0 H(\tilde{x})$. If the optical depth is large out to any frequency with appreciable radiation, photons escape before they diffuse to frequencies where the cloud is optically thin, and we can use the approximation $\lambda_n \ll f \tau_0 H(\tilde{x})$. Thus, the eigenvalues are $\lambda_n \approx \pi n$ with coefficients reducing to $T_n \approx 3 (-1)^{n-1} / \lambda_n$ and $R_n / R \approx T_n - 6[1-(-1)^n] / \lambda_n^3$.

The final solution from Eq.~(\ref{eq:general_full_solution}) is given by
\begin{equation} \label{eq:sphere_J(r,x)}
  J(\tilde{r},\tilde{x}) = \frac{\mathcal{L} \sqrt{6}}{8\pi} e^{-w \tilde{x}} \sum_{n=1}^\infty \frac{Q_n}{\chi_n} \frac{\sin(\lambda_n \tilde{r})}{\tilde{r}} e^{-\chi_n |\tilde{x}|} \, .
\end{equation}
The spectral line profile at the boundary is particularly relevant for observations. From $\sin(\lambda_n) \approx (-1)^{n-1} \lambda_n / f \tau_0 H(\tilde{x})$ we obtain
\begin{equation} \label{eq:sphere_J(x)}
  J(\tilde{x}) = \frac{\mathcal{L} \sqrt{6}}{8\pi} \frac{e^{-w \tilde{x}}}{f \tau_0 H(\tilde{x})} \sum_{n=1}^\infty (-1)^{n-1} \frac{\lambda_n Q_n}{\chi_n} e^{-\chi_n |\tilde{x}|} \, ,
\end{equation}
with normalization
\begin{equation} \label{eq:J_normalization}
   \int_{-\infty}^{\infty} J(x) \text{d}x = \frac{3 \mathcal{L}}{4\pi f} \sum_{n=1}^{\infty} (-1)^{n-1} \frac{Q_n}{\lambda_n} \, .
\end{equation}
We define $f_\text{red}$ as the fraction of flux redward of line centre:
\begin{equation} \label{eq:f_red}
    f_\text{red} \equiv \frac{\int_{-\infty}^{0} J(x) \text{d}x}{ \int_{-\infty}^{\infty} J(x) \text{d}x} = \frac{\sum_{n=1}^\infty (-1)^{n-1} \dfrac{\lambda_n Q_n}{\chi_n (\chi_n -w)}}{2 \sum_{n=1}^\infty (-1)^{n-1} \dfrac{Q_n}{\lambda_n}} \, .
\end{equation}
From Eq.~(\ref{eq:u(r)}), the radiation energy density is
\begin{equation} \label{eq:sphere_u(r)}
  u(\tilde{r}) = \frac{\mathcal{L}}{c} \Gamma\left(\frac{1}{3}\right) \left(\frac{2 a \tau_0}{\sqrt{\pi}}\right)^{1/3} \sum_{n=1}^\infty  Q_n \bar{\chi}_n \frac{\sin(\lambda_n \tilde{r})}{\tilde{r}} \, ,
\end{equation}
from Eq.~(\ref{eq:sec2_Jxk}) the average internal spectrum is
\begin{equation} \label{eq:sphere_Jxk}
  \langle J(\tilde{x}) \rangle = \frac{3\mathcal{L}\sqrt{6}}{8\pi} e^{-w \tilde{x}} \sum_{n=1}^\infty (-1)^{n-1} \frac{Q_n}{\lambda_n \chi_n} e^{-\chi_n |\tilde{x}|} \, ,
\end{equation}
from Eq.~(\ref{eq:sec2_uk}) the average radiation energy density is
\begin{equation} \label{eq:sphere_uk}
  \langle u \rangle = \frac{3\mathcal{L}}{c} \Gamma\left(\frac{1}{3}\right) \left(\frac{2 a \tau_0}{\sqrt{\pi}}\right)^{1/3} \sum_{n=1}^\infty (-1)^{n-1} \frac{Q_n \bar{\chi}_n}{\lambda_n} \, ,
\end{equation}
so the average trapping time becomes (with $t_\text{light} = R / c$) $t_\text{trap}/t_\text{light} = 4 \pi c R^2 \langle u \rangle / 3 \mathcal{L}$. From Eq.~(\ref{eq:sec2_k-weighted-radius}) the characteristic radius is
\begin{equation}
  \frac{r_c}{R} = 1 - 2 \frac{\sum_{n=1}^\infty Q_n \left[(-1)^n - 1\right] \lambda_n^{-3} \bar{\chi}_n}{\sum_{n=1}^\infty Q_n (-1)^n \lambda_n^{-1} \bar{\chi}_n} \, .
\end{equation}
With $4\pi R^2 \int \text{d}\tilde{r} \, \tilde{r}^2  (\text{d}/\text{d}\tilde{r})[\sin(\lambda_n \tilde{r}) / \tilde{r}]  \approx -8 \pi R^2 [1 - (-1)^n] / \lambda_n$, the force multiplier from Eq.~(\ref{eq:sec2_M_F}) becomes
\begin{equation}
  M_\text{F} = 8\pi R^2 \Gamma\left(\frac{4}{3}\right) \left(\frac{2 a \tau_0}{\sqrt{\pi}}\right)^{1/3} \sum_{n=1}^\infty  \frac{\left[1 + (-1)^{n-1}\right]Q_n}{\lambda_n} \bar{\chi}_n \, ,
\end{equation}
and with $4\pi R^2 \int \tilde{r} \sin(\lambda_n \tilde{r}) \text{d}\tilde{r} \approx 4\pi R^2 (-1)^{n-1} / \lambda_n$ the average number of scatterings before escapes from Eq.~(\ref{eq:sec2_N_scat}) is
\begin{equation}
  N_\text{scat} = \sqrt{24} \pi^{3/2}\,\tau_0 R^2 \sum_{n=1}^\infty \frac{(-1)^{n-1} Q_n}{\lambda_n \chi_n} \, .
\end{equation}
In the remaining subsections we consider specific cases for $Q_n$.

\subsubsection{Central point source}
\label{sec:sphere-PS}
\begin{figure}
    \centering
    \includegraphics[width=\columnwidth]{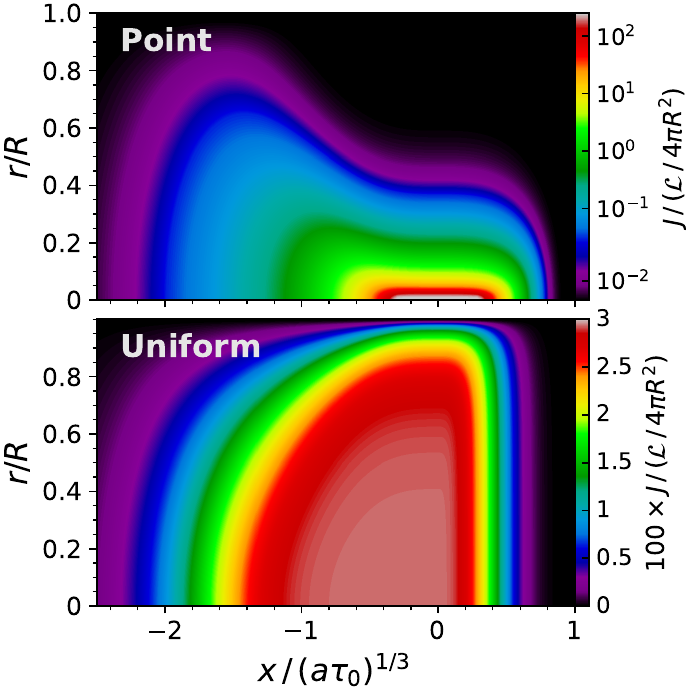}
    \caption{The solution of Eq.~(\ref{eq:final_w}), giving us the internal spectrum $J(r,x)$ as a function of frequency $x\,/\,(a\tau_0)^{1/3}$ and radius $r/R$ for a homogeneous sphere undergoing homologous expansion with $w = 3 V_\text{max} / \sqrt{6} v_\text{th} = 10$. For a point source (top panel, using Eq.~\ref{eq:internal_radiation_field_point}), most of the intensity is clustered around the line centre at $x\,/\,(a\tau_0)^{1/3} = 0$ and near the centre of the sphere. The expansion causes an excess of radiation shifted to the red, especially at relatively higher radii where the photons have traveled a further distance which induces a larger Doppler shift. For a uniform source (bottom panel, using Eq.~\ref{eq:internal_radiation_field_uniform}), The intensity is spread across all radii, corresponding to the emission of photons at all radii, and the red shifted intensity is more intense closer to the centre where the photons undergo a larger Doppler shift.}
    \label{fig:internal_radiation_field}
\end{figure}

\begin{figure}
    \centering
    \includegraphics[width=\columnwidth]{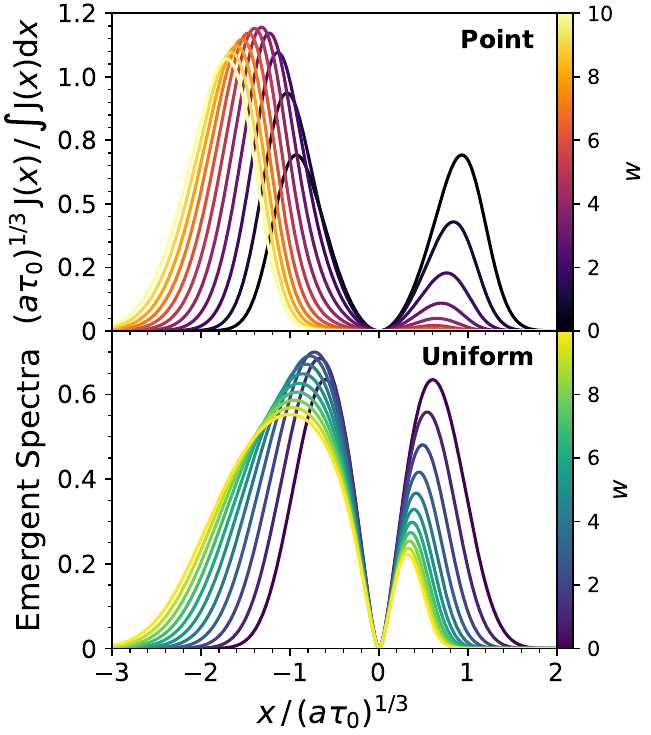}
    \caption{The predicted normalized emergent spectra for a central point source (top panel, using Eq.~\ref{eq:point_series_solution_normalized}) and a uniform source (bottom panel, using Eq.~\ref{eq:uniform_series_solution_normalized}) at the cloud boundary for various values of the expansion parameter, $w \equiv 3 V_\text{max} / \sqrt{6} v_\text{th}$. Qualitatively, both the red and blue peaks shift redward as the velocity increases, as expected due to the Doppler shift. For high enough velocities, the point source flux becomes entirely dominated by the red side of the spectra, while the blue peak persists in the uniform source case. Here the temperature is chosen to be $T \approx 9084$\,K so that $w = V_\text{max} / (10\,\text{km\,s}^{-1})$.}
    \label{fig:analytic_spectra}
\end{figure}

If we place the point source at the spherical centre, i.e. $\eta(r) = \delta(r)/ 4 \pi r^2$, $Q_n = \lambda_n/ 2 \pi R^2$. Eq.~(\ref{eq:sphere_J(r,x)}) becomes
\begin{equation} \label{eq:internal_radiation_field_point}
    J(\tilde{r},\tilde{x}) = \frac{\mathcal{L} \sqrt{6}}{16\pi^2 R^2} e^{-w \tilde{x}} \sum_{n=1}^\infty \frac{\lambda_n}{\chi_n} \frac{\sin(\lambda_n \tilde{r})}{\tilde{r}} e^{-\chi_n |\tilde{x}|} \, ,
\end{equation}
which is shown in Fig.~\ref{fig:internal_radiation_field}. Note that while this series doesn't converge when $\tilde{x} = 0$, the function is smooth around this point, and the limit as $\tilde{x} \rightarrow 0$ is used in the plot. Eq.~(\ref{eq:sphere_J(x)}) is simplified to
\begin{equation} \label{eq:point_source_J(x)}
    J(\tilde{x}) = \frac{\mathcal{L} \sqrt{6}}{16 \pi^2 R^2 } \frac{e^{-w \tilde{x}}}{f \tau_0 H(\tilde{x})} \sum_{n=1}^\infty (-1)^{n-1} \frac{\lambda_n^2 }{\chi_n} e^{-\chi_n |\tilde{x}|} \, ,
\end{equation}
where the normalized solution in the original frequency notation is
\begin{equation} \label{eq:point_series_solution_normalized}
\begin{aligned}
      \frac{J(x)}{\int J(x)\,\text{d}x} =\; &
      \sqrt{\frac{2 \pi}{3}} \frac{x^2}{a \tau_0} \exp \left(-\sqrt{\frac{2 \pi}{27} }\frac{w x^3}{a\tau_0} \right) \\
      &\sum_{n=1}^\infty (-1)^{n-1} \frac{\lambda_n^2 }{\chi_n} \exp \left(-\chi_n \sqrt{\frac{2 \pi}{27} }\frac{|x^3|}{a\tau_0}\right).
\end{aligned}
\end{equation}
In the upper panel of Fig.~\ref{fig:analytic_spectra}, we present several predicted normalized emergent spectra from Eq.~(\ref{eq:point_series_solution_normalized}) for varying values of $w$, as a function of the variable $x / (a\tau_0)^{1/3}$, which serves as a convenient frequency scaling from the static solution. These analytic spectra are presented in the comoving frame before Doppler-shifting to the lab frame, however, in Appendix~\ref{app:labframe} we discuss this in greater detail. We remind the reader that here $w = 3 V_\text{max} / \sqrt{6}v_\text{th}$ is related to the divergence of the velocity field. As the expansion velocity increases, the asymmetry of the line profile increases and the peak locations gradually shift redward ($x < 0$). Eventually, blue photons can no longer escape the cloud, as Doppler shifting dominates over frequency diffusion. To quantify this, the fraction of flux redward of line centre is
\begin{equation} \label{eq:point_f_red}
    f_\text{red} = \sum_{n=1}^\infty (-1)^{n-1} \frac{\lambda_n^2}{\chi_n (\chi_n - w)} \approx \frac{1}{2} + \frac{\text{ln}(2)}{\pi}w \quad (w \ll 1) \, .
\end{equation}
Since the complexity increases due to the dependence of $\chi_n$ on $w$, there is no special function to make the solution more compact. Still, it is clear that the expansion effects are significant once $w > 1$.

For completeness, as in \citet{Nebrin2024} one may also consider a small, but finite emission region with radius $\epsilon < R$. In this case, $\eta(r) = 3/4 \pi \epsilon^3$ inside $\epsilon$ and zero elsewhere. From Eq.~(\ref{eq:general_Qn}), we get
\begin{equation} \label{eq:arbitrary_source_distribution}
    Q_n = \frac{3R}{2 \pi \epsilon^3} \left[ \frac{-\tilde{\epsilon} \lambda_n \text{cos}(\tilde{\epsilon} \lambda_n) + \text{sin}(\tilde{\epsilon} \lambda_n)}{\lambda_n^2}\right] \, ,
\end{equation}
where $\tilde{\epsilon} \equiv \epsilon/R$. As $\epsilon \rightarrow 0$, $Q_n \rightarrow \lambda_n/2 \pi R^2$, reproducing the point-source limit. However, the extended source series converge with much fewer terms, even if the resulting expressions are more cumbersome. More generally, the properties of an emission region of any radius smaller than $R$ can be computed numerically using the general derivations above.

\begin{figure}
    \centering
    \includegraphics[width=\columnwidth]{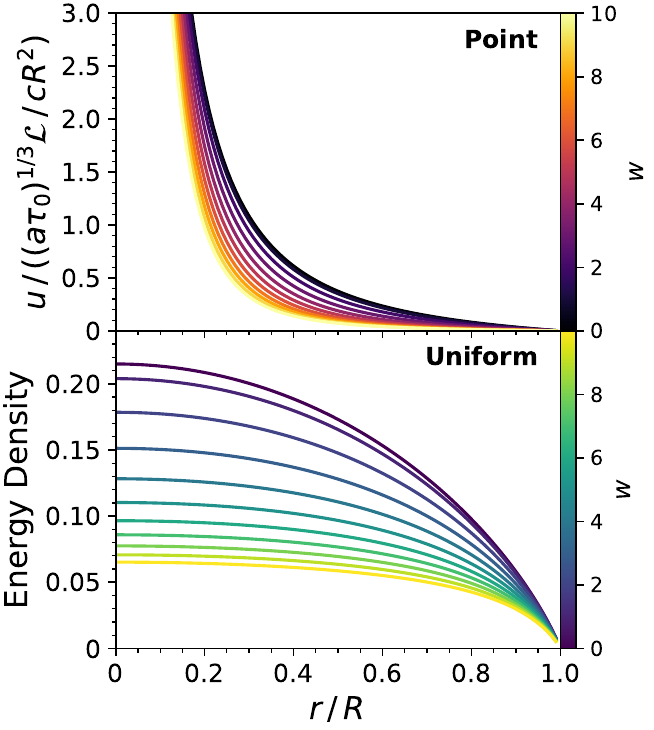}
    \caption{The analytical energy density for a central point source (top panel, using Eq.~\ref{eq:energy_density_point}) and a uniform source (bottom panel, using Eq.~\ref{eq:energy_density_uniform}) as a function of $r/R$ for various values of the expansion parameter, $w \equiv 3 V_\text{max} / \sqrt{6} v_\text{th}$. The point source energy density has a singularity at the origin due to the emission profile, and the series diverges to $+\infty$. The uniform source has an even distribution of energy out to large relative radii. For both sources, the energy density goes to zero at the boundary.}
    \label{fig:analytic_energy_density}
\end{figure}

\begin{figure}
    \centering
    \includegraphics[width=\columnwidth]{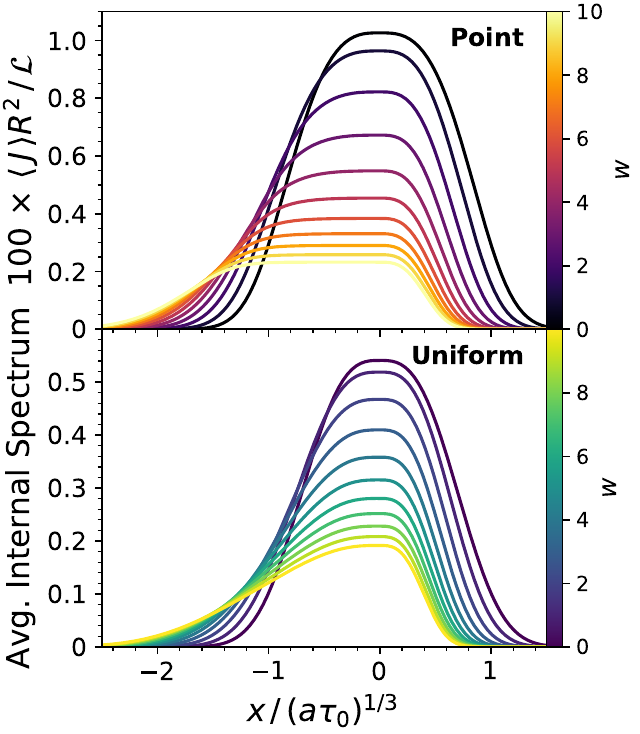}
    \caption{The average internal spectrum for a central point source (top panel, using Eq.~\ref{eq:internal_spectrum_point}) and a uniform source (bottom panel, using Eq.~\ref{eq:internal_spectrum_uniform}) for various values of the expansion parameter, $w \equiv 3 V_\text{max} / \sqrt{6} v_\text{th}$. For both types of sources, the spectrum shifts from symmetry about the line centre to being skewed towards the red as the velocity expansion parameter $w$ increases.}
    \label{fig:analytic_internal_spectrum}
\end{figure}

Another approach that yields an insightful expression is the limit when $w \ll 1$, i.e. the expansion velocity $V_\text{max}$ is much smaller than the thermal velocity $v_\text{th}$. Keeping only the lowest order in $w$ gives
\begin{equation}
    J(\tilde{x}) \approx \frac{\mathcal{L}\sqrt{6} e^{-w \tilde{x}}}{64\pi R^2 f \tau_0 H(\tilde{x})} \, \text{sech}^2 \left(\frac{\pi \tilde{x}}{2}\right) \qquad (w \ll 1) \, ,
\end{equation}
which corresponds to eq.~(87) in \cite{LaoSmith2020}, multiplied by an extra factor of $e^{-w \tilde{x}}$. This correction introduces a velocity-dependent asymmetry, qualitatively illustrating how even a modest velocity gradient skews the spectrum redward. This low-order expansion of the full solution stops being accurate for $w \gtrsim 1$, and is no longer normalizable when $w > \pi$. The normalized version $J(x) / \int J(x)\,\text{d}x$ is
\begin{equation}
      \sqrt{\frac{\pi^3}{24}} \frac{\sin w}{w} \, \exp \left(-\sqrt{\frac{2 \pi}{27} }\frac{w x^3}{a\tau_0} \right) \frac{x^2}{ a\tau_0} \,  \text{sech}^2 \left(\sqrt{\frac{\pi^3}{54}} \frac{|x^3|}{a\tau_0}\right).
\end{equation}
For reference, the radiation energy density, plotted in Fig.~\ref{fig:analytic_energy_density}, is
\begin{equation} \label{eq:energy_density_point}
    u(\tilde{r}) = \frac{\mathcal{L}}{c R^2} \Gamma\left(\frac{1}{3}\right) \left(\frac{a \tau_0}{4 \pi^{7/2}}\right)^{1/3} \sum_{n=1}^{\infty} \lambda_n \bar{\chi}_n \frac{\text{sin}(\lambda_n \tilde{r})}{\tilde{r}} \, .
\end{equation}
The average internal spectrum, plotted in Fig.~\ref{fig:analytic_internal_spectrum}, is
\begin{align} \label{eq:internal_spectrum_point}
    \langle J(\tilde{x}) \rangle = \frac{3\mathcal{L}\sqrt{6}}{16\pi^2 R^2} e^{-w \tilde{x}} \sum_{n=1}^\infty (-1)^{n-1} \frac{e^{-\chi_n |\tilde{x}|}}{\chi_n}  \, .
\end{align}
The ratio between the trapping time $t_\text{trap} \equiv 4 \pi R^3 \langle u \rangle / 3 \mathcal{L}$ and the cloud light crossing time $t_\text{light} = R/c$ is
\begin{equation} \label{eq:trapping_time_point}
    \frac{t_\text{trap}}{t_\text{light}} = 2 \, \Gamma\left(\frac{1}{3}\right) \left(\frac{2 a \tau_0}{ \sqrt{\pi}}\right)^{1/3} [\varsigma_\text{o}(w,0) - \varsigma_\text{e}(w,0)] \, .
\end{equation}
The characteristic radius after some manipulation becomes
\begin{equation}
    \frac{r_c}{R} = 1 - 4 \frac{\varsigma_\text{o}(w,2)}{\varsigma(w,0)} \, .
\end{equation}
The force multiplier is
\begin{equation} \label{eq:force_multiplier_point}
    M_\text{F} = 8 \, \Gamma\left(\frac{4}{3}\right) \left(\frac{2a \tau_0}{\sqrt{\pi}}\right)^{1/3} \varsigma_\text{o}(w,0) \, .
\end{equation}
In the static case ($w = 0$), using properties derived in Appendix~\ref{app:sigma_function}, this reduces to $M_{\rm F} \approx 3.51 (a\tau_0)^{1/3}$ as in \cite{LaoSmith2020}. The average number of scatterings comes out to be
\begin{equation}
    N_\text{scat} = \sqrt{6 \pi} \tau_0 \sum_{n=1}^{\infty} \frac{(-1)^{n-1}}{\chi_n} \, .
\end{equation}

\subsubsection{Uniform source}
If the emissivity traces the scattering coefficient, $\eta(r) = 3 / (4\pi R^3)$. Therefore, the constants $Q_n$ from Eq.~(\ref{eq:general_Qn}) are
\begin{equation}
  Q_n = \frac{3}{2\pi R^2} \int_0^1 \tilde{r} \sin(\lambda_n \tilde{r})\,\text{d}\tilde{r} \approx -\frac{3\cos(\lambda_n)}{2\pi R^2 \lambda_n} \approx \frac{3 (-1)^{n-1}}{2\pi R^2 \lambda_n} \, .
\end{equation}
The general expression from Eq.~(\ref{eq:sphere_J(r,x)}) becomes
\begin{equation} \label{eq:internal_radiation_field_uniform}
  J(\tilde{r},\tilde{x}) = \frac{3 \sqrt{6}\mathcal{L} }{16 \pi^2 R^2} e^{-w \tilde{x}} \sum_{n=1}^\infty (-1)^{n-1} \frac{e^{-\chi_n |\tilde{x}|}}{\lambda_n \chi_n} \frac{\text{sin}(\lambda_n \tilde{r})}{\tilde{r}}  \, .
\end{equation}
and is plotted in Fig.~\ref{fig:internal_radiation_field}. The emergent spectral profile from Eq.~(\ref{eq:sphere_J(x)}) becomes
\begin{equation}
  J(\tilde{x}) = \frac{3 \sqrt{6}\mathcal{L} }{16 \pi^2 R^2} \frac{e^{-w \tilde{x}}}{f \tau_0 H(\tilde{x})} \sum_{n=1}^\infty \frac{e^{-\chi_n |\tilde{x}|}}{\chi_n}  \, ,
\end{equation}
where the normalized version $J(x) / \int J(x)\,\text{d}x$ is
\begin{equation} \label{eq:uniform_series_solution_normalized}
      \sqrt{6 \pi} \frac{x^2}{a \tau_0} \exp \left(-\sqrt{\frac{2 \pi}{27} }\frac{w x^3}{a\tau_0} \right) \sum_{n=1}^\infty \frac{1}{\chi_n} \exp \left(-\chi_n \sqrt{\frac{2 \pi}{27} }\frac{|x^3|}{a\tau_0}\right) \, .
\end{equation}
Again, several examples are shown in the bottom panel of Fig.~\ref{fig:analytic_spectra}. Here, the severity of the red shift and the dominance of the red peak of the flux profile, although certainly present, are considerably less extreme than the point source case. This is explained by the presence of emitted photons close to the edge of the cloud, which results in less of a shift before escape in comparison with the point source. We find the fraction of the flux that is shifted to the red to be
\begin{equation}
    f_\text{red} = \sum_{n=1}^\infty \frac{3}{\lambda_n^2 + w(w - \chi_n)} \approx \frac{1}{2} + \frac{3 \zeta(3) w}{\pi^3} \quad (w \ll 1) \, ,
\end{equation}
where $\zeta(3) \approx 1.202$. The first-order extrapolation to reach $f_\text{red} \approx 1$ is $w \approx 8.6$ compared to $w \approx 4.5$ for the point source. Therefore, the series expansion in the limit when $w \ll 1$ is valid to larger $w$.

For reference, the radiation energy density, plotted in Fig.~\ref{fig:analytic_energy_density} is
\begin{equation} \label{eq:energy_density_uniform}
    u(\tilde{r}) = \frac{3 \mathcal{L}}{c R^2} \Gamma\left(\frac{1}{3}\right) \left(\frac{a \tau_0}{4 \pi^{7/2}}\right)^{1/3} \sum_{n=1}^{\infty} (-1)^{n-1} \frac{\bar{\chi_n}}{\lambda_n} \frac{\text{sin}(\lambda_n \tilde{r})}{\tilde{r}} \, .
\end{equation}
Near the cloud centre, $\text{sin}(\lambda_n \tilde{r})/\tilde{r} \approx \lambda_n$, and the profile is flat with
\begin{equation}
    u(\tilde{r} \ll 1) \approx \frac{3 \mathcal{L}}{c R^2} \Gamma\left(\frac{1}{3}\right) \left(\frac{a \tau_0}{4 \pi^{7/2}}\right)^{1/3}[\varsigma_\text{o}(w,0) - \varsigma_\text{e}(w,0)] \, .
\end{equation}
The average internal spectrum, plotted in Fig.~\ref{fig:analytic_internal_spectrum} is
\begin{align} \label{eq:internal_spectrum_uniform}
    \langle J(\tilde{x}) \rangle = \frac{9\mathcal{L}\sqrt{6}}{16\pi^2 R^2} e^{-w \tilde{x}} \sum_{n=1}^\infty \frac{e^{-\chi_n |\tilde{x}|}}{\lambda_n^2 \chi_n} \, .
\end{align}
The trapping time is $t_\text{trap} = 4 \pi R^3 \langle u \rangle / 3 \mathcal{L}$, or relative to $\langle t_\text{light} \rangle = (3 / 4)\, R / c$ is (see Appendix~\ref{app:sphere_distance} for a brief explanation)
\begin{equation} \label{eq:trapping_time_uniform}
    \frac{t_\text{trap}}{\langle t_\text{light} \rangle} = 8 \, \Gamma\left(\frac{1}{3}\right) \left(\frac{2 a \tau_0}{ \sqrt{\pi}}\right)^{1/3} \varsigma(w,2) \, .
\end{equation}
The characteristic radius after some manipulation becomes
\begin{equation}
    \frac{r_c}{R} = 1 - 4 \frac{\varsigma_\text{o}(w,4)}{\varsigma(w,2)} \, .
\end{equation}
The force multiplier is
\begin{equation} \label{eq:force_multiplier_uniform}
    \frac{M_{\rm F}}{(a \tau_0)^{1/3}} = 24\,\Gamma\left(\frac{4}{3}\right) \left(\frac{2}{\sqrt{\pi}}\right)^{1/3} \varsigma_\text{o}(w,2) \, .
\end{equation}
In the static case ($w = 0$), using properties derived in Appendix~\ref{app:sigma_function}, this reduces to $M_{\rm F} \approx 0.51 (a \tau_0)^{1/3}$ as in \cite{LaoSmith2020}. The average number of scatterings comes out to be
\begin{equation}
    N_\text{scat} = 3 \sqrt{6 \pi} \tau_0 \sum_{n=1}^{\infty} \frac{1}{\lambda_n^2 \chi_n} \approx 3 \sqrt{6 \pi} \tau_0 \left[\frac{\zeta(3)}{\pi^3} - \frac{w^2}{2\pi^5} \zeta(5)\right] \, ,
\end{equation}
which also agrees with \cite{LaoSmith2020} in the case of $w \ll 1$.

\begin{figure*}
    \centering
    \includegraphics[width=1.05\textwidth]
    {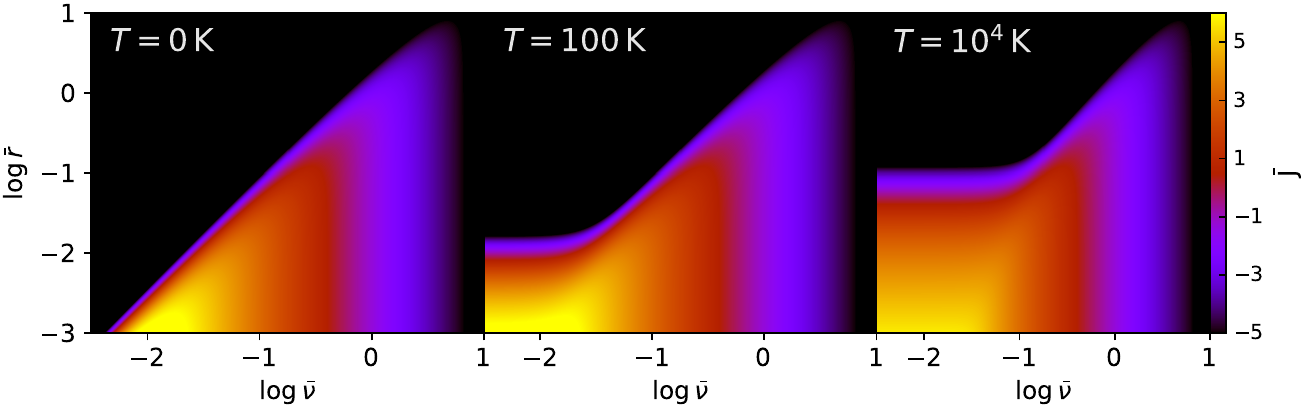}
    \includegraphics[width=\textwidth]{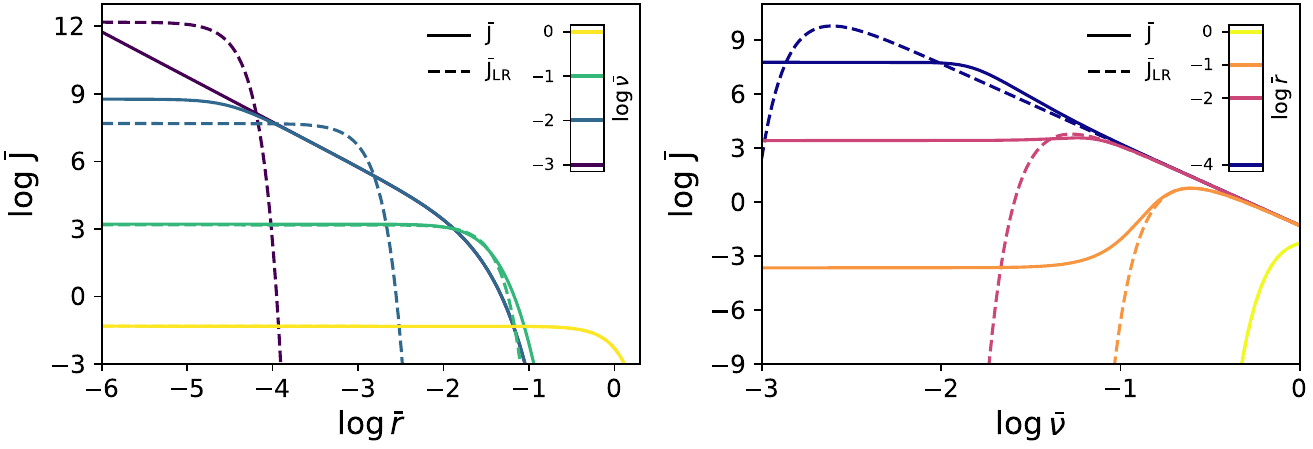}
    \caption{Comparison of the radiative transfer solution $\bar{J}(\bar{r}, \bar{\nu})$ in an infinite expanding Universe, using the parameter-space conventions of \citet{LoebRybicki1999}. \textit{Top Panel}: Illustration of how finite temperature affects the solution, especially closer to line centre, shown for $T = 0\,\text{K}$, $T = 100\,\text{K}$, and $T = 10^4\,\text{K}$. The $T = 0$ case corresponds to Eq.~(\ref{eq:sec4_LRsol}), whereas the $T > 0$ cases require our new solution from Eq.~(\ref{eq:sec4_Jrx}). \textit{Bottom Left (Right) Panel}: Comparisons between the zero- and finite-temperature solutions at $T = 10^4\,\text{K}$ for several slices of constant $\bar{\nu}$ ($\bar{r}$) as a function of $\bar{r}$ ($\bar{\nu}$). The new ($T > 0$) solutions give elevated values at smaller radii or frequencies and fall off much more gradually than the exponential suppression seen in the $T = 0$ limit.}
    \label{fig:cosmological_expansion}
\end{figure*}

\section{Cosmological expansion}
\label{sec:cosmological_expansion}
In this section, we extend our analysis to a scenario analogous to viewing a distant high-redshift galaxy in an expanding Universe. The main difference from previous sections is that the spatial domain is now unbounded and the observer is placed at a large distance, as in the cosmological solution derived by \citet{LoebRybicki1999}. However, unlike their treatment, which assumes a zero-temperature medium by neglecting frequency diffusion, we incorporate the finite temperature and full scattering terms back into the equations. Interestingly, this case leads to a closed form solution that reduces to the less general one in the appropriate limits, and reveals additional insights about radiation transport in expanding media.

Since we are considering a infinite region we adopt different variable transformations that avoid optical depths in normalizations:
\begin{equation}
    \tilde{\bmath{\nabla}} \equiv \frac{\bmath{\nabla}}{k(\bmath{r})} \;\;\; \text{and} \;\;\; \text{d}\tilde{x} = \sqrt{\frac{2}{3}}\frac{\text{d}x}{ H(x)} \;\;\; \text{such that} \;\;\; \tilde{x} \approx \sqrt{\frac{2\pi}{27}} \frac{x^3}{a } \, .
\end{equation}
From Eq.~(\ref{eq:sec2_moving_initial}), with the approximation $3 H^2 \approx \sqrt{6 \pi }\delta(\tilde{x})$, we have
\begin{equation}
    \tilde{\nabla}^2 \tilde{J} + \frac{1}{k} \sqrt{\frac{2}{3}} \frac{\partial \tilde{J}}{\partial \tilde{x}} \bmath{\nabla} \boldsymbol{\cdot} \bmath{u} + \frac{\partial^2 \tilde{J}}{\partial \tilde{x}^2} = -\frac{\eta}{k}\delta(\tilde{x}) \, ,
\end{equation}
where we have applied $J = \tilde{J} \sqrt{6} \mathcal{L}/4 \pi$. To distinguish between the infinite-region case, we set the velocity field $\bmath{v} = \xi \bmath{r}$, where $\xi$ is the Hubble-like constant. The factor appearing in the equation is
\begin{equation}
    \bmath{\nabla} \boldsymbol{\cdot} \bmath{u} = \bmath{\nabla} \boldsymbol{\cdot} \left( \frac{\bmath{v}}{v_{\text{th}}} \right) = \frac{1}{r^2} \frac{\partial }{\partial r} \left(r^2 \frac{\xi r}{v_{\text{th}}}\right) = \frac{3 \xi}{v_{\text{th}}} \, .
\end{equation}
Defining the dimensionless expansion parameter $\tilde{\xi} = \sqrt{6} \xi / k v_{\text{th}}$, and specializing to a point source with $\eta = \delta(\tilde{\bmath{r}})$ where $\tilde{\bmath{r}} = k \bmath{r}$ we obtain:
\begin{equation} \label{eq:sec_eq_rectang_coord}
    \tilde{\nabla}^2 \tilde{J} + \tilde{\xi} \frac{\partial \tilde{J}}{\partial \tilde{x}} + \frac{\partial^2 \tilde{J}}{\partial \tilde{x}^2} = -\delta(\bmath{\tilde{r}}) \delta(\tilde{x}) \, .
\end{equation}
This is our general equation for the cosmological expansion case.
Because the domain is infinite, we apply the boundary condition that $\tilde{J} \rightarrow 0$ as either $\tilde{r} \rightarrow \infty$ or $\tilde{x} \rightarrow \pm \infty$. Under these conditions, a Fourier transform is the natural approach, and we define:
\begin{equation} \label{eq:fourier_transform}
    \mathcal{J}(\bmath{\eta}, \vartheta) = \int \tilde{J}(\bmath{\tilde{r}},\tilde{x}) e^{-i(\bmath{\eta} \cdot \bmath{\tilde{r}} + \vartheta \tilde{x})} \, \text{d}\bmath{\tilde{r}} \, \text{d}\tilde{x} \, .
\end{equation}
where $\bmath{\eta}$ is the wave vector. In these coordinates, Eq.~(\ref{eq:sec_eq_rectang_coord}) gives
\begin{equation} \label{mathcal J solution}
    \mathcal{J} = \left( \bmath{\eta} \boldsymbol{\cdot} \bmath{\eta} + \vartheta^2 - i \vartheta \tilde{\xi} \right)^{-1} \, .
\end{equation}
Taking the inverse Fourier transform, the normalized solution is
\begin{equation}\label{eq:sec4_Jrx}
    \tilde{J}(\tilde{r},\tilde{x}) =\frac{\tilde{\xi}^2}{32 \pi^3} e^{-\frac{1}{2} \tilde{\xi} \tilde{x}} \frac{\mathcal{K}_1 \left(\frac{1}{2} \tilde{\xi} \sqrt{\tilde{r}^2 + \tilde{x}^2}\right)}{\sqrt{\tilde{r}^2 + \tilde{x}^2}} \, ,
\end{equation}
with the normalization $(4 \pi)^2 \int_0^{\infty} \tilde{J}(\tilde{r},\tilde{x}) \tilde{r}^2 \, \text{d}\tilde{r} = 1$. Here $\mathcal{K}_1(z)$ is the modified Bessel function of the second kind.

To connect with previous results, we compare our solution with the zero-temperature limit derived by \citet{LoebRybicki1999}. Using their notation (denoted by quantities with bars), where $\bar{\nu} = (\nu_0 - \nu)/ \nu_*$ is related to redshift by $z = (\nu_*/\nu_\text{obs}) \bar{\nu}$, and $\bar{r} = r/r_*$. The solution in Eq.~(21) of \citet{LoebRybicki1999}, which is valid for $\bar{\nu} > 0$, is
\begin{equation} \label{eq:sec4_LRsol}
    \bar{J}(\bar{r},\bar{\nu}) = \frac{1}{4 \pi} \left(\frac{9}{4 \pi \bar{\nu}^3}\right)^{3/2} \exp \left(- \frac{9 \bar{r}^2}{4 \bar{\nu}^3}\right) \, .
\end{equation}
with the normalization $(4 \pi)^2 \int_0^{\infty} \bar{J}(\bar{r},\bar{x}) \bar{r}^2 \, \text{d}\bar{r} = 1$. In Appendix~\ref{app:cosmological_expansion}, we provide the characteristic radius $r_*$ and frequency $v_*$, relate them to our parameters, and show the steps to perform the inverse Fourier transform, obtain the final normalization, and verify that our solution agrees with Eq.~(\ref{eq:sec4_LRsol}) in the zero-temperature limit.

In Fig.~\ref{fig:cosmological_expansion}, we demonstrate how finite temperature in the high-redshift Universe impacts the intensity. The upper panels illustrate $\bar{J}(\bar{r}, \bar{\nu})$ at temperatures of $T = \{0, 100, 10^4\}\,\text{K}$. The main difference is the power-law radial dependence at small frequencies for the $T > 0$ cases, with the height increasing with temperature. The lower panels show slices of fixed frequency (left) and radius (right), respectively. At larger radii or frequencies (higher $\bar{\nu}$ corresponds to redder photons), the two solutions are approximately equal, reflecting that frequency diffusion becomes less important than Doppler shifting in those regimes. Although our new solution reduces to the previous one in the zero-temperature limit, it is a much more physical solution near the origin ($\bar{r} \ll 1$) and near line centre ($\bar{\nu} \ll 1$).

\section{Gridless Monte Carlo Radiative Transfer}
\label{sec:Gridless_MCRT}
In the previous sections, we derived analytic solutions for several idealized scenarios. We now describe the numerical method we used to validate our new solutions. To ensure the robustness of the results, we employ the MCRT method to solve Eq.~(\ref{eq:general}) directly without the spatial and frequency diffusion approximations leading to Eq.~(\ref{eq:final_w}). Although we focus our discussion on velocity gradients in spherical geometries, our gridless MCRT (GMCRT) method is easily generalized to other applications. The main idea is to perform exact integrations for the optical depth rather than a discretized version based on an arbitrary grid representation. This is particularly useful for idealized models with analytic representations, where it is unnecessary to discretize altogether, or for local gradients within simulation cells.

\subsection{Overview of the GMCRT method}
\begin{figure}
    \centering
    \includegraphics[width=.9\columnwidth]
    {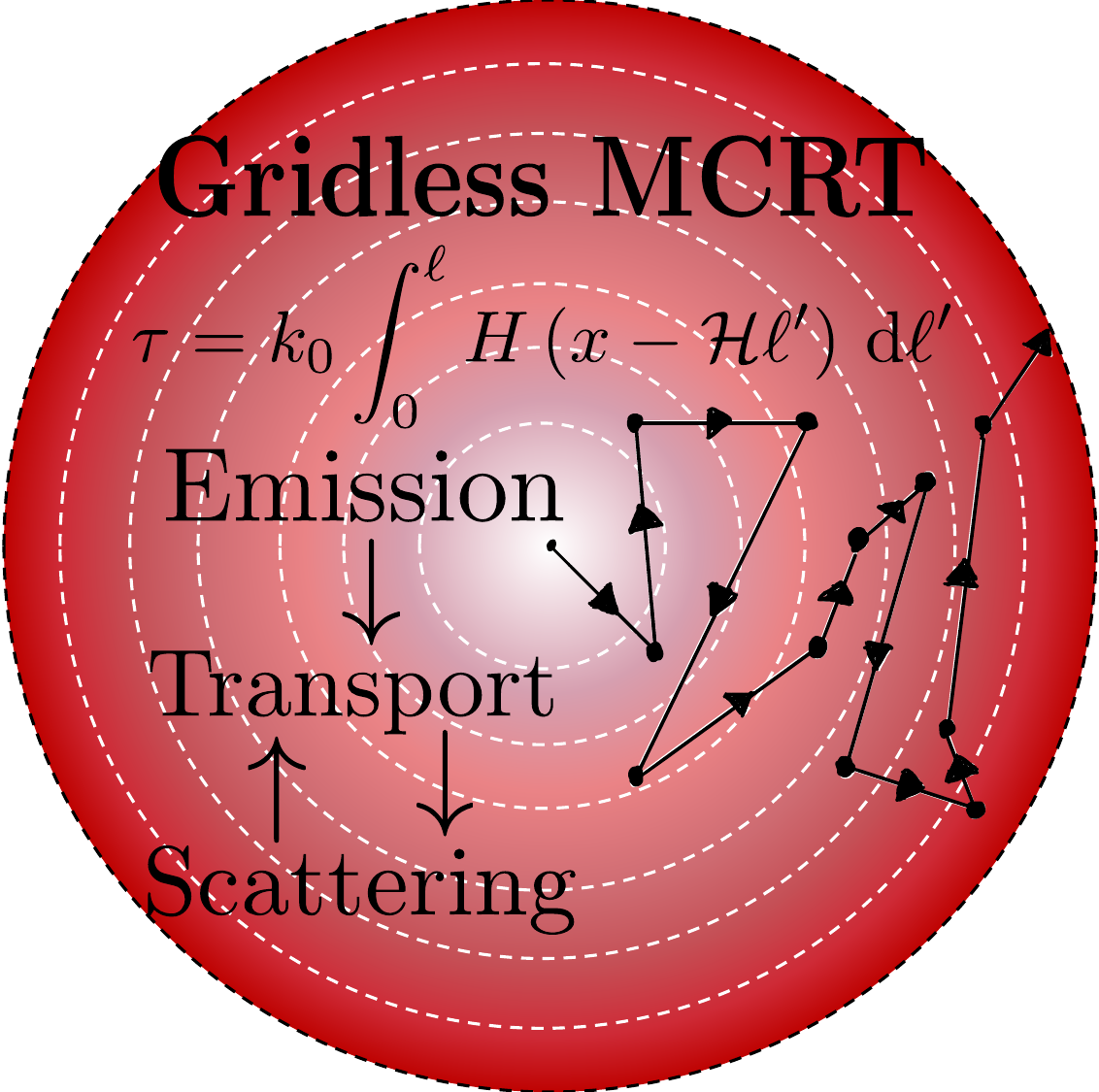}
    \caption{Schematic of Gridless Monte Carlo Radiative Transfer (GMCRT). During transport, the optical depth is integrated exactly accounting for continuous Doppler shifting as opposed to the piecewise-constant static approximation that is commonly employed by MCRT codes but only valid when the change in velocity over a cell is small compared to the thermal velocity ($\nabla u \cdot \Delta\ell \ll v_\text{th}$). A photon is followed from emission and undergoes scattering within concentric shells, as illustrated by the dotted circles. If these represent clouds of increasing size, the photon can contribute statistics to multiple effective simulations with varying prescribed cloud radii ($R$), optical depths ($\tau_0$), and maximum velocities ($V_\text{max}$). This allows us to obtain profiles for various quantities as functions of optical depth from one simulation as opposed to running multiple simulations at different optical depths.}
    \label{fig:gmcrt_schematic}
\end{figure}

MCRT numerically solves the radiative transfer equation by sampling the trajectories of a large number of photon packets to collectively build up statistically convergent estimates of the radiation field and observable properties. In Fig.~\ref{fig:gmcrt_schematic}, we illustrate the main procedures of GMCRT in the idealized case of homologous expansion in spherical geometry without the presence of density gradients. First, photon packets are generated according to the specified emission source distribution. Then, the trajectory is determined by alternating between ray tracing and scattering until the photon escapes the cloud. The Monte Carlo philosophy employs random numbers to decide how far photons move between subsequent scattering events and the change in frequency and direction during each scattering event. By tracking all paths and statistical estimators, we reconstruct global properties such as the emergent spectra.

We have updated the Cosmic Ly$\alpha$ Transfer (\textsc{colt}) code with options for GMCRT with continuous velocity gradients, which we describe herein. Details of the remaining numerical prescriptions can be found in \cite{Smith2015}. To compute quantities such as trapping time and force multiplier, we employ path- and event-based estimators that track radiation energy density and momentum deposition as photons traverse each simulation cell \citep{Smith2017}. In spherical geometry, we divide the domain into concentric shells of increasing radius (the dotted circles in Fig.~\ref{fig:gmcrt_schematic}) to accumulate statistics for the resolved internal structure. We also track the outermost shell visited so far ($i_\text{max}$) for each photon. This allows us to efficiently simulate multiple cloud sizes at once, similar to nested Russian dolls, each with different radii ($R$), optical depths $\tau_0$, and maximum velocities $V_\text{max}$. Once a photon escapes its current outermost shell, we treat it as having escaped from the smaller cloud and increment $i_\text{max}$, ceasing to record contributions to shells at smaller radii. We continue adding contributions to the maximum shell visited, even if the photon subsequently moves inward. As a result, each photon naturally contributes to multiple gridless simulations, and we can later cumulatively sum over inner shells to get the net results for outer ones. We therefore derive continuous profiles of the desired quantities as functions of increasing optical depth from a single run.

\subsection{GMCRT with generic gradients}
We now describe the method for calculating the optical depth traversed taking into account both continuous Doppler shifting due to velocity gradients \citep{Smith2022} and arbitrary density gradients \citep{LaoSmith2020}. The Ly$\alpha$ scattering coefficient is directly proportional to the Hjerting-Voigt function $H(x)$ from Eq.~(\ref{eq:H}). Expanding to second order in $a$ (the damping parameter) gives:
\begin{equation}\label{eq:H_expanded}
    H(x) \approx e^{-x^2} + \frac{2 a}{\sqrt{\pi}} \big[ 2 x F(x) - 1 \big] + a^2 e^{-x^2} ( 1 - 2 x^2 )
\end{equation}
for a fixed $a$.\footnote{The following derivations assume a piecewise-constant temperature, although we allow piecewise-linear velocity and density profiles. Temperature gradients significantly increase the mathematical and computational complexity without substantially improving accuracy for typical $\sim 10^4\,\text{K}$ conditions. However, further modifications may be beneficial for future studies employing GMCRT in environments with strong temperature variations.} The (complex) complementary error function is related to the area under a Gaussian by $\text{erfc}(z) \equiv 1 - 2 \int_0^z e^{-y^2}\text{d}y / \sqrt{\pi}$, and the Dawson integral is $F(x) \equiv \int_0^x e^{y^2 - x^2}\text{d}y$.

In a homologous outflow, the expansion induces a steady Doppler shift so that $\Delta u = \mathcal{H} \Delta\ell$, or more generally $\mathcal{H} \equiv \bmath{n} \bmath{\cdot} \bmath{\nabla}(\bmath{n} \bmath{\cdot} \bmath{u})$ is the velocity gradient in units of the thermal velocity (recall $\bmath{u} \equiv \bmath{v} / v_\text{th}$). The frequency $x$ changes continuously along the path $\ell$ as
\begin{equation}
  \frac{\Delta\lambda}{\lambda} = \frac{\Delta v}{c} \quad \Rightarrow \quad x' = x - \mathcal{H} \Delta\ell \, ,
\end{equation}
where we have used the relation between Doppler frequency and velocity: $x = -\Delta u$. Let $k_0 \equiv n_\text{\HI} \sigma_0$ denote the initial scattering coefficient at line centre, with cross-section of $\sigma_0 = f_{12} \sqrt{\pi} e^2 / (m_e c \Delta \nu_\text{D})$ and oscillator strength of $f_{12} = 0.4162$, and number density of neutral hydrogen $n_\text{HI}$. With a density gradient, $\mathcal{K} \equiv \bmath{n} \bmath{\cdot} \bmath{\nabla}k_0$, the optical depth traversed by a photon beginning with frequency $x$ is
\begin{align} \label{eq:tau_homologous}
  \tau = \int_0^\ell \left(k_0 + \mathcal{K} \ell'\right)\,H\left( x -\mathcal{H}\ell' \right)\,\text{d}\ell' \, .
\end{align}

It is insightful to consider this with increasing complexity, as most MCRT codes do not incorporate density or velocity gradients, i.e. it is assumed that $\tau = k_0 H(x) \ell$. For the case with density gradients, but no velocity gradients ($\mathcal{H} = 0$), we have
\begin{equation} \label{eq:tau_density_only}
  \tau_\text{density} \approx H(x)\int_0^\ell \left(k_0 + \mathcal{K} \ell'\right)\,\text{d}\ell'
  = \left(k_0 \ell + \frac{1}{2} \mathcal{K} \ell^2\right) \, H(x) \, .
\end{equation}
On the other hand, for the case with velocity gradients, but no density gradients ($\mathcal{K} = 0$), we have
\begin{equation} \label{eq:tau_velocity_only}
    \tau_\text{velocity} \approx \frac{k_0}{\mathcal{H}} \big[\Upsilon(x) - \Upsilon(x-\mathcal{H}\ell)\big] \, ,
\end{equation}
where $x' = x - \mathcal{H} \ell'$ is the variable being evaluated and with
\begin{equation} \label{eq:tau_vel_evaluated_part}
    \Upsilon(x) \equiv \int H(x)\,\text{d}x \approx \frac{\sqrt{\pi}}{2} \text{erf}(x) - \frac{2 a}{\sqrt{\pi}} F(x) + a^2 x e^{-x^2} \, .
\end{equation}
The expression in Eq.~(\ref{eq:tau_velocity_only}) uses the second-order expansion from Eq.~(\ref{eq:H_expanded}), which is sufficiently accurate for the Ly$\alpha$ line although it is possible to include higher-order terms in $a$ if desired.

Finally, evaluating the full integral from Eq.~(\ref{eq:tau_homologous}) gives the optical depth in the presence of both velocity and density gradients:
\begin{equation} \label{eq:tau_generic}
    \tau \approx \tau_\text{velocity} + \frac{\mathcal{K}}{\mathcal{H}} \Bigl\{ \frac{1}{\mathcal{H}} \left[ \bar{\Upsilon}(x) - \bar{\Upsilon}(x-\mathcal{H}\ell) \right] - \ell \Upsilon(x - \mathcal{H} \ell) \Bigr\} \, ,
\end{equation}
where we define the second integration of $H(x)$ as
\begin{align} \label{eq:int_upsilon}
    \bar{\Upsilon}(x) &\equiv \int \Upsilon(x) \, \text{d}x \\
    \approx &\frac{\sqrt{\pi}}{2} x\,\text{erf}(x)\,-\, \frac{a x^2}{\sqrt{\pi}} {}_2F_2\left(1 ,1 \, ; \, \frac{3}{2} , 2 \, ; \, -x^2\right) + \frac{(1 - a^2)}{2}e^{-x^2} \, , \notag
\end{align}
where ${}_2F_2$ is a generalized hypergeometric function.

In MCRT, the distance $\ell$ to the next scattering event is sampled by drawing the optical depth $\tau_\text{scat}$ from an exponential distribution, achieved by drawing a uniform random number $\xi \in (0,1)$ with
\begin{equation}
    \tau_\text{scat} = -\text{ln}\,\xi \, .
\end{equation}
We solve for the traversal distance by equating this to the integrated optical depth from Eq.~(\ref{eq:tau_homologous}). In the piecewise-constant static case, the inversion of this equation trivially becomes $\ell = \tau_\text{scat}\,/\,k_0 H(x)$. For numerical stability, whenever $\mathcal{K} \ell \ll 1$ and $\mathcal{H} \ell \ll 1$ it is suitable to use this uniform density and static cell approximation. Similarly, in the static case ($\mathcal{H} \ll 1$) but with general density gradients, we can invert Eq.~(\ref{eq:tau_density_only}) and solve for the distance to scattering as
\begin{equation} \label{eq:l_density_only}
  \ell = \frac{-k_0 + \sqrt{k_0^2 + 2 \tau_\text{scat} \mathcal{K} /H(x)}}{\mathcal{K}} \, .
\end{equation}
However, in both the pure velocity gradient case (Eq.~\ref{eq:tau_velocity_only}) and the general case (Eq.~\ref{eq:tau_generic}), the optical depth integral that defines the scattering distance, $f(\ell) \equiv \int_0^\ell k(\ell')\,\text{d}\ell' - \tau_\text{scat} = 0$, cannot be inverted analytically. We therefore utilize robust root finding techniques to solve for the distance $\ell$ given a value of $\tau_\text{scat} < \tau_\text{face}$ when a scattering event is guaranteed to occur within the cell. Specifically, we employ the following hybrid method in the public version of \textsc{colt}:
\begin{enumerate}
    \item \textit{Static Case:} We can safely ignore gradients if the change in frequency ($\mathcal{H} \Delta\ell$) is smaller than a threshold value, e.g. $<10^{-6}$.
    \item \textit{Partial Inversion:} We make a best estimate by inverting each term, e.g. using the inverse error function for the line component $\ell_\text{line}$. However, any grey components scale linearly with distance so we temper the initial distance to $\ell = \ell_\text{line} \tau_\text{scat} / (\tau_\text{line} + k_\text{grey} \ell_\text{line})$.
    \item \textit{Secure Bracketing:} We initialise a lower bound $\ell_\text{L} = 0$ and an upper bound $\ell_\text{R} = \ell_\text{face}$ of the distance to the next cell interface. Because $k > 0$, $f(\ell_\text{L}) = -\tau_\text{scat} < 0$ and $f(\ell_\text{R} = \tau_\text{face} - \tau_\text{scat} > 0$.
    \item \textit{Aggressive bisection:} To avoid large overshoots and ensure that subsequent iterations start close to the root, we update the upper bound to be $0.9 \ell_\text{L} + 0.1 \ell_\text{R}$ as long as $f(\ell) > 0$.
    \item \textit{Midpoint + Secant Bisection and Halley's Method:} We alternate between a midpoint and secant bisection, i.e. the trail is $(\ell_\text{L} + \ell_\text{R}) / 4 + (\ell_\text{L} \tau_\text{R} - \ell_\text{R} \tau_\text{L})/2(\tau_\text{R} - \tau_\text{L})$ updating the left or right bound depending on the sign of the new value $f(\ell)$. Halley's method is an efficient predictor with cubic convergence if the first and second derivatives are known analytically. We also have an option to fall back onto a secant method in case Halley's method goes out of the bracketed range or to incorporate it into the bisection trial.
    \item \textit{Termination:} The loop repeats until $|f|$ or $|\Delta\ell/\ell| < 10^{-9}$. In most cases, convergence is reached within a handful of iterations.
\end{enumerate}

\begin{figure}
    \centering
    \includegraphics[width=\columnwidth]
    {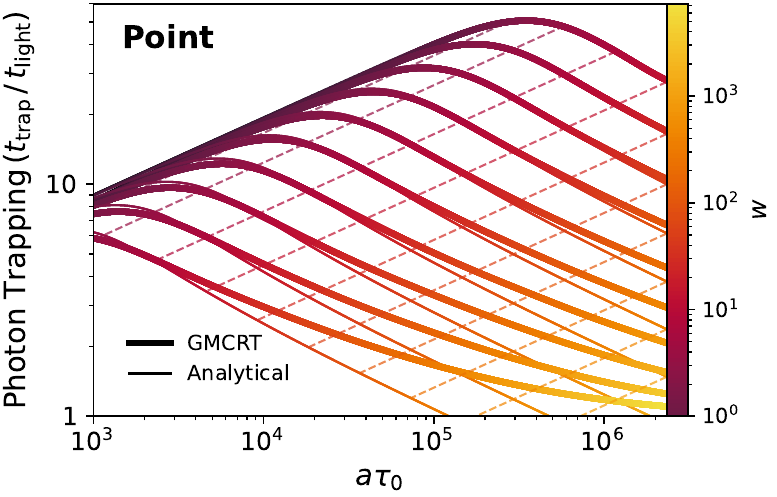}
    \caption{Trapping time ($t_\text{trap}$), normalized by the light-crossing time ($t_\text{light}$), as a function of $a\tau_0$ for a point source with $T = 100\,\text{K}$ and $\tau_{0,\text{max}} = 5 \times 10^8$. The thin lines correspond to the analytical predictions (Eq.~\ref{eq:trapping_time_point}) plotted for different values of $a \tau_0$ and $w \equiv 3 V_\text{max} / \sqrt{6} v_\text{th}$, while the thick lines are from GMCRT simulations with $5 \times 10^4$ photons with the above parameters. The thin dashed lines guide the eye to slices of constant $w$, i.e. constant velocity at a given temperature. For large enough $w$, the simulation data asymptotically approach $t_\text{trap}/t_\text{light} = 1$, reflecting the physical lower limit. The agreement is excellent up to $w \lesssim 60$, with discrepancies at higher $w$ due to the breakdown of the diffusion approximation.}
    \label{fig:trapping_time_point}
\end{figure}

\subsection{GMCRT validation results}
\label{sec:GMCRT-validation}
We now present numerical validations of the analytic solutions derived in Section~\ref{sec:spherical_clouds} using the GMCRT method. The notable simulation parameters are temperature ($T$), which determines the value of $a$, optical depth of the outermost shell ($\tau_{0,\text{max}}$), and maximum expansion velocity at the outermost shell edge ($V_\text{max}$). The homologous expansion is enforced by setting the velocity field as $v(r) = V_\text{max} r / R$. As $\tilde{r} = r / R$ is the parameter that appears in radially dependent solutions, there is no loss in generality for not specifying the non-normalized radius $R$. The dimensionless parameter $w = w(V_\text{max}, T)$ appearing in all analytical solutions is computed from $V_\text{max}$ and $T$ (or equivalently $a$). Our plots use $w$ as the physically relevant control parameter, so, for the most part, we colour-code the resulting properties of the radiation field by $w$ throughout this section. Since the simulation must be initiated with the parameters $V_\text{max}$ and $T$ separately, $w$ must be calculated afterward. Generally, unless noted otherwise, GMCRT results are plotted as thicker lines, and analytical results are plotted as thinner ones.

Each plotted curve corresponds to a single GMCRT simulation at a fixed $\tau_{0,\text{max}}$ and $w \equiv 3 V_\text{max} / \sqrt{6} v_\text{th}$ at the outer radius $R$. Therefore, each curve represents a path through $(\tau_0, w)$ parameter space, as a function of radius within the cloud such that $w(r) = w\,r/R$ and $\tau_0(r) = \tau_{0,\text{max}}\,r/R$. To highlight this dependence, we colour the curves by their local $w$ value. While a 2D representation would be more complete, we chose a format that facilitates direct comparison between GMCRT and analytic solutions along the computationally efficient Russian doll tracks described above.

The main computational difference between the point and uniform source cases is that photons can be emitted in any of the radial shells for a uniform source. To get data points at varying optical depths, we calculate the contributions of photons emitted in inner shells to their emission shell and all outer shells. This leads to a varying number of photons contributing to the statistics of each shell rather than all photons contributing to every shells statics as in the point source case. To alleviate ourselves of some of the consequences of this, we introduce a biasing scheme in which the photons are drawn based on boosted cell luminosities (with the volume boost being $\propto V_\text{cell}^\beta$ for some $\beta < 1$) and the initial photon weights are decreased by a corresponding correction factor to ensure uniformity. This is useful for computing internal properties of the radiation field, however biasing is not needed when computing emergent spectra. Finally, we note that all emitted photons are sampled from a Voigt profile in the local rest-frame of the medium.

\subsubsection{Trapping time}
We begin by analysing the trapping time ($t_\text{trap}$) results. Fig.~\ref{fig:trapping_time_point} compares the GMCRT results for the trapping time with the analytical solution from Eq.~(\ref{eq:trapping_time_point}). The thin dashed lines represent the contours where $w$ is constant, which will vary as the maximum expansion velocity increases. There is excellent agreement, especially for lower values of $w$, where the diffusion approximation is expected to be valid. The disagreement shown in the lower curves or at high enough $w$ is driven by the asymptotic behaviour of $t_\text{trap}/\,t_\text{light}$ toward 1, its physical minimum value. In the regime where the asymptote dominates the value of trapping time, the assumption of diffusion in physical space breaks down, since the velocity is so high that photons free stream through the medium with minimal scattering. The free-streaming behaviour cannot be captured in a solely diffusion regime. The qualitatively derived scaling in Section~\ref{sec:Lyman-alpha radiation transport} is confirmed as $w \equiv 3 \beta / \sqrt{6}$, so the asymptotic scaling is $(a\tau_0/w^2)^{1/3}$, which drives the decrease in trapping time shown in the figure.

\begin{figure}
    \centering
    \includegraphics[width=\columnwidth]
    {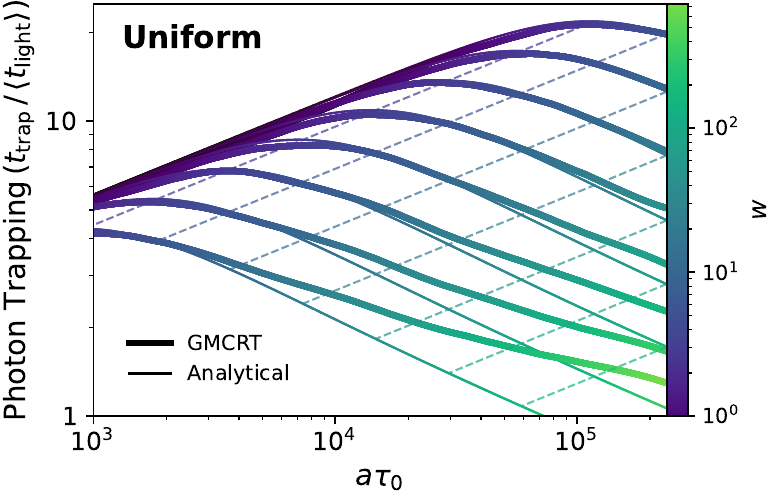}
    \caption{Trapping time ($t_\text{trap}$) for a uniform source normalized by the average light-crossing time $\langle t_\text{light} \rangle = (3/4)\,R/c$, as a function of $a\tau_0$ with $T = 10^4\,\text{K}$ and $\tau_{0,\text{max}} = 5 \times 10^8$. The thin lines are the analytical predictions from Eq.~(\ref{eq:trapping_time_uniform}) plotted for different values of $a \tau_0$ and $w \equiv 3 V_\text{max} / \sqrt{6} v_\text{th}$, while the thick lines are from GMCRT simulations with $5 \times 10^4$ photons and the parameters described above. The thin dashed lines guide the eye to slices of constant $w$, i.e. constant velocity at a give temperature. As before, $t_\text{trap}/\langle t_\text{light} \rangle$ cannot dip below 1, causing an asymptotically driven error at higher velocities. The agreement is excellent up to $w \approx 50$ with all curves with $w \lesssim 25$ (third one from the top) showing nearly perfect agreement.}
    \label{fig:trapping_time_uniform}
\end{figure}

Fig.~\ref{fig:trapping_time_uniform} shows the trapping time for a uniform source and demonstrates a good agreement between the GMCRT results and analytic expression from Eq.~(\ref{eq:trapping_time_uniform}) at lower values of $w$. Yet again, the physical limit of $t_\text{trap}/\langle t_\text{light}\rangle$ drives an error as the simulation results tend towards 1. As explained in Appendix~\ref{app:sphere_distance} ,$1\,/\,\langle t_\text{light}\rangle$ is the proper normalization factor, as the light crossing time varies based on the position in which a photon is emitted. The scaling is also in agreement with the back-of-the-envelope derivation from Section~\ref{sec:Lyman-alpha radiation transport}, with an initial scaling $t_\text{trap}/t_\text{light} \propto (a\tau_0)^{1/3}$ when the velocities are low and $t_\text{trap}/t_\text{light} \propto (a\tau_0/w^2)^{1/3}$ as the effect of velocity gradients starts to dominate. Note that since $\langle t_\text{light} \rangle$ is a constant multiple of $t_\text{light}$, the same scaling relations hold. Here, simulations were performed with $T = 10^4 \, K$ as opposed to $T = 100 \, K$ (as in Fig.~\ref{fig:trapping_time_point}) to ensure that most analytical trapping times would remain above unity to show the success of the analytical solution.

\begin{figure}
    \centering
    \includegraphics[width=\columnwidth]
    {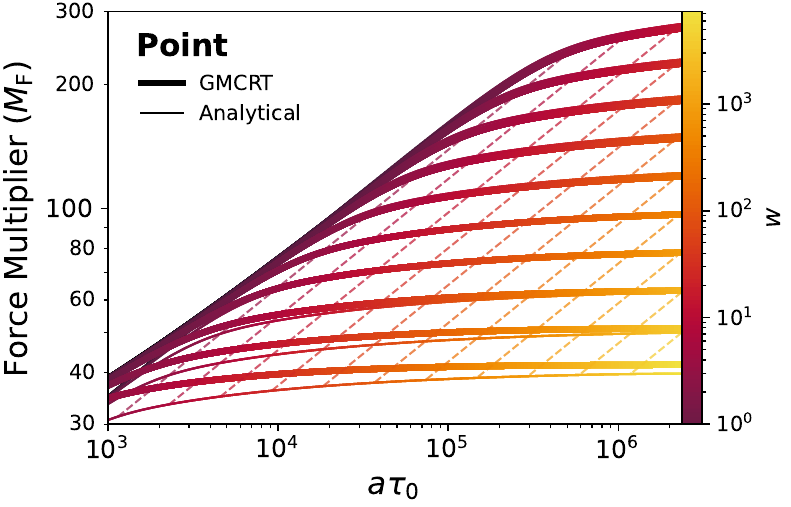}
    \caption{Force multiplier ($M_\text{F}$) as a function of $a \tau_0$, with the same point source setup as described in Fig.~\ref{fig:trapping_time_point}, comparing Eq.~(\ref{eq:force_multiplier_point}) (thin lines) to the results from GMCRT simulations (thick lines). Colours and dashed contours again indicate different $w \equiv 3 V_\text{max} / \sqrt{6} v_\text{th}$ values. The agreement is excellent, with the analytic solution capturing the GMCRT solution almost indistinguishably up to $w \backsimeq 1800$, corresponding to the third curve from the bottom. The disagreement for the lower curves is well understood (see the text for a detailed explanation).}
    \label{fig:force_multiplier_point}
\end{figure}

\subsubsection{Force multiplier}
We now present results for the force multiplier ($M_\text{F}$). Fig.~\ref{fig:force_multiplier_point} shows $M_\text{F}$ for a point source in the same parameter space as Fig.~\ref{fig:trapping_time_point}, showing even better accuracy in this range. For large $a\tau_0 \gg 10^3$ and moderate $w$, the expected static scaling $M_\text{F} \propto (a\tau_0)^{1/3}$ is evident until Doppler shifting causes $M_\text{F}$ to plateau. By analysing the Sigma function variant included in Eq.~(\ref{eq:force_multiplier_point}), we find that the scaling approaches $M_\text{F} \approx$ constant when both $a\tau_0$ and $w$ become large. This is distinctly different from the trapping time scaling derived in Section~\ref{sec:Lyman-alpha radiation transport}, which starts to decrease at a certain point. It is important to note that for physically relevant velocities, the asymptotic scaling of the Sigma function is not enough to combat the $1/3$ scaling in optical depth, meaning that $M_\text{F}$ continues to grow in a homologous expanding cloud.

The error in the lower curves in Fig.~\ref{fig:force_multiplier_point} is again driven by the breakdown of the diffusion regime under high-velocity conditions. For high enough $w$, we do not expect significant spatial diffusion given that the photons are very quickly redshifted away from resonance and with reduced wing scattering they are not restored back to line centre. Coupled with a lower value of $a\tau_0$, the diffusion approximation does not hold. The scaling starts to asymptotically approach 0 before the photons start to diffuse when velocities are high, but under this regime, the $1/3$ scaling of the force multiplier in optical depth space has not yet been set. Thus, one can expect a different behaviour of $M_\text{F}$ here compared to trapping scaling estimates. In fact, we expect the analytical solution to underestimate $M_\text{F}$ since with high expansion velocities, the flattening begins around $a\tau_0 \approx 10^2$ where there is a transition between the diffusion and the free-streaming limits.\footnote{Analytic solutions are known to become less accurate below $a\tau_0 \lesssim 10^3$ \citep[see e.g.][]{Neufeld1990, McClellan2022, Nebrin2024}.}

\begin{figure}
    \centering
    \includegraphics[width=\columnwidth]
    {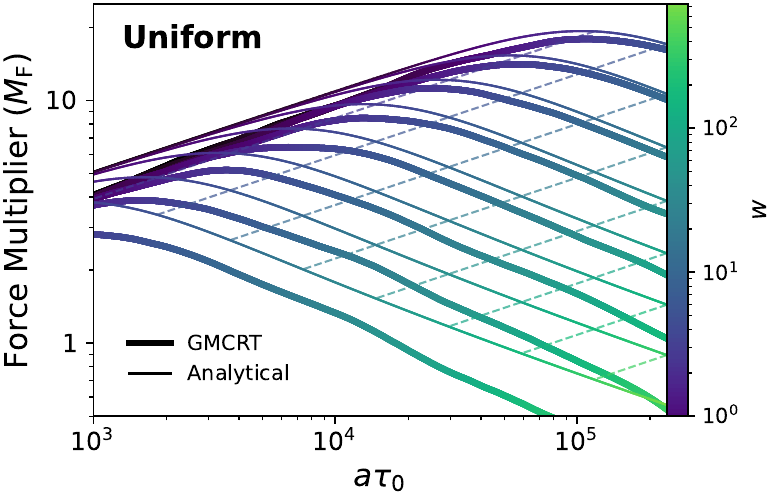}
    \caption{Force multiplier ($M_\text{F}$) for a uniform source as a function of $a \tau_0$ with the same setup as described in Fig.~\ref{fig:trapping_time_uniform}, comparing Eq.~(\ref{eq:force_multiplier_uniform}) (thin lines) to the results from GMCRT simulations (thick lines). Colours and dashed contours again indicate different $w \equiv 3 V_\text{max} / \sqrt{6} v_\text{th}$ values. There is some disagreement between the analytical and numerical results, however the source of the discrepancy is known and explained in the body of the text. Further, in the regime where the diffusion approximation becomes more accurate (when $a\tau_0 \rightarrow \infty$), the agreement continues to improve as seen for the lower velocities in the upper right corner of the plot.}
    \label{fig:force_multiplier_uniform}
\end{figure}

Fig.~\ref{fig:force_multiplier_uniform} shows $M_\text{F}$ for a uniform source as a function of $a\tau_0$. Although, it seems that the simulation results diverge from the analytical solution given by Eq.~(\ref{eq:force_multiplier_uniform}), this is the result of a convergence issue. The force multiplier in this extended source setup is particularly sensitive to the number of scattering events that a photon undergoes in a simulation, so much so that the variance of the total number of scatterings can be used to measure the convergence of the force multiplier quantity. For a uniform source, the possibility of photons being emitted near the outer shell implies that certain photons could escape without undergoing many scattering events, while photons near the cloud centre may undergo many interactions before escape. To even this out, significantly more photons are needed for convergence.

Another issue that affects the results is the use of core skipping. Core skipping is a performance enhancement technique used in radiative transfer simulations wherein the perpendicular velocities of scattering atoms are chosen to be outside a fixed critical frequency\footnote{With a dynamical core-skipping approach, $x_\text{crit}$ does not have to be fixed. However, while scaling by the minimum optical depth to escape produces accurate emergent spectra, it fails to preserve the internal radiation field.} $x_\text{crit}$ \citep[see][]{Smith2015}. Without core skipping, photons are often trapped in resonance with the line and can undergo millions of scatterings before escaping into the wing, whereas most physical properties, such as position and frequency, do not change in a meaningful way. Certainly the total path length traversed by a photon only accumulates significantly when the photon is in the wing, however it is possible that the scatterings impart momentum that does not exactly cancel while the photon is stuck near the core, explaining the discrepancy in Fig.~\ref{fig:force_multiplier_uniform} even when Fig.~\ref{fig:trapping_time_uniform} is in agreement (trapping time is sensitive to the variance of the photon path lengths, and this quantity is not affected strongly by core skipping). Unfortunately, the computational cost of performing numerous simulations without core skipping while also using enough photons to ensure convergence is too high to be justified for the scope of this paper, so we opted to use $x_\text{crit} = 1$, which balances computational expense and accuracy, which compromised the numerical results for the force multiplier in this section. In \cite{Nebrin2024}, we found that as we increased the optical depth, the analytical solution converges to the GMCRT results, and in certain tests that did not use core skipping, we found excellent agreement with the expected static scaling of $M_\text{F} \propto (a\tau_0)^{1/3}$. It is also crucial to mention that we see the scaling in the presence of velocity to be identical to the trapping time scaling in the case of force multiplier, a significant distinction from the point source case. An in-depth study of where this scaling shifts is done in \cite{Nebrin2024} utilising Eq.~(\ref{eq:arbitrary_source_distribution}), and the distinction is in the scaling shift of the Sigma function demonstrated in Fig.~\ref{fig:sigma_function_scaling}. The issue of convergence of MCRT methods will be the subject of a forthcoming study.

\begin{figure}
    \centering
    \includegraphics[width=\columnwidth]{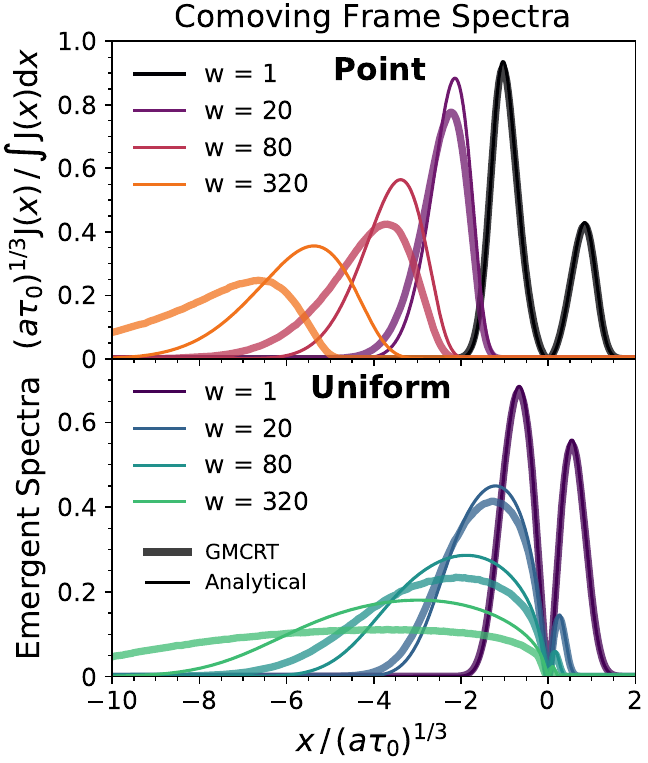}
    \caption{Emergent spectra presented in the internal comoving frame for several expansion values of $w \equiv 3 V_\text{max} / \sqrt{6} v_\text{th} = \{1, 20, 80, 320\}$. The thin lines represent the analytical solutions for a central point source (top panel, Eq.~\ref{eq:point_series_solution_normalized}) and a uniform source (bottom panel, Eq.~\ref{eq:uniform_series_solution_normalized})  while the thick lines are GMCRT simulations ran with $5 \times 10^6$ photons with $\tau_0 = 5 \times 10^8$ with a temperature $T \approx 9084\,\text{K}$ so that $w = V_\text{max} / (10\,\text{km\,s}^{-1})$. The agreement is excellent at lower values of $w$, and gradually the distributions start to have more disagreement. The analytical solution does a good job of capturing the suppression of blueshifted flux as the expansion velocity increases.}
    \label{fig:comoving_frame_spectra}
\end{figure}

\begin{figure}
    \centering
    \includegraphics[width=\columnwidth]{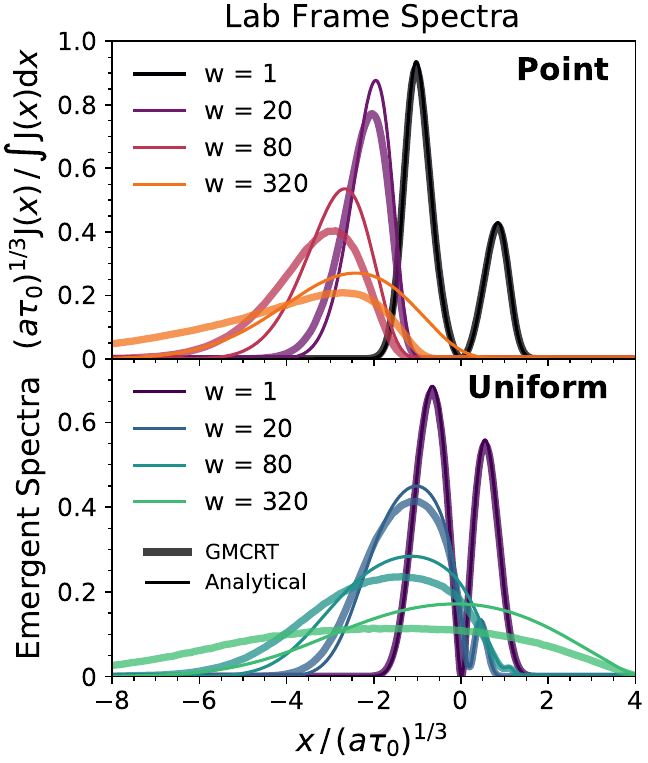}
    \caption{Emergent spectra in the external lab frame after Doppler-shifting (see Appendix~\ref{app:labframe} for a full discussion). The same setup is used as in Fig.~\ref{fig:comoving_frame_spectra} and again, the disagreement is gradually increasing.}
    \label{fig:lab_frame_spectra}
\end{figure}

\subsubsection{Emergent spectra}
To compare analytical emergent spectra (Eq.~\ref{eq:point_series_solution_normalized} for the point source, and Eq.~\ref{eq:uniform_series_solution_normalized} for the uniform source) with GMCRT results we use two frames of reference; the internal comoving frame shown in Fig.~\ref{fig:comoving_frame_spectra} and the external lab frame shown in Fig.~\ref{fig:lab_frame_spectra}, where an extended discussion of how we shifted the comoving-frame spectra derived in Section~\ref{sec:spherical_clouds} to the lab frame, which is utilised for most MCRT experiments, see Appendix~\ref{app:labframe}. We choose the temperature here to be $T \approx 9084\,\text{K}$ such that $w = V_\text{max}/(10\,\text{km/s})$, and present the results for a few values of $w$. The agreement at lower velocities is excellent, and the shifting of the flux to the red is clearly evident as the velocity increases, even for the point source, where blueshifted flux is completely suppressed. However, the disagreement with higher expansion parameters stems from a smaller degree of diffusion as $w$ increases. We expect that if the optical depth was increased, then the emergent spectra would converge to the analytical solution.

For a point source, \citet{Laursen2009} pointed out that as the cloud becomes optically thin, the peak will eventually shift back towards the line centre. At what velocity this happens depends on the optical depth of the cloud, and the effect is observable only in the lab frame after Doppler-shifting. The leakage of some flux blueward of line centre for $w = 320$ in Fig.~\ref{fig:lab_frame_spectra} is due to the deviation of the angular dependence of the emergent spectra from the case of Thomson scattering. As long as the cloud has an adequate optical depth, the photons still diffuse in both frequency and physical space, and the shift back towards line centre is nullified until the velocity term in Eq.~(\ref{eq:final_w}) dominates completely.

For a uniform source, the behaviour of the results is different from the point source case in that the profile becomes very wide. This is caused by photons that are emitted near the edge of the sphere. After Doppler-shifting to the lab frame (see Appendix~\ref{app:labframe}), intermediate to high expansion velocities experience leakage of resonant photons (near $x = 0$). Furthermore, since these photons are emitted at $r_\text{em} \approx R$, the lab-frame velocity of this photon will be $v_\text{em} \approx V_\text{max}$. Physical emission mechanisms, such as recombination and collisional excitation, emit photons near line centre in their rest-frame, resulting in a blue shifted photon with respect to the lab frame, generating a systemic velocity offset. Assuming our cloud has an adequate optical depth, if the expansion is characterised by an intermediate velocity (e.g., $w = 20$ in the bottom panel of Fig.~\ref{fig:lab_frame_spectra}), frequency diffusion will still be strong enough so that leakage of photons at the effective line centre will be strongly suppressed. At higher velocities, the Doppler shift dominates, and wing photons emitted near the edge of the cloud are less likely to scatter before escaping. This effect is illustrated clearly by the curves with $w = 80$ (roughly the point where the Doppler shift starts to dominate at this optical depth, as characterised by the flat effective line centre) and $w = 320$ (where wing photons are essentially free streaming) from the bottom panel of Fig.~\ref{fig:lab_frame_spectra}. Although interesting as an idealised test setup, this probably does not arise as often in nature.

\begin{figure}
    \centering
    \includegraphics[width=\columnwidth]
    {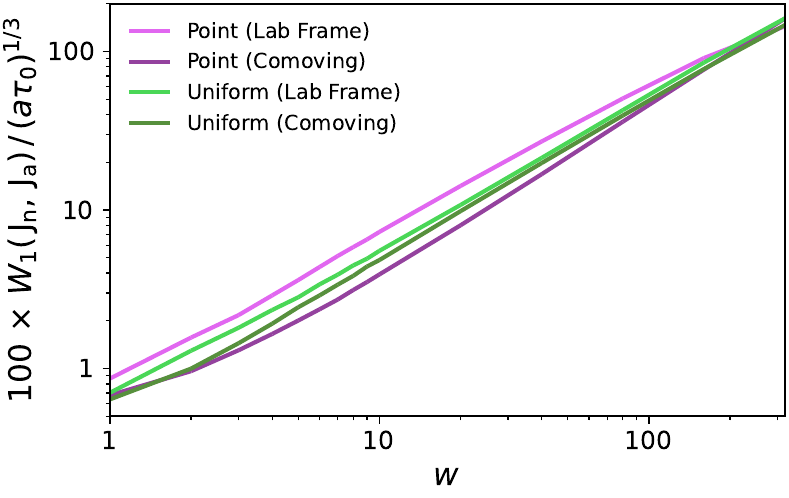}
    \caption{One-dimensional Wasserstein distance (Eq.~\ref{eq:wasserstein_distance}) between the normalized emergent spectra from GMCRT simulations ($J_\text{n}$, i.e. the histogram of escaping frequencies) and analytic solutions ($J_\text{a}$, from Eq.~\ref{eq:point_series_solution_normalized} for the point source results at Eq.~\ref{eq:uniform_series_solution_normalized} for the uniform results), for the same setup as in Fig.~\ref{fig:comoving_frame_spectra}. After scaling by $100 / (a\tau_0)^{1/3}$, this metric represents the percent error in units of $x\,/\,(a\tau_0)^{1/3}$, the dependent variable in the distributions. The distance follows a linear growth with increasing $w \equiv 3 V_\text{max} / \sqrt{6} v_\text{th}$, indicating gradual divergence between the analytic and numerical spectra. At lower values of $w$, the comoving-frame spectra achieve approximately twice the accuracy of the lab-frame spectra for point sources, while the difference is less distinguishable for a uniform source. Once the distributions reach $\sim100\%$ error, the performance difference becomes negligible.}
    \label{fig:emd_plot}
\end{figure}

To further analyse the accuracy of our solution, we utilize the one-dimensional Wasserstein (earth mover's) distance defined as
\begin{equation} \label{eq:wasserstein_distance}
    W_1(u, v) \equiv \underset{\pi \in \Gamma(u,v)}{\text{inf}} \int_{\mathbb{R} \times \mathbb{R}} |x - y| \, \text{d}\pi(x,y) \, ,
\end{equation}
where $u, v$ are distributions on $\mathbb{R}$, and $\Gamma(u,v)$ is the set of joint distributions with marginal distributions $u, v$ respectively \citep{Ramdas2015}. Qualitatively, this measures the amount of ``work'' needed to transform one distribution into another, with units matching those of the $x$-axis. In our context, the distributions are the normalised spectra with $u = J_\text{n}$ representing the numerical results of the GMCRT simulations, while $v = J_\text{a}$ is from Eq.~(\ref{eq:point_series_solution_normalized}). We plot the value of $100\,\times\,W_1(J_\text{n}, J_\text{a})\,/\,(a\tau_0)^{1/3}$ in Fig.~\ref{fig:emd_plot}, which is the percentage error in a Wasserstein distance sense, in units of $x\,/\,(a\tau_0)^{1/3}$. As expected, $W_1$ increases with $w$, reflecting the growing discrepancies in more dynamic regimes. Increasing the optical depth $\tau_0$ at fixed temperature $T$ results in lower Wasserstein distances for each fixed $w$, suggesting that the GMCRT spectra converge to the analytical solution in a Wasserstein metric space. Exact simulated spectra are available in both the comoving and lab frames (see the discussion in Appendix~\ref{app:labframe}). However, the proper frame for comparison with analytic solutions is the comoving frame, where the angular dependence of escaping photons does not need to be accounted for. To shift the analytical spectra to the lab frame without the use of numerical knowledge (to keep the analytics separate from numerical results) requires assumptions about the angular distribution of emergent spectra. For now we adopt the Thomson phase function and leave it to future work to determine whether the angular dependence is analytically tractable. This explains why the comoving frame is consistently more accurate than the lab-frame spectra.

\begin{figure}
    \centering
    \includegraphics[width=\columnwidth]{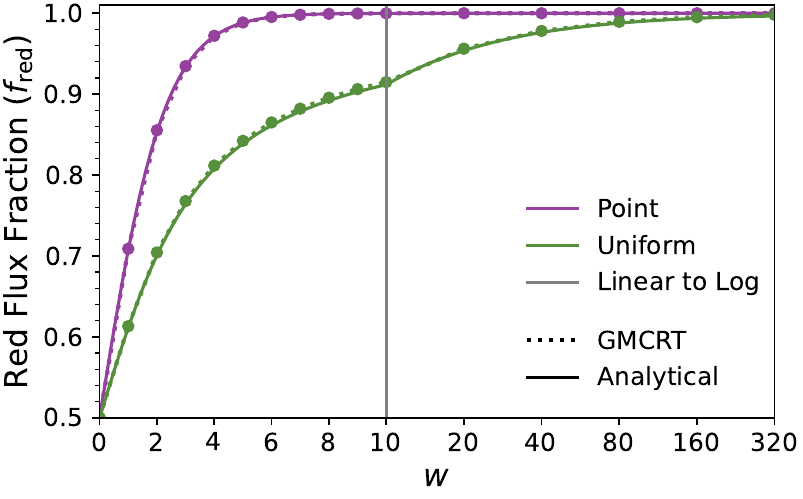}
    \caption{Fraction of the total flux that is redward of line centre in the comoving frame, $f_\text{red} \equiv \int_{-\infty}^0 J(x)\,\text{d}x / \int_{-\infty}^\infty J(x)\,\text{d}x$. The solid lines represent the analytic predictions from Eq.~(\ref{eq:point_f_red}) (point source red fraction) and Eq.~(\ref{eq:f_red}) (uniform source red fraction). The dotted lines interpolate the simulation data points, indicated by the dots. The analytical solutions closely match the GMCRT data, accurately capturing the shift to red-peak dominance of the overall flux as $w \equiv 3 V_\text{max} / \sqrt{6} v_\text{th}$ increases, even for large expansion parameters.}
    \label{fig:red_fraction}
\end{figure}

Fig.~\ref{fig:red_fraction} shows the fraction of the total flux that is redshifted relative to line centre in the comoving frame, highlighting the suppression of blue flux ($f_\text{blue} = 1 - f_\text{red}$) as $w$ increases. Here we leave the $w$-axis in linear scaling since suppression occurs quickly, especially in the case of the point source where $f_\text{red} \approx 1$ is reached at $w \approx 5$. The analytical predictions from Eqs.~(\ref{eq:point_f_red}) and (\ref{eq:f_red}) reproduce the GMCRT simulation data points almost exactly, demonstrating the effectiveness and versatility of our analytic solutions in capturing key spectral features even in rapidly expanding media.

\section{Summary and Discussion}
\label{sec:summary}
Resonant scattering of Ly$\alpha$ photons provides a powerful tracer of the physical conditions in a wide range of astrophysical environments, including high-redshift galaxies. Although numerical simulations are essential to fully capture the complexity of Ly$\alpha$ radiative transfer, analytic models offer more straightforward insights into the roles of geometry, velocity fields, dust absorption, emission sources, and opacity distributions. In this paper, we build on previous studies and further explore concepts from our companion works \citet{LaoSmith2020} and \citet{Nebrin2024}, focusing on the impact of homologous expansion on resonant-line transport. We now summarise our main findings and discuss their broader implications.

\begin{enumerate}
  \item \textit{Velocity-induced effects and novel analytic solutions:} In Section~\ref{sec:Lyman-alpha radiation transport}, we introduced generalized scaling relations and PDE formalisms that incorporate velocity gradients into the usual diffusion approximation of Ly$\alpha$ transport. These scaling arguments provide an intuitive explanation for how velocity gradients can facilitate the escape of Ly$\alpha$ photons and modify the emergent line profiles. In Section~\ref{sec:homologous_expansion}, we presented our new analytic solutions for homologous-like expansion in optically thick spherical clouds. Our explicit series expressions include the full radiation field, emergent line profile, fraction of redward flux, internal spectrum, energy density, trapping time, characteristic radius, force multiplier, and number of scatterings before escape for central point and uniform sources. For clouds with $\beta = V_\text{max} / v_\text{th}$ and $\bar{\beta} = \beta/2 + \sqrt{1 + \beta^2 / 4}$, these solutions are accurate in the dynamically optically-thick regime of $a\tau_0/\bar{\beta}^2 \gtrsim 10^3$. \\

  \item \textit{Cosmological expansion with finite temperature:} In Section~\ref{sec:cosmological_expansion}, we extended the classic zero-temperature solution of \citet{LoebRybicki1999} to include finite-temperature effects and frequency diffusion. The resulting radiation field demonstrates how thermal line broadening and partial frequency redistribution avoids the frequency singularity in the previous solution and produces additional trapping at small radii relative to the pure Hubble flow limit. Our solution reduces to the original version when $T \to 0$, providing a framework for further generalizations for Ly$\alpha$ transport in cosmological settings. \\

  \item \textit{Gridless MCRT for continuous velocity gradients:} Recognizing that the diffusion-based approximations can fail in highly dynamic or rapidly expanding systems, we introduced a novel Gridless Monte Carlo Radiative Transfer (GMCRT) method in Section~\ref{sec:Gridless_MCRT}. By integrating optical depths exactly in the comoving frame, including the continuous Doppler shifting of photon frequencies, GMCRT circumvents the need for finer meshes to explicitly resolve velocity and opacity gradients. We applied GMCRT to the spherical expansions described in our analytic solutions and found excellent agreement in regimes where diffusion approximations hold. At high velocities or modest optical depths, some discrepancies emerge, underscoring where full numerical treatments become necessary. These results highlight the importance of identifying precisely when analytic and numerical solutions remain valid. We leave it to future studies to best determine how to incorporate velocity gradients into dynamical core-skipping schemes \citep{Smith2015}. \\

  \item \textit{Comparisons with earlier moving media studies:} Early MCRT work by \citet{Bonilha1979} indicated a trapping time scaling of $t_\text{trap}/t_\text{light} \propto (V_\text{max}/v_\text{th})^{-3/2}$ in expanding slabs. More recent discussions \citep[e.g.][]{Bithell1990, Omukai2001, Oh2002} built on these scalings. Our analytic solutions, in agreement with our GMCRT tests, suggest a much weaker velocity dependence, $t_\text{trap}/t_\text{light} \propto (V_\text{max}/v_\text{th})^{-2/3}$ in highly optically thick clouds and even more so for the force multiplier with $M_\text{F} \propto (V_\text{max}/v_\text{th})^{-1/3}$, implying that expansion may not reduce Ly$\alpha$ trapping as drastically as previously thought from the standard static $\propto (a\tau_0)^{1/3}$ case. This difference likely reflects the advances in MCRT methodologies in the last few decades. We also emphasize that these scalings also represent a significant improvement in intuition compared to misleading estimates based on an initial frequency offset within static clouds. \\

  \item \textit{Physical implications and astrophysical contexts:} While we have focused on idealized spherical geometries with uniform density and point or uniform emission, the derived solutions demonstrate how velocity gradients and optical depth work together to shape emergent line profiles and internal properties of the radiation field. These processes are relevant for interpreting Ly$\alpha$ observations, especially in scenarios where outflows can strongly regulate Ly$\alpha$ escape. Further explorations also incorporating dust absorption and photon destruction mechanisms, as in \citet{Nebrin2024}, will provide insights into escape fractions and Ly$\alpha$ radiation pressure feedback, e.g. to inform subgrid prescriptions in hydrodynamical simulations.
\end{enumerate}

As with most idealized models, the solutions presented here do not address certain complexities, such as dust absorption, turbulence, and clumpy or anisotropic distributions. These effects are known to influence Ly$\alpha$ escape and spectra in realistic environments \citep[e.g.][]{Laursen2013, GronkeDijkstra2016, Kimm2019, Smith2019, Smith2022disk, Blaizot2023, Garel2024}. Incorporating them analytically typically requires introducing additional terms or altered boundary conditions in the PDE formalism, whereas numerical schemes like GMCRT can capture them by appropriately sampling scattering events. We are intrigued by the prospect of applying the fully second-order GMCRT methodology to three-dimensional hydrodynamical simulations where local velocity and density gradients are under-resolved, thereby complementing or surpassing conventional MCRT solvers in both accuracy and performance. Additionally, given our interest in Ly$\alpha$ feedback, we will more thoroughly explore the convergence properties of MCRT for force calculations. In certain cases it may be more practical to use diffusion, moment-based, or discrete ordinate solvers.

This paper has demonstrated how velocity-induced effects can be self-consistently included in analytic Ly$\alpha$ radiative transfer solutions. Taken alongside \citet{LaoSmith2020} and \citet{Nebrin2024}, these results expand our capability to interpret line profiles and predict radiation trapping and feedback coupling from first principles. Future extensions will target more realistic astrophysical conditions, including turbulence and dust, building on these common frameworks. Ultimately, these developments are complementary to high-resolution numerical simulations and can guide the physical intuition that underpins the next generation of Ly$\alpha$ models for star-forming environments throughout cosmic history.

\section*{Acknowledgements}
We thank the broader Ly$\alpha$ community for many fruitful discussions, including participants of the Kochel Cosmic Lyman Alpha Workshop.
O.N. acknowledges support from the Swedish Research Council grant 2020-04691.
This research was supported in part by grant NSF PHY-2309135 to the Kavli Institute for Theoretical Physics (KITP).

\section*{Data Availability}
The data underlying this paper will be shared on reasonable request to the corresponding author.




\bibliographystyle{mnras}
\bibliography{biblio}

\begin{thebibliography}{}
\makeatletter
\relax
\def\mn@urlcharsother{\let\do\@makeother \do\$\do\&\do\#\do\^\do\_\do\%\do\~}
\def\mn@doi{\begingroup\mn@urlcharsother \@ifnextchar [ {\mn@doi@}
  {\mn@doi@[]}}
\def\mn@doi@[#1]#2{\def\@tempa{#1}\ifx\@tempa\@empty \href
  {http://dx.doi.org/#2} {doi:#2}\else \href {http://dx.doi.org/#2} {#1}\fi
  \endgroup}
\def\mn@eprint#1#2{\mn@eprint@#1:#2::\@nil}
\def\mn@eprint@arXiv#1{\href {http://arxiv.org/abs/#1} {{\tt arXiv:#1}}}
\def\mn@eprint@dblp#1{\href {http://dblp.uni-trier.de/rec/bibtex/#1.xml}
  {dblp:#1}}
\def\mn@eprint@#1:#2:#3:#4\@nil{\def\@tempa {#1}\def\@tempb {#2}\def\@tempc
  {#3}\ifx \@tempc \@empty \let \@tempc \@tempb \let \@tempb \@tempa \fi \ifx
  \@tempb \@empty \def\@tempb {arXiv}\fi \@ifundefined
  {mn@eprint@\@tempb}{\@tempb:\@tempc}{\expandafter \expandafter \csname
  mn@eprint@\@tempb\endcsname \expandafter{\@tempc}}}

\bibitem[\protect\citeauthoryear{{Adams}}{{Adams}}{1972}]{Adams1972}
{Adams} T.~F.,  1972, \mn@doi [\apj] {10.1086/151503}, \href
  {https://ui.adsabs.harvard.edu/abs/1972ApJ...174..439A} {174, 439}

\bibitem[\protect\citeauthoryear{Adams}{Adams}{1975}]{Adams1975}
Adams T.~F.,  1975, \mn@doi [ApJ] {10.1086/153891}, 201, 350

\bibitem[\protect\citeauthoryear{Ahn, Lee  \& Lee}{Ahn et~al.}{2002}]{Ahn2002}
Ahn S.-H.,  Lee H.-W.,   Lee H.~M.,  2002, \mn@doi [ApJ] {10.1086/338497}, 567,
  922

\bibitem[\protect\citeauthoryear{{Basko}}{{Basko}}{1981}]{Basko1981}
{Basko} M.~M.,  1981, \mn@doi [Astrophysics] {10.1007/BF01014298}, \href
  {https://ui.adsabs.harvard.edu/abs/1981Ap.....17...69B} {17, 69}

\bibitem[\protect\citeauthoryear{{Behrens}, {Dijkstra}  \&
  {Niemeyer}}{{Behrens} et~al.}{2014}]{Behrens2014}
{Behrens} C.,  {Dijkstra} M.,   {Niemeyer} J.~C.,  2014, \mn@doi [\aap]
  {10.1051/0004-6361/201322949}, \href
  {https://ui.adsabs.harvard.edu/abs/2014A&A...563A..77B} {563, A77}

\bibitem[\protect\citeauthoryear{{Bithell}}{{Bithell}}{1990}]{Bithell1990}
{Bithell} M.,  1990, \mnras, \href
  {https://ui.adsabs.harvard.edu/abs/1990MNRAS.244..738B} {244, 738}

\bibitem[\protect\citeauthoryear{{Blaizot} et~al.,}{{Blaizot}
  et~al.}{2023}]{Blaizot2023}
{Blaizot} J.,  et~al., 2023, \mn@doi [\mnras] {10.1093/mnras/stad1523}, \href
  {https://ui.adsabs.harvard.edu/abs/2023MNRAS.523.3749B} {523, 3749}

\bibitem[\protect\citeauthoryear{{Bonilha}, {Ferch}, {Salpeter}, {Slater}  \&
  {Noerdlinger}}{{Bonilha} et~al.}{1979}]{Bonilha1979}
{Bonilha} J.~R.~M.,  {Ferch} R.,  {Salpeter} E.~E.,  {Slater} G.,
  {Noerdlinger} P.~D.,  1979, \mn@doi [\apj] {10.1086/157426}, \href
  {https://ui.adsabs.harvard.edu/abs/1979ApJ...233..649B} {233, 649}

\bibitem[\protect\citeauthoryear{{Castor}}{{Castor}}{2004}]{Castor2004}
{Castor} J.~I.,  2004, {Radiation Hydrodynamics}

\bibitem[\protect\citeauthoryear{{Chandrasekhar}}{{Chandrasekhar}}{1945}]{Chandrasekhar1945}
{Chandrasekhar} S.,  1945, \mn@doi [\apj] {10.1086/144771}, \href
  {https://ui.adsabs.harvard.edu/abs/1945ApJ...102..402C} {102, 402}

\bibitem[\protect\citeauthoryear{{Chung}, {Dijkstra}, {Ciardi}  \&
  {Gronke}}{{Chung} et~al.}{2016}]{Chung2016}
{Chung} A.~S.,  {Dijkstra} M.,  {Ciardi} B.,   {Gronke} M.,  2016, \mn@doi
  [\mnras] {10.1093/mnras/stv2340}, \href
  {https://ui.adsabs.harvard.edu/abs/2016MNRAS.455..884C} {455, 884}

\bibitem[\protect\citeauthoryear{{Dijkstra}}{{Dijkstra}}{2014}]{Dijkstra2014}
{Dijkstra} M.,  2014, \mn@doi [\pasa] {10.1017/pasa.2014.33}, \href
  {https://ui.adsabs.harvard.edu/abs/2014PASA...31...40D} {31, e040}

\bibitem[\protect\citeauthoryear{{Dijkstra}, {Haiman}  \& {Spaans}}{{Dijkstra}
  et~al.}{2006}]{Dijkstra2006}
{Dijkstra} M.,  {Haiman} Z.,   {Spaans} M.,  2006, \mn@doi [\apj]
  {10.1086/506243}, \href
  {https://ui.adsabs.harvard.edu/abs/2006ApJ...649...14D} {649, 14}

\bibitem[\protect\citeauthoryear{{Field}}{{Field}}{1959}]{Field1959}
{Field} G.~B.,  1959, \mn@doi [\apj] {10.1086/146654}, \href
  {https://ui.adsabs.harvard.edu/abs/1959ApJ...129..551F} {129, 551}

\bibitem[\protect\citeauthoryear{{Garavito-Camargo}, {Forero-Romero}  \&
  {Dijkstra}}{{Garavito-Camargo} et~al.}{2014}]{Garavito2014}
{Garavito-Camargo} J.~N.,  {Forero-Romero} J.~E.,   {Dijkstra} M.,  2014,
  \mn@doi [\apj] {10.1088/0004-637X/795/2/120}, \href
  {https://ui.adsabs.harvard.edu/abs/2014ApJ...795..120G} {795, 120}

\bibitem[\protect\citeauthoryear{{Garel}, {Michel-Dansac}, {Verhamme},
  {Mauerhofer}, {Katz}, {Blaizot}, {Leclercq}  \& {Salvignol}}{{Garel}
  et~al.}{2024}]{Garel2024}
{Garel} T.,  {Michel-Dansac} L.,  {Verhamme} A.,  {Mauerhofer} V.,  {Katz} H.,
  {Blaizot} J.,  {Leclercq} F.,   {Salvignol} G.,  2024, \mn@doi [\aap]
  {10.1051/0004-6361/202450654}, \href
  {https://ui.adsabs.harvard.edu/abs/2024A&A...691A.213G} {691, A213}

\bibitem[\protect\citeauthoryear{{Ge} \& {Wise}}{{Ge} \&
  {Wise}}{2017}]{GeWise2017}
{Ge} Q.,  {Wise} J.~H.,  2017, \mn@doi [\mnras] {10.1093/mnras/stx2074}, \href
  {https://ui.adsabs.harvard.edu/abs/2017MNRAS.472.2773G} {472, 2773}

\bibitem[\protect\citeauthoryear{{Gray}}{{Gray}}{1973}]{Gray1973}
{Gray} D.~F.,  1973, \mn@doi [\apj] {10.1086/152344}, \href
  {https://ui.adsabs.harvard.edu/abs/1973ApJ...184..461G} {184, 461}

\bibitem[\protect\citeauthoryear{{Gronke} \& {Dijkstra}}{{Gronke} \&
  {Dijkstra}}{2016}]{GronkeDijkstra2016}
{Gronke} M.,  {Dijkstra} M.,  2016, \mn@doi [\apj]
  {10.3847/0004-637X/826/1/14}, \href
  {https://ui.adsabs.harvard.edu/abs/2016ApJ...826...14G} {826, 14}

\bibitem[\protect\citeauthoryear{{Gronke}, {Bull}  \& {Dijkstra}}{{Gronke}
  et~al.}{2015}]{GronkeShell2015}
{Gronke} M.,  {Bull} P.,   {Dijkstra} M.,  2015, \mn@doi [\apj]
  {10.1088/0004-637X/812/2/123}, \href
  {https://ui.adsabs.harvard.edu/abs/2015ApJ...812..123G} {812, 123}

\bibitem[\protect\citeauthoryear{{Hansen} \& {Oh}}{{Hansen} \&
  {Oh}}{2006}]{HansenOh2006}
{Hansen} M.,  {Oh} S.~P.,  2006, \mn@doi [\mnras]
  {10.1111/j.1365-2966.2005.09870.x}, \href
  {https://ui.adsabs.harvard.edu/abs/2006MNRAS.367..979H} {367, 979}

\bibitem[\protect\citeauthoryear{{Harrington}}{{Harrington}}{1973}]{Harrington1973}
{Harrington} J.~P.,  1973, \mn@doi [\mnras] {10.1093/mnras/162.1.43}, \href
  {https://ui.adsabs.harvard.edu/abs/1973MNRAS.162...43H} {162, 43}

\bibitem[\protect\citeauthoryear{{Higgins} \& {Meiksin}}{{Higgins} \&
  {Meiksin}}{2012}]{Higgins2012}
{Higgins} J.,  {Meiksin} A.,  2012, \mn@doi [\mnras]
  {10.1111/j.1365-2966.2012.21917.x}, \href
  {https://ui.adsabs.harvard.edu/abs/2012MNRAS.426.2380H} {426, 2380}

\bibitem[\protect\citeauthoryear{Hummer}{Hummer}{1962}]{Hummer1962}
Hummer D.~G.,  1962, \mn@doi [MNRAS] {10.1093/mnras/125.1.21}, 125, 21

\bibitem[\protect\citeauthoryear{{Kakiichi} \& {Gronke}}{{Kakiichi} \&
  {Gronke}}{2021}]{Kakiichi2021}
{Kakiichi} K.,  {Gronke} M.,  2021, \mn@doi [\apj] {10.3847/1538-4357/abc2d9},
  \href {https://ui.adsabs.harvard.edu/abs/2021ApJ...908...30K} {908, 30}

\bibitem[\protect\citeauthoryear{{Kimm}, {Haehnelt}, {Blaizot}, {Katz},
  {Michel-Dansac}, {Garel}, {Rosdahl}  \& {Teyssier}}{{Kimm}
  et~al.}{2018}]{Kimm2018}
{Kimm} T.,  {Haehnelt} M.,  {Blaizot} J.,  {Katz} H.,  {Michel-Dansac} L.,
  {Garel} T.,  {Rosdahl} J.,   {Teyssier} R.,  2018, \mn@doi [\mnras]
  {10.1093/mnras/sty126}, \href
  {https://ui.adsabs.harvard.edu/abs/2018MNRAS.475.4617K} {475, 4617}

\bibitem[\protect\citeauthoryear{{Kimm}, {Blaizot}, {Garel}, {Michel-Dansac},
  {Katz}, {Rosdahl}, {Verhamme}  \& {Haehnelt}}{{Kimm} et~al.}{2019}]{Kimm2019}
{Kimm} T.,  {Blaizot} J.,  {Garel} T.,  {Michel-Dansac} L.,  {Katz} H.,
  {Rosdahl} J.,  {Verhamme} A.,   {Haehnelt} M.,  2019, \mn@doi [\mnras]
  {10.1093/mnras/stz989}, \href
  {https://ui.adsabs.harvard.edu/abs/2019MNRAS.486.2215K} {486, 2215}

\bibitem[\protect\citeauthoryear{{Kimm}, {Bieri}, {Geen}, {Rosdahl}, {Blaizot},
  {Michel-Dansac}  \& {Garel}}{{Kimm} et~al.}{2022}]{Kimm2022}
{Kimm} T.,  {Bieri} R.,  {Geen} S.,  {Rosdahl} J.,  {Blaizot} J.,
  {Michel-Dansac} L.,   {Garel} T.,  2022, \mn@doi [\apjs]
  {10.3847/1538-4365/ac426d}, \href
  {https://ui.adsabs.harvard.edu/abs/2022ApJS..259...21K} {259, 21}

\bibitem[\protect\citeauthoryear{{Lao} \& {Smith}}{{Lao} \&
  {Smith}}{2020}]{LaoSmith2020}
{Lao} B.-X.,  {Smith} A.,  2020, \mn@doi [\mnras] {10.1093/mnras/staa2198},
  \href {https://ui.adsabs.harvard.edu/abs/2020MNRAS.497.3925L} {497, 3925}

\bibitem[\protect\citeauthoryear{Laursen, Razoumov  \& Sommer-Larsen}{Laursen
  et~al.}{2009}]{Laursen2009}
Laursen P.,  Razoumov A.~O.,   Sommer-Larsen J.,  2009, \mn@doi [ApJ]
  {10.1088/0004-637X/696/1/853}, 696, 853

\bibitem[\protect\citeauthoryear{{Laursen}, {Duval}  \& {{\"O}stlin}}{{Laursen}
  et~al.}{2013}]{Laursen2013}
{Laursen} P.,  {Duval} F.,   {{\"O}stlin} G.,  2013, \mn@doi [\apj]
  {10.1088/0004-637X/766/2/124}, \href
  {https://ui.adsabs.harvard.edu/abs/2013ApJ...766..124L} {766, 124}

\bibitem[\protect\citeauthoryear{{Loeb} \& {Rybicki}}{{Loeb} \&
  {Rybicki}}{1999}]{LoebRybicki1999}
{Loeb} A.,  {Rybicki} G.~B.,  1999, \mn@doi [\apj] {10.1086/307844}, \href
  {https://ui.adsabs.harvard.edu/abs/1999ApJ...524..527L} {524, 527}

\bibitem[\protect\citeauthoryear{{McClellan}, {Davis}  \& {Arras}}{{McClellan}
  et~al.}{2022}]{McClellan2022}
{McClellan} B.~C.,  {Davis} S.~W.,   {Arras} P.,  2022, \mn@doi [\apj]
  {10.3847/1538-4357/ac7724}, \href
  {https://ui.adsabs.harvard.edu/abs/2022ApJ...934...37M} {934, 37}

\bibitem[\protect\citeauthoryear{Mihalas \& Auer}{Mihalas \&
  Auer}{2001}]{mihalas2001laboratory}
Mihalas D.,  Auer L.,  2001, Journal of Quantitative Spectroscopy and Radiative
  Transfer, 71, 61

\bibitem[\protect\citeauthoryear{{Nebrin}, {Smith}, {Lorinc}, {H{\"o}rnquist},
  {Larson}, {Mellema}  \& {Giri}}{{Nebrin} et~al.}{2025}]{Nebrin2024}
{Nebrin} O.,  {Smith} A.,  {Lorinc} K.,  {H{\"o}rnquist} J.,  {Larson}
  {\r{A}}.,  {Mellema} G.,   {Giri} S.~K.,  2025, \mn@doi [\mnras]
  {10.1093/mnras/staf038}, \href
  {https://ui.adsabs.harvard.edu/abs/2025MNRAS.537.1646N} {537, 1646}

\bibitem[\protect\citeauthoryear{{Neufeld}}{{Neufeld}}{1990}]{Neufeld1990}
{Neufeld} D.~A.,  1990, \mn@doi [\apj] {10.1086/168375}, \href
  {https://ui.adsabs.harvard.edu/abs/1990ApJ...350..216N} {350, 216}

\bibitem[\protect\citeauthoryear{{Neufeld} \& {McKee}}{{Neufeld} \&
  {McKee}}{1988}]{Neufeld1988}
{Neufeld} D.~A.,  {McKee} C.~F.,  1988, \mn@doi [\apjl] {10.1086/185241}, \href
  {https://ui.adsabs.harvard.edu/abs/1988ApJ...331L..87N} {331, L87}

\bibitem[\protect\citeauthoryear{{Oh} \& {Haiman}}{{Oh} \&
  {Haiman}}{2002}]{Oh2002}
{Oh} S.~P.,  {Haiman} Z.,  2002, \mn@doi [\apj] {10.1086/339393}, \href
  {https://ui.adsabs.harvard.edu/abs/2002ApJ...569..558O} {569, 558}

\bibitem[\protect\citeauthoryear{{Omukai}}{{Omukai}}{2001}]{Omukai2001}
{Omukai} K.,  2001, \mn@doi [\apj] {10.1086/318296}, \href
  {https://ui.adsabs.harvard.edu/abs/2001ApJ...546..635O} {546, 635}

\bibitem[\protect\citeauthoryear{Osterbrock}{Osterbrock}{1962}]{Osterbrock1962}
Osterbrock D.~E.,  1962, \mn@doi [ApJ] {10.1086/147258}, 135, 195

\bibitem[\protect\citeauthoryear{{Padmanabhan} \& {Loeb}}{{Padmanabhan} \&
  {Loeb}}{2024}]{Padmanabhan2024}
{Padmanabhan} H.,  {Loeb} A.,  2024, \mn@doi [\jcap]
  {10.1088/1475-7516/2024/10/059}, \href
  {https://ui.adsabs.harvard.edu/abs/2024JCAP...10..059P} {2024, 059}

\bibitem[\protect\citeauthoryear{{Partridge} \& {Peebles}}{{Partridge} \&
  {Peebles}}{1967}]{Partridge1967}
{Partridge} R.~B.,  {Peebles} P.~J.~E.,  1967, \mn@doi [\apj] {10.1086/149079},
  \href {https://ui.adsabs.harvard.edu/abs/1967ApJ...147..868P} {147, 868}

\bibitem[\protect\citeauthoryear{{Phillips} \& {Meszaros}}{{Phillips} \&
  {Meszaros}}{1986}]{Phillips1986}
{Phillips} K.~C.,  {Meszaros} P.,  1986, \mn@doi [\apj] {10.1086/164682}, \href
  {https://ui.adsabs.harvard.edu/abs/1986ApJ...310..284P} {310, 284}

\bibitem[\protect\citeauthoryear{{Planck Collaboration} et~al.,}{{Planck
  Collaboration} et~al.}{2020}]{Planck2020}
{Planck Collaboration} et~al., 2020, \mn@doi [\aap]
  {10.1051/0004-6361/201833910}, \href
  {https://ui.adsabs.harvard.edu/abs/2020A&A...641A...6P} {641, A6}

\bibitem[\protect\citeauthoryear{{Ramdas}, {Garcia}  \& {Cuturi}}{{Ramdas}
  et~al.}{2015}]{Ramdas2015}
{Ramdas} A.,  {Garcia} N.,   {Cuturi} M.,  2015, \mn@doi [arXiv e-prints]
  {10.48550/arXiv.1509.02237}, \href
  {https://ui.adsabs.harvard.edu/abs/2015arXiv150902237R} {p. arXiv:1509.02237}

\bibitem[\protect\citeauthoryear{{Rybicki}}{{Rybicki}}{2006}]{Rybicki2006}
{Rybicki} G.~B.,  2006, \mn@doi [\apj] {10.1086/505327}, \href
  {https://ui.adsabs.harvard.edu/abs/2006ApJ...647..709R} {647, 709}

\bibitem[\protect\citeauthoryear{Rybicki \& Dell'Antonio}{Rybicki \&
  Dell'Antonio}{1994}]{Rybicki1994}
Rybicki G.~B.,  Dell'Antonio I.~P.,  1994, \mn@doi [ApJ] {10.1086/174170}, 427,
  603

\bibitem[\protect\citeauthoryear{{Rybicki} \& {Lightman}}{{Rybicki} \&
  {Lightman}}{1979}]{RybickiBook}
{Rybicki} G.~B.,  {Lightman} A.~P.,  1979, {Radiative processes in
  astrophysics}.
John Wiley \& Sons, Ltd

\bibitem[\protect\citeauthoryear{{Semelin}, {Combes}  \& {Baek}}{{Semelin}
  et~al.}{2007}]{Semelin2007}
{Semelin} B.,  {Combes} F.,   {Baek} S.,  2007, \mn@doi [\aap]
  {10.1051/0004-6361:20077965}, \href
  {https://ui.adsabs.harvard.edu/abs/2007A&A...474..365S} {474, 365}

\bibitem[\protect\citeauthoryear{{Seon} \& {Kim}}{{Seon} \&
  {Kim}}{2020}]{Seon2020}
{Seon} K.-i.,  {Kim} C.-G.,  2020, \mn@doi [\apjs] {10.3847/1538-4365/aba2d6},
  \href {https://ui.adsabs.harvard.edu/abs/2020ApJS..250....9S} {250, 9}

\bibitem[\protect\citeauthoryear{{Smith}, {Safranek-Shrader}, {Bromm}  \&
  {Milosavljevi{\'c}}}{{Smith} et~al.}{2015}]{Smith2015}
{Smith} A.,  {Safranek-Shrader} C.,  {Bromm} V.,   {Milosavljevi{\'c}} M.,
  2015, \mn@doi [\mnras] {10.1093/mnras/stv565}, \href
  {https://ui.adsabs.harvard.edu/abs/2015MNRAS.449.4336S} {449, 4336}

\bibitem[\protect\citeauthoryear{{Smith}, {Bromm}  \& {Loeb}}{{Smith}
  et~al.}{2017}]{Smith2017}
{Smith} A.,  {Bromm} V.,   {Loeb} A.,  2017, \mn@doi [\mnras]
  {10.1093/mnras/stw2591}, \href
  {https://ui.adsabs.harvard.edu/abs/2017MNRAS.464.2963S} {464, 2963}

\bibitem[\protect\citeauthoryear{{Smith}, {Tsang}, {Bromm}  \&
  {Milosavljevi{\'c}}}{{Smith} et~al.}{2018}]{Smith2018}
{Smith} A.,  {Tsang} B. T.~H.,  {Bromm} V.,   {Milosavljevi{\'c}} M.,  2018,
  \mn@doi [\mnras] {10.1093/mnras/sty1509}, \href
  {https://ui.adsabs.harvard.edu/abs/2018MNRAS.479.2065S} {479, 2065}

\bibitem[\protect\citeauthoryear{{Smith}, {Ma}, {Bromm}, {Finkelstein},
  {Hopkins}, {Faucher-Gigu{\`e}re}  \& {Kere{\v{s}}}}{{Smith}
  et~al.}{2019}]{Smith2019}
{Smith} A.,  {Ma} X.,  {Bromm} V.,  {Finkelstein} S.~L.,  {Hopkins} P.~F.,
  {Faucher-Gigu{\`e}re} C.-A.,   {Kere{\v{s}}} D.,  2019, \mn@doi [\mnras]
  {10.1093/mnras/sty3483}, \href
  {https://ui.adsabs.harvard.edu/abs/2019MNRAS.484...39S} {484, 39}

\bibitem[\protect\citeauthoryear{{Smith}, {Kannan}, {Garaldi}, {Vogelsberger},
  {Pakmor}, {Springel}  \& {Hernquist}}{{Smith} et~al.}{2022a}]{Smith2022}
{Smith} A.,  {Kannan} R.,  {Garaldi} E.,  {Vogelsberger} M.,  {Pakmor} R.,
  {Springel} V.,   {Hernquist} L.,  2022a, \mn@doi [\mnras]
  {10.1093/mnras/stac713}, \href
  {https://ui.adsabs.harvard.edu/abs/2022MNRAS.512.3243S} {512, 3243}

\bibitem[\protect\citeauthoryear{{Smith} et~al.,}{{Smith}
  et~al.}{2022b}]{Smith2022disk}
{Smith} A.,  et~al., 2022b, \mn@doi [\mnras] {10.1093/mnras/stac2641}, \href
  {https://ui.adsabs.harvard.edu/abs/2022MNRAS.517....1S} {517, 1}

\bibitem[\protect\citeauthoryear{{Song}, {Seon}  \& {Hwang}}{{Song}
  et~al.}{2020}]{Song2020}
{Song} H.,  {Seon} K.-I.,   {Hwang} H.~S.,  2020, \mn@doi [\apj]
  {10.3847/1538-4357/abac02}, \href
  {https://ui.adsabs.harvard.edu/abs/2020ApJ...901...41S} {901, 41}

\bibitem[\protect\citeauthoryear{Tasitsiomi}{Tasitsiomi}{2006a}]{Tasitsiomi2006}
Tasitsiomi A.,  2006a, \mn@doi [ApJ] {10.1086/504460}, 645, 792

\bibitem[\protect\citeauthoryear{{Tasitsiomi}}{{Tasitsiomi}}{2006b}]{Tasitsiomi2006b}
{Tasitsiomi} A.,  2006b, \mn@doi [\apj] {10.1086/505682}, \href
  {https://ui.adsabs.harvard.edu/abs/2006ApJ...648..762T} {648, 762}

\bibitem[\protect\citeauthoryear{{Tomaselli} \& {Ferrara}}{{Tomaselli} \&
  {Ferrara}}{2021}]{Tomaselli2021}
{Tomaselli} G.~M.,  {Ferrara} A.,  2021, \mn@doi [\mnras]
  {10.1093/mnras/stab876}, \href
  {https://ui.adsabs.harvard.edu/abs/2021MNRAS.504...89T} {504, 89}

\bibitem[\protect\citeauthoryear{Unno}{Unno}{1952}]{Unno1952}
Unno W.,  1952, PASJ, 4, 100

\bibitem[\protect\citeauthoryear{{Verhamme}, {Schaerer}  \&
  {Maselli}}{{Verhamme} et~al.}{2006}]{Verhamme2006}
{Verhamme} A.,  {Schaerer} D.,   {Maselli} A.,  2006, \mn@doi [\aap]
  {10.1051/0004-6361:20065554}, \href
  {https://ui.adsabs.harvard.edu/abs/2006A&A...460..397V} {460, 397}

\bibitem[\protect\citeauthoryear{{Yajima}, {Li}, {Zhu}  \& {Abel}}{{Yajima}
  et~al.}{2012}]{Yajima2012}
{Yajima} H.,  {Li} Y.,  {Zhu} Q.,   {Abel} T.,  2012, \mn@doi [\mnras]
  {10.1111/j.1365-2966.2012.21228.x}, \href
  {https://ui.adsabs.harvard.edu/abs/2012MNRAS.424..884Y} {424, 884}

\bibitem[\protect\citeauthoryear{Zheng \& Miralda-Escud\'{e}}{Zheng \&
  Miralda-Escud\'{e}}{2002}]{Zheng2002}
Zheng Z.,  Miralda-Escud\'{e} J.,  2002, \mn@doi [ApJ] {10.1086/342400}, 578,
  33

\bibitem[\protect\citeauthoryear{{Zheng} \& {Wallace}}{{Zheng} \&
  {Wallace}}{2014}]{Zheng2014}
{Zheng} Z.,  {Wallace} J.,  2014, \mn@doi [\apj] {10.1088/0004-637X/794/2/116},
  \href {https://ui.adsabs.harvard.edu/abs/2014ApJ...794..116Z} {794, 116}

\makeatother
\end{thebibliography}




\appendix

\section{Properties of the Sigma Function} \label{app:sigma_function}
Because the Sigma function $\varsigma(w,s)$ enters many of the physical quantities derived in this paper, we explore its properties in detail. We restrict $s$ to be a real number in this discussion. We have been using $\lambda_n = \pi n$, $\chi_n = \sqrt{w^2 + \lambda_n^2}$, and $\bar{\chi}_n = \frac{1}{2} \lambda_n^{-2/3} \chi_n^{-1} [(\chi_n + w)^{1/3} + (\chi_n - w)^{1/3}]$. For reference, we repeat $\varsigma(w,s)$ from Eq.~(\ref{eq:sigma_function}) as
\begin{equation}
    \varsigma(w,s) \equiv \sum_{n=1}^{\infty} \bar{\chi}_n \lambda_n^{-s} = \sum_{n=1}^{\infty} \frac{(\chi_n + w)^{1/3} + (\chi_n - w)^{1/3}}{2 \lambda_n^{s + 2/3} \chi_n} \, .
\end{equation}
Defining $\hat{\chi}_n \equiv \frac{1}{6} \lambda_n^{-2/3} [(\chi_n + w)^{1/3} - (\chi_n - w)^{1/3}]$, the derivative with respect to $w$ is: $\partial \varsigma / \partial w = \sum_{n=1}^{\infty} (\hat{\chi}_n - w \bar{\chi}_n) / \lambda_n^2 \chi_n^2$.

\textbf{Convergence and symmetry:} We begin by noting, for any fixed $s \in \mathbb{R}$, $\varsigma(w,s) = \varsigma(-w,s)$. Using the Integral convergence test, we find that $\varsigma$ converges as long as $s > -1/3$. As each term in the sum is positive, it is also clear that the sub-series $\varsigma_\text{o}$ and $\varsigma_\text{e}$ (consisting of odd and even $n$ terms, respectively) also converge, hence their difference $\varsigma_\text{o} - \varsigma_\text{e} = \sum_{n=1}^{\infty} (-1)^{n - 1} \bar{\chi}_n \lambda_n^{-s}$ converges. Additionally, to validate that the equations for physical quantities in Section~\ref{sec:spherical_clouds} make sense, it is important that $\varsigma_\text{o} > \varsigma_\text{e}$, guaranteeing certain series are strictly positive. In summary: $\varsigma > \varsigma_\text{o} > \varsigma_\text{e} > 0$.

\textbf{Parameter behaviour:} Fig.~\ref{fig:sigma_function_scaling} in the main text illustrates $\varsigma(w,s)$ for several values of $s$, including those used in this paper. In the spherical cloud solutions of Section~\ref{sec:spherical_clouds}, the difference between point and uniform source solutions for a given physical quantity is often an integer offset of 2 in $s$, i.e. $s_\text{uniform} - s_\text{point} = 2$ with perhaps the addition of an alternating sign in the series. It is interesting to note that $\varsigma(w, 0) \gg \varsigma(w, 2) \gg \varsigma(w, 4)$, suggesting that the presence of strong velocity fields has a more pronounced impact when $s$ is smaller (point source case) than when $s$ is larger (uniform case). In the case of uniform emission within some finite sphere inside the cloud, $s$ would lie somewhere between the values of the point and uniform sources, which for a particular emission ratio $r_\text{s}/R$ corresponds to the values of 1 and 3 shown in Fig.~\ref{fig:sigma_function_scaling}. Numerically, we see a steep fall-off in $\varsigma$ with increasing $s$, corresponding to the predictions tested with MCRT in \cite{Nebrin2024}.

\textbf{Limiting behaviour for large $\boldsymbol{w}$:} The scaling of $\varsigma$ for large $|w|$ is of particular interest because it determines the scaling of physical quantities such as the trapping time $t_\text{trap}$ and force multiplier $M_\text{F}$ in rapidly moving media. When $s = 0$, relevant for the point source case, the asymptotic expansion of $\varsigma(w,0)$ as $w \rightarrow \infty$ can be found by converting the series into an integral over $z = n \pi /w$. This is appropriate since $z$ becomes essentially a continuous variable as $w \rightarrow \infty$. Upon evaluating the resulting integral numerically, we find that
\begin{equation}
    \lim_{w\rightarrow\infty} \varsigma(w,0) \approx \dfrac{1.594}{w^{1/3}} \, .
    \label{Sigma function limit s = 0}
\end{equation}
However, this limit only becomes very accurate for extremely large velocities, typically $w \gtrsim 10^4 - 10^5$, far outside the physically relevant range. Thus, a slightly smaller pre-factor can give better results for $w \lesssim 100$ \citep[even this may be outside the range for which the analytical solution is expected to hold, as discussed in appendix A in][]{Nebrin2024}. In Fig.~\ref{fig:sigma_function_scaling} we show that $\varsigma(w,0) \approx 1.52 \,w^{-1/3}$ give a satisfactory fit for $30 \lesssim w \leq \text{few} \times 100$ (and is close enough to the exact asymptotic limit in Eq. \ref{Sigma function limit s = 0} for larger $w$). Physically, this implies that the maximum saturation value for $M_\text{F}$ from a point source is rarely attained, even for extremely large expansion rates.

More robustly, for values of $s > 1/3$, corresponding to sources of non-zero radius, we get the asymptotic limit:
\begin{equation}
    \lim_{w \rightarrow \infty} \varsigma(w,s) \approx  \dfrac{\zeta\left(s + \frac{2}{3} \right)}{ 2^{2/3} \pi^{s + 2/3}} \, w^{-2/3} \, .
\end{equation}
Instances of this are plotted in Fig.~\ref{fig:sigma_function_scaling}, and it is seen that the series approach this asymptote quickly compared to the $s = 0$ case.

\textbf{Limiting behaviour for small $\boldsymbol{w}$:}
A rather non-insightful expansion in terms of Riemann Zeta functions can be used to approximate $\varsigma(w,s)$ for $|w| \lesssim 5$. Define $p = s + 4/3$. Then
\begin{equation}
    \varsigma(w,s) \approx \frac{\zeta(p)}{\pi^p} - \frac{4}{9\pi^{p + 2}}\zeta(p + 2) w^2 + \frac{80}{243\pi^{p + 4}}\zeta(p + 4) w^4 \, .
\end{equation}
Finally, useful relations to validate static solutions ($w = 0$) are
\begin{equation}
    \varsigma(0,s) = \sum_{n=1}^{\infty} \frac{1}{(\pi n)^{s+4/3}} = \pi^{-s-4/3} \, \zeta\left(s + \frac{4}{3}\right) \, ,
\end{equation}
such that $\varsigma_\text{e}(0,s) = 2^{-s-4/3}\,\varsigma(0,s)$ and as always $\varsigma_\text{o} = \varsigma - \varsigma_\text{e}$.

\section{Lab-frame spectra}
\label{app:labframe}

\begin{figure*}
    \centering
    \includegraphics[width=\linewidth]
    {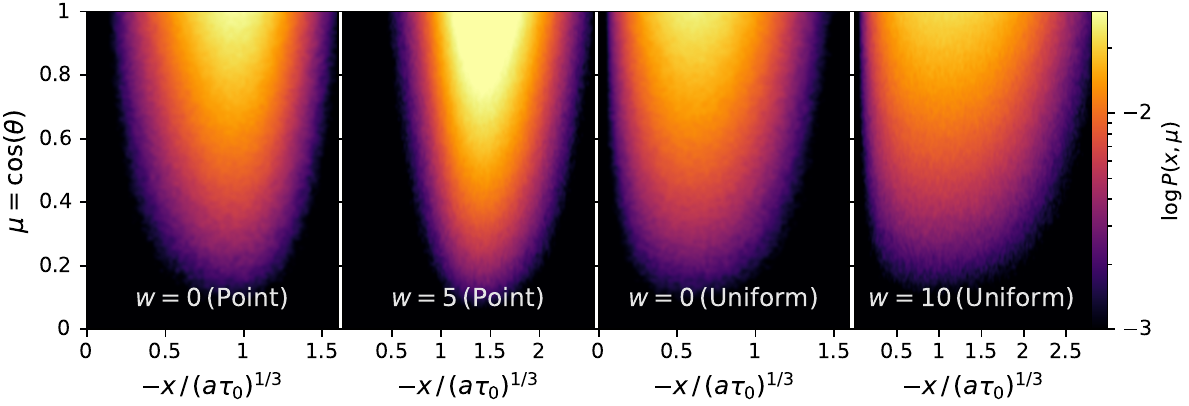}
    \caption{Red flux portions ($x < 0$) of the joint distributions $P(x, \mu)$ for a point source (two leftmost panels) and a uniform source (two rightmost panels) for various expansion parameters $w \equiv 3 V_\text{max} / \sqrt{6} v_\text{th}$. Here $\mu = \text{cos}(\theta)$ is the angle of escape and $x$ is the comoving-frame dimensionless frequency offset normalized by the constant value $(a\tau_0)^{1/3}$ with a temperature $T \approx 9084 \, K$ and $\tau_0 = 5 \times 10^8$ as in Section~\ref{sec:Gridless_MCRT}. The uniform source distribution becomes more broad as the expansion parameter increases, while the point source distribution gets shifted overall.}
    \label{fig:joint_angular_pdf}
\end{figure*}

To compare analytical spectra derived in the comoving frame with MCRT results in the lab frame, one must properly transform the former. The lab frame intensity can be obtained using standard transformation laws \citep[e.g.][]{mihalas2001laboratory}. Let $\boldsymbol{n}_{\rm LF}$ and $\nu_{\rm LF}$ be the direction unit vector and frequency of a photon in the lab frame (LF), respectively. The comoving-frame values are $\boldsymbol{n}$ and $\nu$. Then, the specific intensity in the LF ($I^{\rm LF}_\nu$) and comoving frame ($I_\nu$) are related via $I^{\rm LF}_\nu( \boldsymbol{n}_{\rm LF}) = (\nu_{\rm LF}/\nu)^3 I_\nu(\boldsymbol{n})$, and the solid angle elements by $\text{d}\Omega_{\rm LF} = \text{d}\Omega \, (\nu/\nu_{\rm LF})^2$. Taken together, the mean intensity in the lab frame is
\begin{equation}
    J_\nu^{\rm LF} = \int_{4\pi} \dfrac{\text{d}\Omega}{4 \pi} \, \left(\dfrac{\nu_{\rm LF}}{\nu} \right) I_\nu(\boldsymbol{n}) \, .
\end{equation}
In the non-relativistic limit, the comoving frequency is related to the (fixed) LF frequency via $\nu = \nu_{\rm LF} (1 - \boldsymbol{v} \boldsymbol{\cdot} \boldsymbol{n} / c)$. Assuming radial expansion or contraction, $\boldsymbol{v} = V_\text{max} \boldsymbol{r} / R$, so that $\boldsymbol{v} \boldsymbol{\cdot} \boldsymbol{n} = \mu V_\text{max} r / R$, where $\mu = \cos\theta$. Converting to dimensionless frequency units and defining $\beta = V_\text{max} / v_\text{th}$ gives $x = x_\text{LF} - \beta \mu r / R$. Thus, the LF mean intensity (assuming isotropy) at frequency $x_{\rm LF}$ within the cloud can then be written as:
\begin{equation}
    J_x^{\rm LF}[x = x_{\rm LF}] = \dfrac{1}{2}\int_{-1}^{1} \text{d}\mu \, (1 + \beta \mu) I_x[x = x_\text{LF} - \beta \mu r / R] \, ,
\end{equation}
where $x_{\rm LF}$ is fixed in the integrand. For the emergent flux at the cloud's edge ($r = R$) we drop the small amplitude correction and note that photons emerging with angles $\mu \in (0,1]$ contribute according to $I(\mu) \mu \text{d}\mu$. Hence,
\begin{equation}
    F_x^{\rm LF}[x = x_{\rm LF}] \propto \int_{0}^{1} \text{d}\mu \, P(\mu) J_x[x = x_\text{LF} - \beta \mu] \, ,
\end{equation}
where $P(\mu)$ is the angular distribution of escaping photons. For Thomson scattering in optically thick media, one often takes $P(\mu) = (6/7) \mu (1 + 2 \mu)$, while in the Eddington approximation $P(\mu) = \mu (1 + 3 \mu / 2)$. Both distributions are expected to be good approximations to the true angular distribution in optically thick clouds \citep{Phillips1986, Ahn2002, Tasitsiomi2006b}. In the optically thin limit, however, the escaping intensity is more forward-peaked, eventually tending toward $P(\mu) \rightarrow \delta_{\rm D}(\mu -1)$. The analytical solutions presented in the main text assumed the Eddington approximation, which is expected to become less accurate for larger velocities, where the cloud is effectively optically thin, and $P(\mu)$ becomes strongly forward-peaked.

\begin{figure}
    \centering
    \includegraphics[width=\columnwidth]{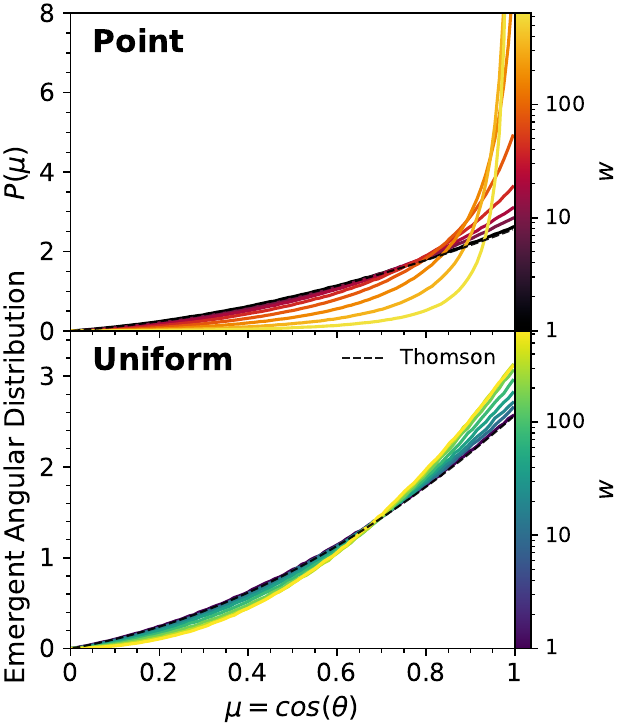}
    \caption{Angular distribution $P(\mu)$ of emerging photons with $\mu \equiv \boldsymbol{\hat{r}} \boldsymbol{\cdot} \boldsymbol{n}$ from the GMCRT runs presented in Section ~\ref{sec:Gridless_MCRT}, plotted for various expansion parameters $w \equiv 3 V_\text{max} / \sqrt{6} v_\text{th}$. We compare to the predictions from Thomson scattering (dashed curves) with almost perfect agreement at lower $w$. For higher velocity gradients, the cloud becomes more free-streaming and photons escape in a more forward-peaked manner.}
    \label{fig:angular_pdf}
\end{figure}

To understand the frequency dependence, in Fig.~\ref{fig:joint_angular_pdf}, we illustrate the joint distributions $P(x, \mu)$ for a point source (two leftmost panels) and a uniform source (two rightmost panels), where $x$ is in the comoving frame before the final Doppler shift to the lab frame. However, the clouds are optically thick so the marginalized distributions $P(\mu|x)$ are nearly independent of frequency. We therefore turn to characterising the velocity dependence.

In Fig.~\ref{fig:angular_pdf}, we confirm these expectations by plotting $P(\mu)$ from our GMCRT simulations for various $w$-values. Nearly static clouds ($w=0$) show excellent agreement with the Thompson approximation, consistent with previous studies \citep[e.g.][]{Ahn2002, Tasitsiomi2006b, Laursen2009, Garavito2014}. As $w$ increases, the angular distribution becomes more forward-peaked. This effect is less pronounced for clouds with uniform sources, since the uniform emission gives rise to a nearly isotropic intensity, more consistent with the Eddington approximation. Thus, the analytical solution for the uniform source remains accurate to higher values of $w$ than the solution for the central point source.

\begin{figure}
    \centering
    \includegraphics[width=\columnwidth]{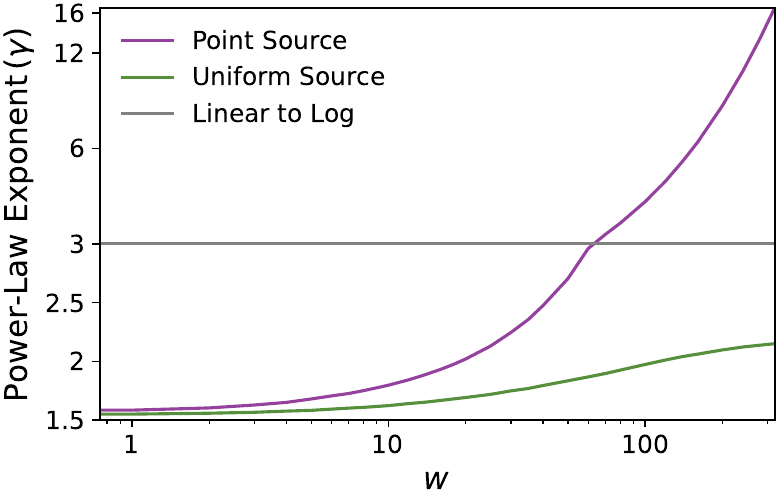}
    \caption{Power-law exponents from a Least Squares fit with the normalised fitting curve $P(\mu) = (1 + \gamma)\,\mu^\gamma$ for the GMCRT angular distributions from Fig.~\ref{fig:angular_pdf} as a function of $w \equiv 3 V_\text{max}/\sqrt{6} v_\text{th}$. As expected, the power-law slope increases gradually for a uniform source, while the growth is exponential for a point source indicating the more rapid transformation into a Dirac Delta function.}
    \label{fig:angular_power_laws}
\end{figure}

From Fig.~\ref{fig:angular_pdf}, it is evident that $P(\mu)$, i.e. marginalized over $x$, is well approximated by a power law. In Fig.~\ref{fig:angular_power_laws}, we present power-law fits to the comoving-frame GMCRT angular distributions using the normalised fitting curve $f_{\gamma}(\mu) = (1 + \gamma) \, \mu^\gamma$ where $\gamma$ is the exponent (or power-law slope) that mainly depends on the expansion parameter $w$. In both cases, the static limit corresponds to Thomson scattering with $\gamma \approx 1.54$. As $w$ increases the point source dependence steepens exponentially, while the uniform case flattens off around $\gamma \approx 2.2$. Overall, these frame transformations and corresponding angular distributions provide intuition and guidance for a more direct comparison between analytic solutions and MCRT observables.

\section{Uniform light-crossing time}
\label{app:sphere_distance}
The uniform source has a reduced average light-crossing time compared to a point source as most photons originate closer to the sphere edge than the centre. To derive the appropriate free-streaming limit, we consider a sphere of radius $R$. We wish to calculate the isotropically-averaged distance $\langle d \rangle$ from the point $P=(0,0,-a)$ to escape the sphere. In spherical coordinates with $(\theta,\phi)$ measured with respect to $P$ towards the origin, the distance to any point on the sphere for a fixed direction is
\begin{equation}
  d(\theta) = a\,\cos\theta + \sqrt{R^2 - a^2\sin^2\theta} \, .
\end{equation}
Thus, the isotropic average over all directions, which after noticing the first term integrates to zero, becomes
\begin{equation}
\begin{aligned}
  \langle d \rangle &= \frac{1}{2} \int_{0}^{\pi} \sqrt{R^2 - a^2 \sin^2\theta}\,\sin\theta\,\text{d}\theta \\
  &= \frac{R}{2} \, + \, \frac{R^2 - a^2}{2\,a}\,
\ln\left(\frac{a + R}{\sqrt{R^2 - a^2}}\right) \, .
\end{aligned}
\end{equation}
This has the limits of $\langle d\rangle \rightarrow \{R, R/2\}$ as $a \rightarrow \{0, R\}$, respectively. Taking the volume average for a source of extent $r_\text{s} < R$ we obtain
\begin{equation}
  \langle d \rangle_\text{s} = \frac{\int_0^{r_\text{s}} \langle d \rangle a^2\,\text{d}a}{\int_0^{r_\text{s}} a^2\,\text{d}a} = \frac{3 R}{8 \tilde{r}_\text{s}^3} \left[ \tilde{r}_\text{s} + \tilde{r}_\text{s}^3 - \left( \tilde{r}_\text{s}^2 - 1 \right)^2 \tan^{-1}\tilde{r}_\text{s} \right] \, ,
\end{equation}
where $\tilde{r}_\text{s} \equiv r_\text{s} / R$. For the uniform source we use $r_\text{s} = R$ and therefore derive the average light-crossing time as $\langle t_\text{light} \rangle = (3 / 4) R / c$.

\section{Cosmological Expansion}
\label{app:cosmological_expansion}
\subsection{Inverse Fourier transform}
Here we show the detailed steps for performing the inverse Fourier transform in Eq.~(\ref{eq:fourier_transform}), i.e. $\tilde{J} = (2\pi)^{-4} \int \mathcal{J} e^{i(\bmath{\eta} \boldsymbol{\cdot} \bmath{\tilde{r}} + \vartheta \tilde{x})} \, \textrm{d}\bmath{r} \, \textrm{d} \vartheta$, where $\bmath{\eta} = (\lambda, \mu, \nu)$ and $\bmath{\tilde{r}} = (\tilde{X}, \tilde{Y}, \tilde{Z})$. Substituting Eq.~(\ref{mathcal J solution}) for $\mathcal{J}$ into this expression yields
\begin{equation}
\begin{aligned}
    \tilde{J} &= \dfrac{1}{(2 \pi)^4} \int \dfrac{e^{i(\bmath{\eta} \boldsymbol{\cdot} \bmath{\tilde{r}} + \vartheta \tilde{x})}}{\bmath{\eta} \boldsymbol{\cdot} \bmath{\eta} + (\vartheta  - \frac{1}{2}i \tilde{\xi} )^2 + (\frac{1}{2} \tilde{\xi} )^2} \, \textrm{d}\bmath{\eta} \, \textrm{d}\vartheta \\
    &= \dfrac{e^{-\frac{1}{2} \tilde{\xi} \tilde{x}}}{(2 \pi)^4 } \int \dfrac{ e^{i(\bmath{\eta} \boldsymbol{\cdot} \bmath{\tilde{r}} + \vartheta \tilde{x})}}{\bmath{\eta} \boldsymbol{\cdot} \bmath{\eta} + \vartheta'^2 + (\frac{1}{2} \tilde{\xi} )^2} \, \textrm{d} \bmath{\eta} \, \textrm{d}\vartheta' \, ,
\end{aligned}
\end{equation}
where we have introduced $\vartheta' = \vartheta - \frac{1}{2}i \tilde{\xi}$. In general, $\vartheta'$ would lie on the line in the complex plane from $-\infty-i\tilde{\xi}/2$ to $+\infty-i\tilde{\xi}/2$. However, the factor $\tilde{\xi}$ is small under most of the situations we consider, so we can treat the integral as effectively on the real axis since it does not add or remove any poles.

Defining the four-vectors $\tilde{\bmath{X}} = (\tilde{X},\tilde{Y},\tilde{Z},\tilde{x})$ and $\bmath{\lambda} = (\lambda,\mu,\nu,\vartheta')$, we can rewrite the expression as
\begin{equation}
    \tilde{J} = \dfrac{e^{-\frac{1}{2}\xi \tilde{x}}}{(2 \pi)^4} \int \dfrac{ e^{i \bmath{\lambda} \boldsymbol{\cdot} \tilde{\bmath{X}} }}{|\bmath{\lambda}|^2 + (\frac{1}{2} \tilde{\xi} )^2} \, \textrm{d} \bmath{\lambda} \, ,
\end{equation}
where the $\bmath{\lambda} \boldsymbol{\cdot} \tilde{\bmath{X}}$ represents the Euclidean inner product. This is a four-dimensional integral over the parameter space $\bmath{\lambda}$. Converting to 4D spherical coordinates with $\tilde{\bmath{X}}$ along the polar axis ($\lambda$ direction), we have $\bmath{\lambda} \boldsymbol{\cdot} \tilde{\bmath{X}} = \varrho |\tilde{\bmath{X}}| \cos \phi_1$. One parametrization is: $\{\lambda, \mu, \nu, \theta'\} = \varrho\,\{\cos \phi_1, \sin \phi_1 \cos \phi_2, \sin \phi_1 \sin \phi_2 \cos \phi_3, \sin \phi_1 \sin \phi_2 \sin \phi_3\}$, where $\varrho \ge 0$, $\{\phi_1, \phi_2\} \in [0,\pi]$ and $\phi_3 \in [0, 2 \pi)$. In these coordinates, the integral reduces to
\begin{equation}
\begin{aligned}
    \tilde{J} &= \dfrac{e^{-\frac{1}{2} \tilde{\xi} \tilde{x}}}{(2 \pi)^4}  \int \dfrac{ e^{i |\tilde{\bmath{X}}| \varrho \cos \phi_1}}{\varrho^2 + (\frac{1}{2} \tilde{\xi} )^2} \varrho^3 \sin^2 \phi_1 \sin \phi_2 \,\, \textrm{d}\varrho\,\textrm{d}\phi_1  \textrm{d}\phi_2  \textrm{d}\phi_3 \\
    & = \dfrac{e^{-\frac{1}{2}\tilde{\xi} \tilde{x}}}{(2 \pi)^3}  \dfrac{ \pi  \tilde{\xi } }{|\tilde{\bmath{X}}|} \mathcal{K}_1 \left(\frac{1}{2} \tilde{\xi}  |\tilde{\bmath{X}}|\right) \, ,\\
\end{aligned}
\end{equation}
where $\mathcal{K}_1$ is the modified Bessel function of the second kind.

\subsection{Normalization}
To compare with observational data, it is useful to have a normalized version of $\tilde{J}$. Based on the discussion in \cite{LoebRybicki1999}, we want the solution to be normalized at each slice with fixed radius so we perform the integral over the spatial volume only. Noting that
\begin{equation}
    \dfrac{\mathcal{K}_1 (z)}{z} = \dfrac{1}{4} \int_0^{\infty} \dfrac{\textrm{d}t}{t^2} \exp \left(-t-\dfrac{z^2}{4 t}\right) \, ,
\end{equation}
we can rewrite the spatial integral (omitting constant factors) as
\begin{equation}
    \begin{aligned}
    &\int_{0}^{\infty} \tilde{r}^2 \,\text{d}\tilde{r} \,e^{-\frac{1}{2}\tilde{\xi} \tilde{x}} \dfrac{\mathcal{K}_1 \left(\frac{1}{2} \tilde{\xi} |\tilde{\bmath{X}}|\right)}{\frac{1}{2} \tilde{\xi}  |\tilde{\bmath{X}}|} \\
    &= \dfrac{e^{-\frac{1}{2}\tilde{\xi}\tilde{x}}}{4} \int_0^{\infty} \tilde{r}^2 \, \text{d}\tilde{r} \,\int_{0}^{\infty} \dfrac{\text{d}t}{t^2}  \exp \left(- t - \tilde{\xi}^2 \, \dfrac{\tilde{r}^2 + \tilde{x}^2}{16 t}\right) \\
    &= \dfrac{4 \pi}{ \tilde{\xi}^3} \exp \left(- \dfrac{1}{2} \tilde{\xi} \tilde{x} - \dfrac{1}{2} \tilde{\xi} |\tilde{x}| \right) \, ,
    \end{aligned}
\end{equation}
where $\tilde{r} = \sqrt{\tilde{X}^2 + \tilde{Y}^2 + \tilde{Z}^2}$, and we have exchanged the order of integration. Since we only consider $\tilde{x} < 0$, this result is independent of $\tilde{r}$. Hence, the final expression is identical to Eq.~(\ref{eq:sec4_Jrx}) with the normalization $(4 \pi)^2 \int_0^{\infty} \tilde{J} \tilde{r}^2\,\text{d}\tilde{r} = 1$.

\subsection{Parameter choices}
To compare with the solution of \cite{LoebRybicki1999}, we adopt standard cosmological parameters \citep[e.g.\ from][]{Planck2020}, including the Hubble constant $H_0$, gravitational constant $G$, primordial mass fraction of hydrogen $X$, and current relative matter and baryon densities $\Omega_{\text{b},0}$ and $\Omega_{\text{m},0}$. In addition, the present-day critical density is $\rho_{\text{crit},0} = 3 H_0^2 / 8 \pi G$, the physical hydrogen number density at redshift $z$ is $n_\text{H}(z) = X \Omega_{\text{b},0} \rho_{\text{crit},0} (1 + z)^3 / m_\text{H}$, and the Hubble parameter (equivalent to $\xi$ in Section~\ref{sec:cosmological_expansion}) at redshift $z$ is $H(z) = H_0 \sqrt{\Omega_{\text{m},0}} (1 + z)^{3/2}$. The Ly$\alpha$ absorption cross-section at line centre ($\nu_0$) is $\sigma_0 = (f_{12} \sqrt{\pi} e^2) / (m_e c \Delta \nu_\text{D})$, with $k_0 = n_\text{\HI} \sigma_0$, and assuming a pre-reionized intergalactic medium gives $n_\text{\HI} \approx n_\text{H}$. Following \cite{LoebRybicki1999}, we define
\begin{equation}
    \alpha \equiv \frac{H(z) \nu_0}{c} \quad \text{and} \quad \beta \equiv \left( \frac{3 c^2 \Lambda^2}{32 \pi^3 \nu_0^2} \right) n_\text{H}(z) \, ,
\end{equation}
which yield the characteristic frequency and radius scales
\begin{equation} \label{eq:star_quantities}
    \nu_* \equiv \frac{\beta}{\alpha} \quad \text{and} \quad r_* \equiv \frac{\beta}{\alpha^2} \, .
\end{equation}
We then express our definitions in a form directly comparable to \citet{LoebRybicki1999}, using a bar above their corresponding quantities. Specifically, with $J = \tilde{J} \sqrt{6} \mathcal{L}/4 \pi$ and noting their $\text{photons~s}^{-1}$ convention, the relationships between our two notations are
\begin{equation}
    \tilde{J} = \frac{\bar{J}}{(k r_\ast)^3} \, , \quad \tilde{r} = k\,r_*\,\bar{r} \, , \quad
    \tilde{\xi} = \frac{\sqrt{6}H(z)}{k v_\text{th}}  \, , \;\; \text{and} \;\;
    x = - \frac{\bar{\nu} \nu_\ast}{\Delta\nu_\text{D}} \,
\end{equation}
where $k = \sqrt\pi \beta/[a(\Delta\nu_D)^2]$ is consistent with $k = k_0 = n_\text{\HI} \sigma_0$ since we assumed constant density. Further details (including constants) are found in the main text.

\subsection{Zero-temperature limit}
Finally, we show that in the zero-temperature limit, our solution in Eq.~(\ref{eq:sec4_Jrx}) is equivalent to the solution of \cite{LoebRybicki1999} in Eq.~(\ref{eq:sec4_LRsol}). Our general expression,
\begin{equation} \label{eq:appendix_Jt(rt,xt)}
    \tilde{J}(\tilde{r},\tilde{x}) = \frac{\tilde{\xi}^2}{32 \pi^3} e^{-\frac{1}{2} \tilde{\xi} \tilde{x}} \frac{\mathcal{K}_1 \left(\frac{1}{2} \tilde{\xi} \sqrt{\tilde{r}^2 + \tilde{x}^2}\right)}{\sqrt{\tilde{r}^2 + \tilde{x}^2}} \, ,
\end{equation}
features the dimensionless parameter $\tilde{\xi}$ that is independent of temperature, since it contains the product of $k v_\text{th}$. To analyze the temperature dependence in other parameters, we recall that $\tilde{r} = k r$ to rewrite the argument of $\mathcal{K}_1$ in Eq.~(\ref{eq:appendix_Jt(rt,xt)}) as $\frac{1}{2} \tilde{\xi} k  \sqrt{r^2 + \tilde{x}^2 / k^2}$, where
\begin{equation}
    \frac{\tilde{x}^2}{k^2} = \frac{1}{k^2} \left(\sqrt{\frac{2 \pi}{27}} \frac{x^3}{a}\right)^2 = \frac{2 \pi }{27} \left[ \frac{2 (\nu - \nu_0)^3}{k \Delta \nu_D^2 \Delta \nu_L} \right]^2 \propto T^{-1} \, ,
\end{equation}
which therefore dominates at low temperatures. Using the asymptotic form at infinity, $\mathcal{K}_1 (z) \sim \sqrt{\pi/2} \, \exp (-z)/\sqrt{z}$, and keeping only the lowest-order terms in $z = \frac{1}{2} \tilde{\xi} |\tilde{x}| \sqrt{1 + (\tilde{r}/|\tilde{x}|)^2} \approx -\tilde{\xi} |\tilde{x}|/2 - \tilde{\xi} \tilde{r}^2 / 4|\tilde{x}|$, where using $|\tilde{x}|$ tracks positivity as $\tilde{x} < 0$, one obtains
\begin{align}
    \tilde{J}(\tilde{r}, \tilde{x}) &\approx \frac{\tilde{\xi}^{3/2}}{32 \pi^{5/2}} \exp \left(- \frac{1}{2} \tilde{\xi} \tilde{x} - \frac{1}{2} \tilde{\xi} |\tilde{x}| - \dfrac{\tilde{\xi} \tilde{r}^2}{4 |\tilde{x}|}\right) ( \tilde{r}^2 + \tilde{x}^2 )^{-3/4} \notag \\
    &\approx \frac{1}{32 \pi^{5/2}} \frac{1}{|\tilde{x}/\tilde{\xi}|^{3/2}} \exp \left(-\frac{\tilde{\xi} \tilde{r}^2}{4|\tilde{x}|}\right) \, .
\end{align}
normalized according to $(4 \pi)^2 \int_0^{\infty} \tilde{J} \tilde{r}^2 \text{d}\tilde{r} = 1$. By carefully comparing the constants in the above expression with the solution in Eq.~(\ref{eq:sec4_LRsol}), one can verify that they are the same. We thus recover the solution of \cite{LoebRybicki1999} by taking the zero-temperature limit.

\section{Calculation of $\bar{\Upsilon}$}
\label{app:fit}

We use the following approximation for the generalized hypergeometric function $x^2{}_2F_2(1,1; \frac{3}{2},2; -x^2) / \sqrt{\pi}$ for $\bar{\Upsilon}(x)$ in Eq.~(\ref{eq:int_upsilon}):
\begin{equation} \label{eq:fit1}
  \frac{A_0 + z (A_1 + \ldots + z (A_{n-1} + z A_n))}{B_0 + z (B_1 + \ldots + z (B_{n-1} + z B_n))} \, ,
\end{equation}
where $z$ is either $|x|$ or $x^2$ depending on the region and the constants $A_i$ and $B_i$ are given in Table~\ref{tab:fit}. The approximation is nearly double precision accurate for all $x$ values, but multiple regions with high-order polynomials are required to achieve this. The final region is an asymptotic expansion with the following form with $z = |x|^{-2}$:
\begin{align}  \label{eq:fit2}
  A_0 &\ln|x| + |x|^{-1}\,e^{-x^2} (A_1 + z (A_2 + z (A_3 + z (A_4 + z (A_5 - A_6 z))))) \notag \\
  &+ B_0 + z (B_1 + z (B_2 + z (B_3 + z (B_4 + z (B_5 - B_6 z))))) \, .
\end{align}

\begin{table}
  \centering
  \caption{Coefficients for the rational function approximation to the generalized hypergeometric function within different regions (see Eqs.~\ref{eq:fit1}--\ref{eq:fit2}).}
  \label{tab:fit}
  \addtolength{\tabcolsep}{3pt}
  \begin{tabular}{lcc}
    \hline
    $i$ & $A_i$ & $B_i$ \\
    \hline
    \multicolumn{3}{l}{\textbf{Region 1:} \quad $0 < x < 2.14$ \quad ($z = x^2$) \quad Eq.~(\ref{eq:fit1})} \\
    \hline
    0 & $0$ & $1$ \\
    1 & $0.56418958354775629$ & $0.48528497539007305$ \\
    2 & $0.085729533651389712$ & $0.11202066087251571$ \\
    3 & $0.022086832488217919$ & $0.016271004013903894$ \\
    4 & $0.0017036307776357589$ & $0.0016562768063564546$ \\
    5 & $0.00018771767997732845$ & $0.00012438873114514399$ \\
    6 & $8.5231641525552708e-6$ & $7.0533651782670942e-6$ \\
    7 & $5.0427978836558623e-7$ & $3.0296362830876420e-7$ \\
    8 & $1.3331831289246041e-8$ & $9.6956364133569300e-9$ \\
    9 & $4.1502346771379203e-10$ & $2.2108600436326525e-10$ \\
    10 & $5.1601562894552471e-12$ & $3.2481690234488706e-12$ \\
    11 & $5.7441158232281185e-14$ & $2.3420543014955824e-14$ \\
    \hline
    \multicolumn{3}{l}{\textbf{Region 2:} \quad $2.14 < x < 5.15$ \quad ($z = |x|$) \quad Eq.~(\ref{eq:fit1})} \\
    \hline
    0 & $-0.042894679076630557$ & $1$ \\
    1 & $0.22824496819198929$ & $-1.4228913485269134$ \\
    2 & $0.025188478146879472$ & $1.422586587092723$ \\
    3 & $-0.073680556438870441$ & $-0.90352979918845972$ \\
    4 & $0.015478791578754775$ & $0.40905729430981564$ \\
    5 & $0.03578546465825914$ & $-0.11897768600690815$ \\
    6 & $-0.027136284977882523$ & $0.020260960808591361$ \\
    7 & $0.00988058128216986$ & $-0.00051357699758117836$ \\
    8 & $-0.0018266571476397632$ & $-0.00042088721875418447$ \\
    9 & $0.00015895087049937769$ & $0.000070884265471591177$ \\
    10 & $2.2012003108278176e-6$ & $4.4276369978816760e-7$ \\
    \hline
    \multicolumn{3}{l}{\textbf{Region 3:} \quad $5.15 < x < 20$ \quad ($z = |x|$) \quad Eq.~(\ref{eq:fit1})} \\
    \hline
    0 & $0.43682128011845283$ & $1$ \\
    1 & $1.1350393218808412$ & $-1.4655446738737176$ \\
    2 & $-2.6886477055342774$ & $-1.4032687362246785$ \\
    3 & $0.38197597791453486$ & $0.88017831100711925$ \\
    4 & $0.50175054988846498$ & $0.20315464152004808$ \\
    5 & $-0.038945006088876756$ & $-0.050541821629539143$ \\
    6 & $-0.01960980532428994$ & $-0.010129700645083951$ \\
    7 & $-0.0012352801187522119$ & $-0.00047171453699007742$ \\
    8 & $-0.000021929464819640861$ & $-6.5774668524211879e-6$ \\
    9 & $-8.9740333784715440e-8$ & $-2.0308831722073883e-8$ \\
    10 & $0$ & $3.0992253950009977e-12$ \\
    \hline
    \multicolumn{3}{l}{\textbf{Region 4:} \quad $x > 20$ \quad ($z = |x|^{-2}$) \quad Eq.~(\ref{eq:fit2})} \\
    \hline
    0 & $0.56418958354775629$ & $0.55389595193643551$ \\
    1 & $0.5$ & $-0.14104739588693907$ \\
    2 & $-0.25$ & $-0.1057855469152043$ \\
    3 & $0.375$ & $-0.17630924485867384$ \\
    4 & $-0.9375$ & $-0.46281176775401883$ \\
    5 & $3.28125$ & $-1.6661223639144678$ \\
    6 & $-14.765625$ & $-7.6363941679413107$ \\
    \hline
  \end{tabular}
  \addtolength{\tabcolsep}{-3pt}
\end{table}



\label{lastpage}
\end{document}